\documentclass[twocolumn,epjc3]{svjour3}
\usepackage{graphics}
\usepackage{microtype}
\usepackage{bbm}
\usepackage{amsmath}
\usepackage{mathtools}
\usepackage{caption}
\usepackage{subcaption}
\usepackage{amssymb}
\usepackage{mathrsfs}       
\usepackage[pdftex]{graphicx}
\usepackage{pgfplots}
\pgfplotsset{width=10cm,compat=1.9}
\usepackage{xargs}       
\usepackage{wasysym}
\usepackage{engrec}
\usepackage{enumerate}
\usepackage{appendix}
\usepackage{empheq}
\usepackage{float}
\usepackage{stmaryrd}
\usepackage[parfill]{parskip}
\usepackage[pdftex,breaklinks]{hyperref}
\usepackage{titletoc}
\usepackage{makecell}
\usepackage{lscape}
\usepackage{algorithm}
\usepackage{algorithmicx}
\usepackage{tensor}
\usepackage{algpseudocode}
\usepackage{blkarray}
\usepackage{multirow}
\usepackage{bigstrut}
\usetikzlibrary{plotmarks}

\newcommand{\inv}[1]{\frac 1{#1}}
\newcommand{\ket}[1]{|#1\rangle}
\newcommand{\bra}[1]{\langle#1|}
\newcommand{\braket}[2]{\langle#1|#2\rangle}
\newcommand{\ud}{\mathrm{d}}
\newcommand{\nucl}[2]{\({}^{#2}\mathrm{#1}\)}
\newcommand{\cg}[6]{C_{#1 #2 #3 #4}^{#5 #6}}
\newcommand{\threeJ}[6]{
  \begin{pmatrix}
    #1&#3&#5\\#2&#4&#6
  \end{pmatrix}
}

\newenvironment{customlegend}[1][]{%
    \begingroup
    \csname pgfplots@init@cleared@structures\endcsname
    \pgfplotsset{#1}%
}{%
    \csname pgfplots@createlegend\endcsname
    \endgroup
}%
\def\addlegendimage{\csname pgfplots@addlegendimage\endcsname}

\newcommand{\addlegendimageintext}[1]{%
    \tikz {
        \begin{customlegend}[
            legend entries={\empty},
            legend style={
                draw=none,
                inner sep=0pt,
                column sep=0pt,
                nodes={inner sep=0pt}}]
        \addlegendimage{#1}
        \end{customlegend}
    }%
}

\newcommand{\full}{\hspace{-2.5mm}
    \raisebox{1pt}{
  \addlegendimageintext{color = black, mark = -, mark options = {scale=2, solid}, line width = 2.5pt}
  }
  \hspace{-4mm}
}
\newcommand{\hosd}{\!
  \addlegendimageintext{color = black, mark = triangle*, mark options = {scale=1, solid}, only marks, line width = 1.5pt}\!
}
\newcommand{\phfb}{\!
  \addlegendimageintext{color = black, mark = x, mark options = {scale=1.5, solid}, only marks, line width = 1.5pt}\!
}
\newcommand{\shfb}{\!
  \addlegendimageintext{color = black, mark = +, mark options = {scale=1.5, solid}, only marks, line width = 1.5pt}\!
}
\newcommand{\mbpt}{\!
  \addlegendimageintext{color = black, mark = star, mark options = {scale=1.5, solid}, only marks, line width = 1.5pt}\!
}
\newcommand{\pgcm}{\!
  \addlegendimageintext{color = black, mark = *, mark options = {scale=1, solid}, only marks, line width = 1.5pt}\!
}

\newcommand{\etmax}{e_{\mathrm{3max}}}

\newcommand{\norm}[1]{\left\lVert#1\right\rVert}

\begin{document}
\title{In-medium $k$-body reduction of $n$-body operators}
\subtitle{A flexible symmetry-conserving approach based on the sole one-body density matrix}
\author{M. Frosini\thanksref{ad:saclay} \and T. Duguet\thanksref{ad:saclay,ad:kul} \and B. Bally\thanksref{ad:dft} \and Y. Beaujeault-Taudi\`ere \thanksref{ad:dam,ad:fakedam} \and J.-P. Ebran\thanksref{ad:dam,ad:fakedam} \and V. Som\`a\thanksref{ad:saclay} }

\institute{\label{ad:saclay}
IRFU, CEA, Universit\'e Paris-Saclay, 91191 Gif-sur-Yvette, France 
\and
\label{ad:kul}
KU Leuven, Department of Physics and Astronomy, Instituut voor Kern- en Stralingsfysica, 3001 Leuven, Belgium 
\and
\label{ad:dft}
Departamento de F\'isica Te\'orica, Universidad Aut\'onoma de Madrid, 28049 Madrid, Spain 
\and
\label{ad:dam}
CEA, DAM, DIF, 91297 Arpajon, France
\and
\label{ad:fakedam}
Universit\'e Paris-Saclay, CEA, Laboratoire Mati\`ere en Conditions Extr\^emes, 91680, Bruy\`eres-le-Ch\^atel, France}

\date{Received: \today{} / Revised version: date}

\maketitle
%
%
\begin{abstract}
The computational cost of ab initio nuclear structure calculations is rendered particularly acute by the presence of (at least) three-nucleon interactions. This feature becomes especially critical now that many-body methods aim at extending their reach beyond mid-mass nuclei. Consequently, state-of-the-art ab initio calculations are typically performed while approximating three-nucleon interactions in terms of effective, i.e. system-dependent, zero-, one- and two-nucleon operators. While straightforward in doubly closed-shell nuclei, existing approximation methods based on normal-ordering techniques involve either two- and three-body density matrices or a symmetry-breaking one-body density matrix in open-shell systems. In order to avoid such complications, a simple, flexible, universal and accurate approximation technique involving the convolution of the initial operator with a sole symmetry-invariant one-body matrix is presently formulated and tested numerically. Employed with a low-resolution Hamiltonian, the novel approximation method is shown to induce errors below $2-3\%$ across a large range of nuclei, observables and many-body methods.
\end{abstract}

\section{Introduction}
\label{intro}

Dealing fully with three-, possibly four-, nucleon interactions is non-trivial but tractable in a self-consistent mean-field Hartree-Fock (HF) or Hartree-Fock Bogoliubov (HFB) calculation. However, it becomes extremely cumbersome, if not impossible, beyond a certain nuclear mass when solving the $A$-body  Schr{\"o}dinger equation to good-enough accuracy beyond the mean field. Consequently, ab initio calculations of mid-mass nuclei are typically performed on the basis of the so-called \textit{normal-ordered two-body} (NO2B) \textit{approximation} that captures dominant effects of three-nucleon interactions while effectively working with two-nucleon operators~\cite{RoBi12,Gebrerufael:2015yig}. In large-scale no-core shell-model calculations, the error induced by the NO2B approximation of the Hamiltonian was estimated to be of the order of $1$-$3\%$ up to the oxygen region\footnote{For low-resolution Hamiltonians obtained via, e.g., the application of similarity renormalization group (SRG) transformations~\cite{Bogner:2009bt}, the efficiency of the NO2B approximation can be understood on the basis of phase-space arguments in the calculation of homogeneous infinite nuclear matter~\cite{Dyhdalo:2017gyl}. In particular, the analysis of Ref.~\cite{Dyhdalo:2017gyl} makes clear that the quality of the approximation can only improve as the density (mass)  of matter (nuclei) increases.}.

The NO2B approximation was originally designed by normal ordering the Hamiltonian with respect to a Slater determinant through standard Wick's theorem~\cite{Wick50theorem}. The procedure involved the contraction of the three-body operator with the {\it one-body} density matrix of that product-state Slater determinant. The approximate Hamiltonian resulting from the NO2B approximation was consistently employed in many-body methods applicable to closed-shell nuclei expanding the exact solution with respect to a {\it symmetry-conserving}\footnote{In the present work, a symmetry-conserving state represents a state whose associated one-body density matrix is symmetry-invariant, i.e. belongs to the {\it trivial} irreducible representation of the symmetry group of the Hamiltonian. While for the SU(2) group it makes necessary for the many-body state itself to be symmetry invariant, i.e. to be a $J=0$ state, for the U(1) group this condition is automatically satisfied for the {\it normal} one-body density matrix.}, e.g. $J^{\Pi}=0^+$, Slater determinant. In this context, the approximate NO2B Hamiltonian naturally displays the same symmetries as the original one. However, the naive extension of the NO2B approximation to methods applicable to open-shell systems via the use of symmetry-breaking reference states poses a difficulty in that respect. Indeed, ignoring the normal-ordered three-body term delivers in such a situation an approximate operator that itself \emph{explicitly} breaks the corresponding symmetry(ies) of the full Hamiltonian. This feature is unwelcome as it lacks the transparency of restricting the symmetry breaking to approximations of the many-body state, especially in view of the eventual restoration of the symmetry(ies). 

Within the frame of many-body methods breaking U(1) symmetry associated with particle-number conservation \cite{Soma11GGFform,Soma:2020dyc,Sign14BogCC,Tichai18BMBPT,Tichai2020review} via the use of Bogoliubov reference states, \linebreak[4] a \textit{particle-number-conserving} normal-ordered $k$-body \linebreak[4] (PNOkB) approximation of an arbitrary $n$-body operator was recently formulated and validated numerically \cite{Ripoche2020}. Using the PNOkB approximation, ab initio calculations on singly open-shell nuclei based on U(1)-breaking and restored formalisms can thus be safely performed. As one currently becomes interested in methods (further) breaking SU(2) symmetry associated with angular-momentum conservation \cite{Novario:2020kuf,frosini20a} to describe doubly open-shell nuclei, the symmetry-conserving NOkB approximation should be extended to this symmetry group, which happens to be neither easy nor transparent. 

The difficulty is bypassed from the outset when describing open-shell systems through a so-called multi-reference method based on an explicitly correlated and symmetry-conserving reference state, e.g. in the multi-reference in-medium similarity renormalization group method~\cite{Herg16PR,Yao:2019rck}. In this context, it is natural to approximate the three-body interaction through its normal-ordering with respect to the correlated reference state on the basis of Kutzelnigg-Mukherjee's Wick theorem~\cite{kut97a}. The benefit however comes with the prize of having to contract the three-body operator not only with the one-body, but also with the two-body and three-body density matrices of the correlated state. A somewhat similar situation occurs within self-consistent Green's function (SCGF) theory that can be formulated in terms of effective $k$-body vertices obtained by contracting initial $n$-body operators ($n \geq k$) with fully correlated $(n-k)$-body density matrices~\cite{Carbone13}.

In conclusion, several approaches exist to produce so-called {\it effective}, i.e. nucleus-dependent, interactions. The aim is to eventually discard the effective operator(s) of highest $n$-body character(s) whose contribution to, e.g., ground-state energies is (are) expected\footnote{Once again, this property can be justified for low-scale Hamiltonians on the basis of phase-space arguments~\cite{Dyhdalo:2017gyl}.} to be much smaller than for the original operator(s) carrying the same $n$-body character(s). Such a procedure always involves a contraction of the original operator(s) with a (set of) density matrix (matrices) reflecting (i) the symmetries and (ii) the correlations of the many-body state it (they) originates from and that is typically the reference state or the fully correlated state at play in the many-body method of interest. 

In this context, the present work introduces a novel method to build a set of effective $k$-body interactions in view of approximating the initial Hamiltonian. While the Hamiltonian is indeed our primary target, the procedure can in principle be applied to any observable. Our goal is thus to formally justify and test numerically a novel approximation method that 
\begin{enumerate}
\item only invokes contractions with a one-body density matrix,
\item uses a symmetry-invariant one-body density matrix,
\item is flexible regarding the many-body state used to compute that one-body density matrix,
\item re-expresses the approximate Hamiltonian in normal-ordered form with respect to the particle vacuum.
\end{enumerate}
The benefits are that 
\begin{enumerate}
\item the method does not involve $l$-body density matrices with $l>1$,
\item the approximate Hamiltonian resulting from omitting certain effective $k$-body terms always possesses the same symmetry group as the original one,
\item the method does not necessarily have to employ the one-body density matrix associated with the many-body (reference) state at play in the method used to solve Schr\"odinger's equation,
\item the resulting Hamiltonian is explicitly expressed in the original single-particle basis such that it can naturally be employed as the starting point of any many-body method.
\end{enumerate}

Points (3) and (4) underline the fact that the approximation of the Hamiltonian and the resolution of the Schr\"odinger equation, although not unrelated, constitute two different problems and do not necessarily have to be dealt with on the basis of the same many-body scheme. 

Per se, the method is applicable independently of the closed- or (doubly) open-shell character of the system as well as of the ground or excited nature of the targeted state. Still, point (2) implies that only one-body densities deriving from a $J^{\Pi} = 0^{+}$ state can be employed in the approximation procedure, which obviously implies that the employed one-body density matrix does not necessarily derive from the targeted state/nucleus. With excited states of even-even nuclei in mind, one can most naturally approximate the Hamiltonian through the use of a one-body density matrix associated with the ground state. With odd-even or odd-odd systems in mind, one can employ the symmetry-invariant density matrix associated with a {\it fake} odd system described in terms of, e.g., a statistical mixture~\cite{Duguet01a,PerezMartin08a}. In the present paper, the focus is on even-even systems.

The paper is organized as follows. Sec.~\ref{sec:1} is dedicated to the formulation of the method and its relation to existing ones. After explicating in Sec.~\ref{methodsanddensities} the hierarchy of one-body density matrices and many-body methods presently employed to test the approximation method, the corresponding numerical results are presented in Sec.~\ref{sec:res}. While conclusions are provided in Sec.~\ref{conclusions}, several appendices complement the paper with useful technical details.

\section{Formalism}
\label{sec:1}

\subsection{Definitions}

\subsubsection{Operators}

An arbitrary particle-number-conserving operator \(O\) can be written as
\begin{equation} 
	O\equiv\sum_{n=0}^N {O^{nn}} \, , \label{generalop}
\end{equation}
where each \(n\)-body component reads in an arbitrary basis $\{c^{\dagger}_a, c_a\}$ of the one-body Hilbert space ${\cal H}_1$ as
\begin{equation} 
	O^{nn}\equiv \frac{1}{n!} \frac{1}{n!}  
	\sum_{\substack{a_1\cdots a_n\\b_1\cdots b_n}}
	o^{a_1\cdots a_n}_{b_1\cdots b_n} \, 
	A^{a_1\cdots a_n}_{b_1\cdots b_n} \, , \label{defnbodyopvacuum}
\end{equation}
where
\begin{equation} 
A^{a_1\cdots a_n}_{b_1\cdots b_n}\equiv c^\dag_{a_1}\cdots c^\dag_{a_n}c_{b_n}\cdots c_{b_1}
\end{equation}
denotes a string of $n$ one-particle creation and $n$ one-particle annihilation operators such that $(A^{a_1\cdots a_n}_{b_1\cdots b_n})^{\dagger}=A_{a_1\cdots a_n}^{b_1\cdots b_n}$. This string is in normal order with respect to the particle vacuum \(\ket 0\), i.e. 
\begin{equation} 
N(A^{a_1\cdots a_n}_{b_1\cdots b_n})=A^{a_1\cdots a_n}_{b_1\cdots b_n} \, ,
\end{equation}
where $N(\ldots)$ denotes the normal ordering with respect to \(\ket 0\).  

In Eq.~\eqref{defnbodyopvacuum}, the $n$-body matrix elements $\{o^{a_1\cdots a_n}_{b_1\cdots b_n}\}$ constitute a mode-$2n$ tensor denoted as $o^{(n)}$, i.e. a data array carrying $2n$ indices associated with the $n$ ($n$) particle creation (annihilation) operators they multiply. The $n$-body matrix elements are fully anti-symmetric under the exchange of any pair of upper or lower indices, i.e.
\begin{equation}
  o^{a_1\cdots a_n}_{b_1\cdots b_n} = \epsilon(\sigma_u) \epsilon(\sigma_l)  \, o^{\sigma_u(a_1\cdots a_n)}_{\sigma_l(b_1\cdots b_n)} \, ,
\end{equation}
where $\epsilon(\sigma_u)$  ($\epsilon(\sigma_l)$) refers to the signature of the permutation $\sigma_u(\ldots)$ ($\sigma_l(\ldots)$) of the $n$ upper (lower) indices.

\subsubsection{Density matrices}

The $l$-body density matrix associated with a many-body state \(\ket \Theta \) constitutes a mode-$2l$ tensor defined through\footnote{Conventionally, Eq.~\eqref{deflbodydensmat} is consistently extended to $l=0$ via $\rho^{(0)\Theta} \equiv 1$.}
\begin{equation}
	\left[{\rho^{(l)\Theta}}\right]^{b_1\cdots b_l}_{a_1\cdots a_l} \equiv \frac{\langle \Theta | A^{a_1\cdots a_l}_{b_1\cdots b_l} | \Theta \rangle}{\langle \Theta | \Theta \rangle} \, . \label{deflbodydensmat}
\end{equation}
In the following, the superscripts $l$ and $\Theta$ are omitted for $l=1$ and when dealing with a generic density matrix, respectively. The elements of $\rho^{(l)\Theta}$ inherit from $A^{a_1\cdots a_l}_{b_1\cdots b_l}$ a full anti-symmetry under the exchange of any pair of upper or lower indices along with a hermitian character, i.e.
\begin{equation}
	\left[{\rho^{(l)\Theta}}\right]^{b_1\cdots b_l}_{a_1\cdots a_l} = \left(\left[{\rho^{(l)\Theta}}\right]_{b_1\cdots b_l}^{a_1\cdots a_l}\right)^{\ast}\, . \label{lbodydensmathermit}
\end{equation}

Given two density matrices $\rho^{(l)\Theta}$ and  $\rho^{(k)\Psi}$, their tensor product 
\begin{align}
\rho^{(l)\Theta\otimes (k)\Psi} &\equiv  \rho^{(l)\Theta} \otimes \rho^{(k)\Psi}
\end{align}
defines a direct-product $(l\!+\!k)$-body density matrix through the mode-$2(l\!+\!k)$ tensor whose elements are given by
\begin{align}
    \left[\rho^{(l)\Theta \otimes (k)\Psi}\right]^{b_1\cdots b_{l+k}}_{a_1\cdots a_{l+k}} &\equiv \left[{\rho^{(l)\Theta}}\right]^{b_1\cdots b_l}_{a_1\cdots a_l} \left[{\rho^{(k)\Psi}}\right]^{b_{l+1}\cdots b_{l+k}}_{a_{l+1}\cdots a_{l+k}} \, ,
\end{align}
and display the hermitian property characterized in Eq.~\eqref{lbodydensmathermit}. Because of the direct-product character of $\rho^{(l)\Theta\otimes (k)\Psi}$, its elements are only partially anti-symmetrized, i.e. they are anti-symmetric under the exchange of any pair of the first $l$ (or last $k$) upper or lower indices.

In case one considers the $m$-fold tensor product of the same $l$-body density matrix  $\rho^{(l)\Theta}$, the notation can be further simplified according to $\rho^{\otimes(ml)\Theta} \equiv \rho^{(l)\Theta} \otimes \ldots \otimes \rho^{(l)\Theta}$. In particular, the $m$-fold tensor product of the generic one-body density matrix $\rho$ defines a mode-$2m$ tensor whose elements are
\begin{align}
    \left[\rho^{\otimes(m)}\right]^{b_1\cdots b_m}_{a_1\cdots a_m} &\equiv  \rho^{b_1}_{a_1} \cdots  \rho^{b_m}_{a_m} \, .
\end{align}
Because of its pure direct-product character, the elements of $\rho^{\otimes(m)}$ display no property under the exchange of any pair of upper or lower indices but inherit the hermitian property characterized by Eq.~\eqref{lbodydensmathermit}.

\subsubsection{Distance}

In the following, the extent to which two one-body density matrices $\rho$ and $\rho'$ deviate from one another will need to be characterized. The distance
\begin{align}
d(\rho,\rho') \equiv \norm{\rho - \rho'} \, , \label{distance}
\end{align}
provides such a diagnostic, with $\norm{.}$ the \emph{Frobenius norm} reading for an arbitrary mode-$n$ tensor $T$ as
\begin{align}
\norm{ T } \equiv \sqrt{\sum_{i_1...i_n} T_{i_1 ... i_n} T^{\ast}_{i_1 ... i_n}}\, ,  \label{norm}
\end{align}
where the superscript denotes elementwise complex conjugation.

\subsubsection{Convolution}

The convolution of the mode-$2n$ tensor $o^{(n)}$ associated with a $n$-body operator $O^{nn}$ with the mode-$2m$ tensor ($m \leq n$) defining a $m$-body density matrix $\rho^{(m)}$ generates the mode-$2(n\!-\!m)$ tensor $o^{(n)}\!\cdot\!\rho^{(m)}$ with elements
\begin{align}
		\left[o^{(n)}\!\cdot\!\rho^{(m)}\right]^{a_1\cdots a_{n-m}}_{
    b_1\cdots b_{n-m}}
    &\equiv \hspace*{-0.30cm} \sum_{
      \substack {a_{n-m+1} , \cdots , a_n \\ b_{n-m+1} , \cdots , b_n}} \hspace*{-0.35cm}
		o^{a_1\cdots a_n}_{b_1\cdots b_n} \left[\rho^{(m)}\right]^{b_{n-m+1}\ldots b_{n}}_{a_{n-m+1}\ldots a_{n}}  \, .
\end{align}
The tensor  $o^{(n)}\!\cdot\!\rho^{(m)}$ is obviously a pure number whenever $m=n$ and nothing but the initial tensor $o^{(n)}$ whenever $m=0$. 

Given two density matrices $\rho^{(l)\Theta}$ and  $\rho^{(k)\Psi}$, it is straightforward to check that the convolution is such that the following identity holds
\begin{align}
 \left(o^{(n)}\!\cdot\!\rho^{(m)\Theta} \right) \!\cdot\!  \rho^{(l)\Psi} &=  \left(o^{(n)}\!\cdot\!  \rho^{(l)\Psi}  \right) \!\cdot\! \rho^{(m)\Theta}\nonumber \\
 &=o^{(n)}\!\cdot\! \left(\rho^{(m)\Theta} \otimes \rho^{(l)\Psi}\right)\, . \label{propconvolution}
\end{align}

\subsection{Standard NOkB approximation}

\subsubsection{Wick's theorem}

Let us consider a symmetry-conserving product state \(\ket \Phi\), i.e. a $J^{\Pi} = 0^{+}$ Slater determinant. Standard Wick's theorem~\cite{Wick50theorem} with respect to \(\ket \Phi\) entails four elementary contractions
\begin{subequations}
\label{contractions1}
\begin{align}
c^\dag_{a}c^\dag_{b} - :c^\dag_{a}c^\dag_{b}: &= 0 \, , \label{contractions1A} \\
c^\dag_{a}c_{b} - :c^\dag_{a}c_{b}: &= {\rho^{\Phi}}^{b}_{a} \, , \label{contractions1B} \\
c_{a}c^\dag_{b} - :c_{a}c^\dag_{b}: &= \delta_{ab}-{\rho^{\Phi}}^{a}_{b} \, , \label{contractions1C} \\
c_{a}c_{b} - :c_{a}c_{b}: &= 0 \, , \label{contractions1D}
\end{align}
\end{subequations}
where $:\ldots:$ denotes the normal ordering with respect to \(\ket \Phi\). 

Applying Wick's theorem, the operator $O$ defined in Eq.~\eqref{generalop} is rewritten as
\begin{equation}
\label{eq:NOop}
  O = \sum_{k=0}^N \mathbf O^{kk}[\rho^{\Phi}] \, ,
\end{equation}
where $\mathbf O^{kk}[\rho^{\Phi}]$ is a \(k\)-body operator in normal-ordered form\footnote{In the present paper, a $k$-body operator and the tensor representing it are written with a standard font (bold font), e.g. $O^{kk}$ ($\mathbf O^{kk}$), if the operator is in normal order with respect to the particle vacuum $| 0 \rangle$ (a many-body state \(\ket \Phi\)).} with respect to \(\ket \Phi\)
\begin{equation} 
	\mathbf O^{kk}[\rho^{\Phi}] \equiv \frac{1}{k!} \frac{1}{k!}  \sum_{
		\substack{a_1\cdots a_k\\b_1\cdots b_k}
		}
	\mathbf o^{a_1\cdots a_k}_{b_1\cdots b_k}[\rho^{\Phi}] \, 
	:A^{a_1\cdots a_k}_{b_1\cdots b_k}: \, . \label{defkbodyopSD}
\end{equation}
Considering $O^{nn}$ ($n\leq N$) and $k\leq n$, there are 
\begin{equation}
(n\!-\!k)!\binom{n}{n-k}\binom{n}{n-k}
\end{equation}
ways to perform $(n\!-\!k)$ non-zero contractions. Consequently, the matrix elements of $\mathbf O^{kk}[\rho^{\Phi}]$ are related to those defining the original contributions to $O$ through
\begin{equation}
\label{eqn:wick}
	\mathbf o^{a_1\cdots a_k}_{b_1\cdots b_k}[\rho^{\Phi}] = \sum_{n=k}^N \inv{(n-k)!} \left[o^{(n)}\!\cdot\!{\rho^\Phi}^{\otimes(n-k)} \right]^{a_1\cdots a_k}_{b_1\cdots b_k}\, .
\end{equation}

\subsubsection{Approximation}

The normal-ordered $k$-body (NOkB) approximation $O^{\text{NOkB}}[\rho^\Phi]$ to the operator $O$ proceeds by truncating the sum in Eq.~\eqref{eq:NOop} to the desired maximum value $k$. While the original operator is obviously independent of $\rho^\Phi$, $O^{\text{NOkB}}[\rho^\Phi]$ does acquire such a dependence as soon as $k<N$. 

For example, the standard NO2B approximation consists of ignoring beyond normal-ordered $2$-body terms to define the approximate Hamiltonian as~\cite{RoBi12,Gebrerufael:2015yig}
\begin{equation}\label{eq:NO2Bop}
	H^{\text{NO2B}}[\rho^\Phi] \equiv 
	\mathbf H^{00}[\rho^\Phi]
	+\mathbf H^{11}[\rho^\Phi]
	+\mathbf H^{22}[\rho^\Phi] \, .
\end{equation}

Generalizing the approach to a U(1)-breaking product state, i.e. a Bogoliubov reference state, standard Wick's theorem gives rise to non-zero {\it anomalous} contractions (Eqs.~\eqref{contractions1A} and~\eqref{contractions1D}) such that the truncation procedure generates a particle-number-breaking operator. A different truncation scheme was thus formulated to design a particle-number conserving normal-ordered $k$-body (PNOkB) approximation in Ref.~\cite{Ripoche2020}. A similar problem arises when using a SU(2) non-invariant Slater determinant, i.e. whenever \(\ket \Phi\) is not a $J^{\Pi} = 0^{+}$ state. Indeed, the standard NOkB approximation delivers an operator that is not rotationally invariant in such a case. Rather than extending the tedious approach designed in Ref.~\cite{Ripoche2020} for the U(1) case, a novel method is proposed in Sec.~\ref{novelapprox} that avoids such complications from the outset by involving  the one-body density matrix stemming from a symmetry-conserving many-body state.  

\subsubsection{Approximate operator in standard form}

Starting from Eq.~\eqref{eq:NOop}, it is interesting to re-express the operator back into a normal-ordered form with respect to the particle vacuum (Eq.~\eqref{defnbodyopvacuum}). Doing so requires to apply Wick's theorem backward, i.e. with respect to \(\ket 0\). To do so, the only required non-zero contraction is given by 
\begin{align}
:c^\dag_{a}c_{b}: - N(:c^\dag_{a}c_{b}:) &=\, :c^\dag_{a}c_{b}: - c^\dag_{a}c_{b} = - {\rho^{\Phi}}^{b}_{a} \, ,
\end{align}
which is nothing but the opposite of the elementary contraction at play in the first step. The original \(n\)-body part of \(O\) is obtained back in terms of the various contributions entering Eq.~\eqref{eq:NOop} such that the connection between their matrix elements is given by
\begin{equation}
\label{eqn:wick_r}
o^{a_1\cdots a_n}_{b_1\cdots b_n}= \sum_{l=n}^N
	\frac{(-1)^{l-n}}{(l-n)!} \left[\mathbf o^{(l)}[\rho^{\Phi}] \! \cdot \! {\rho^\Phi}^{\otimes(l-n)}\right]^{a_1\cdots a_n}_{b_1\cdots b_n} \, . 
\end{equation}
Truncating Eq.~\eqref{eq:NOop} according to the NOkB approximation and inserting the result into Eq.~\eqref{eqn:wick_r} delivers the matrix elements of the approximate \(n\)-body part of $O$ in normal order with respect to the particle vacuum.

\subsection{Generalized $k$-body approximation}
\label{novelapprox}

The standard NOkB approximation relies on standard Wick's theorem and is thus strictly defined with respect to a symmetry-conserving many-body {\it product} state. Because this restriction is too severe in open-shell systems, a generalization of the procedure is now envisioned such that the involved one-body density matrix can originate from a more general many-body state.

\subsubsection{Two-step procedure}

Given the operator $O$ and the one-body density matrix \(\rho\) associated with an {\it arbitrary}  $J^{\Pi} = 0^{+}$ state, one first {\it defines} the set of anti-symmetrized matrix elements
\begin{equation}
\label{eqn:wick2}
\mathbf o^{a_1\cdots a_k}_{b_1\cdots b_k}[\rho] \equiv \sum_{n=k}^N \inv{(n-k)!} \left[o^{(n)}\!\cdot\!  {\rho}^{\otimes(n-k)} \right]^{a_1\cdots a_k}_{b_1\cdots b_k} \, ,
\end{equation}
in strict {\it analogy} with Eq.~\eqref{eqn:wick} but relaxing the necessity for the density matrix to originate from a Slater determinant\footnote{The matrix elements introduced in Eq.~\eqref{eqn:wick2} are not obtained through a set of algebraic operations on the original operator but via a straight convolution of tensors. Still, the present procedure could be formulated within the frame of the {\it quasi}-normal ordering of Ref.~\cite{kong10a}, which is itself an extension of Kutzelnigg and Mukherjee's universal normal-ordering involving the sole one-body density matrix. In this context, it becomes possible to associate an actual quasi-normal-ordered operator to the tensor $\mathbf o^{(k)}[\rho]$. However, given that such a quasi-normal-ordered operator is of no use in the present context, there is no need to invoke it.}. 

The key point of the present development relates to the fact that, independently of the nature of $\rho$, the inverse operation embodied by Eq.~\eqref{eqn:wick_r} remains valid in the present context and recovers the original operator's matrix elements, i.e.
\begin{equation}
\label{eqn:nowick_r}
 o^{a_1\cdots a_n}_{b_1\cdots b_n} = \sum_{l=n}^N
	\frac{(-1)^{l-n}}{(l-n)!} \left[\mathbf o^{(l)}\left[\rho\right] \cdot \rho^{\otimes(l-n)}\right]^{a_1\cdots a_n}_{b_1\cdots b_n} \, .
\end{equation}
This identity is proven in App.~\ref{twostepsprocedure}. One can thus conclude that the {\it combined} operations embodied by Eqs.~\eqref{eqn:wick} and~\eqref{eqn:wick_r} are actually valid outside the reach of standard Wick's theorem. Indeed, the two steps are utterly general operations, i.e. tensor products that are inverse from one another, holding independently of the nature of $\rho$ (i.e. whether it stems from a Slater determinant or not). Standard Wick's theorem is recovered as a particular case of the general tensor  identities~\eqref{eqn:wick2}-\eqref{eqn:nowick_r}, namely when the one-body density matrix does originate from a Slater determinant.

Thus, the purpose of Eqs.~\eqref{eqn:wick2} and~\eqref{eqn:nowick_r} is to start from the set of tensors defining  each $n$-body contribution to the original operator $O$ in Eqs.~\eqref{generalop}-\eqref{defnbodyopvacuum} and to recover it after having gone through an intermediate set defined in strict analogy with the tensors generated via the single-reference normal ordering. While there is no benefit in applying the two-step procedure \textit{per se}, it ensures that the original operator is exactly recovered when doing so. Based on this property, the method provides a useful way to produce nucleus-dependent approximations to the operator through the truncation of the intermediate set of tensors.

\subsubsection{Approximation}

In close analogy with the NOkB approximation, the \(k\)-body approximation of \( O\) is now introduced. First, the set of tensors defined through Eq.~\eqref{eqn:wick2} is truncated according to
\begin{subequations}
\label{eq:truncO}
  \begin{align}
    {}\mathbf {\bar o}^{(l)}[\rho] &\equiv \mathbf o^{(l)}\left[\rho\right] \text{ for } l \le k \, ,\\
    {}\mathbf {\bar o}^{(l)}[\rho] &\equiv 0 \text{ for } l > k \, .
  \end{align}
\end{subequations}
Second, inserting Eq.~\eqref{eq:truncO} into Eq.~\eqref{eqn:nowick_r} generates the set of tensors ${\bar o}^{(n)}[\rho]$ defining the \(k\)-body approximation of \( O\) in normal order with respect to the particle vacuum according to
\begin{equation} 
	O^{kB}[\rho]\equiv\sum_{n=0}^k	{{\bar O}^{nn}}\left[\rho\right] \, , \label{generalop2}
\end{equation}
where the truncation of the sum naturally derives from Eq.~\eqref{eq:truncO}. While the original operator $O$ is independent of $\rho$, $O^{kB}[\rho]$ does acquire such a dependence as a result of the truncation characterized by Eq.~\eqref{eq:truncO}.

While $O^{kB}$ can be built on the basis of an arbitrarily correlated (symmetry-conserving) state, it does not require the use of $\rho^{(l)}$ with $l>1$. As a result, the procedure is significantly simpler than the one associated with the application of Kutzelnigg-Mukherjee's Wick theorem or the one at play in SCGF theory. The practicality of the approach also relates to the fact that the effective Hamiltonian is expressed in normal-ordered form with respect to the particle vacuum in the  working single-particle basis. As a result, $H^{kB}[\rho]$ can be straightforwardly used in place of $H$ as the input to any many-body method of interest.

\subsubsection{Example}

One is typically interested in the \(2\)-body approximation $H^{2B}[\rho]$ of an initial Hamiltonian containing a 3-body interaction
\begin{align}
    H \equiv& T + V + W  \notag\\
    \equiv& \frac{1}{(1!)^2} \sum_{\substack{a_1\\b_1}} t^{a_1}_{b_1} \, A^{a_1}_{b_1} \notag \\
    &+\frac{1}{(2!)^2} \sum_{\substack{a_1a_2\\b_1b_2}} v^{a_1a_2}_{b_1b_2} \, A^{a_1a_2}_{b_1b_2} \notag \\
    &+\frac{1}{(3!)^2}  \sum_{\substack{a_1a_2a_3\\b_1b_2b_3}} w^{a_1a_2a_3}_{b_1b_2b_3} \, A^{a_1a_2a_3}_{b_1b_2b_3} \, ,  \label{originalH}
\end{align}
where $t^{a_1}_{b_1} $ denotes matrix elements of the kinetic energy whereas $v^{a_1a_2}_{b_1b_2}$ and $w^{a_1a_2a_3}_{b_1b_2b_3}$ denote anti-symmetric matrix elements of two- and three-body interactions, respectively. 

Setting
\begin{align*}
O^{00} &\longrightarrow 0 \, , \\
O^{11} &\longrightarrow T \, , \\
O^{22} &\longrightarrow V \, , \\
O^{33} &\longrightarrow W \, , 
\end{align*}
Eq.~\eqref{eq:truncO} gives for $k=2$
\begin{subequations}
\label{eq:NO2Bopterms}
 \begin{align}
\mathbf {\bar h}^{(0)}[\rho] &\equiv t^{(1)} \! \cdot \! \rho + \frac{1}{2!} v^{(2)} \! \cdot \! \rho^{\otimes(2)} + \frac{1}{3!} w^{(3)} \! \cdot \! \rho^{\otimes(3)} \, , \\
\mathbf {\bar h}^{(1)}[\rho] &\equiv t^{(1)} + v^{(2)} \! \cdot \!\rho + \frac{1}{2!} w^{(3)} \! \cdot \! \rho^{\otimes(2)} \, , \\
\mathbf {\bar h}^{(2)}[\rho] &\equiv v^{(2)} + w^{(3)} \! \cdot \!\rho \, , \\
\mathbf {\bar h}^{(3)}[\rho] &\equiv 0\, .
 \end{align}
\end{subequations}
Except for the key fact that $\rho$ does not necessarily relate to a Slater determinant, Eq.~\eqref{eq:NO2Bopterms} is formally identical to Eq.~\eqref{eq:NO2Bop} defining $H^{\text{NO2B}}$.  Inserting Eq.~\eqref{eq:NO2Bopterms} into Eq.~\eqref{eqn:nowick_r}, one eventually obtains the three tensors 
\begin{subequations}
\label{eq:im-int}
\begin{align}
{\bar h}^{(0)}[\rho] &\equiv \frac{1}{3!} w^{(3)} \! \cdot \! \rho^{\otimes(3)} \, , \label{eq:im-intA} \\
{\bar h}^{(1)}[\rho] &\equiv t^{(1)} - \frac{1}{2!} w^{(3)} \! \cdot \!\rho^{\otimes(2)} \, , \label{eq:im-intB}\\
{\bar h}^{(2)}[\rho] &\equiv v^{(2)} + w^{(3)} \! \cdot \!\rho \, , \label{eq:im-intC}
\end{align}
\end{subequations}
defining the normal-ordered contributions to $H^{2B}\left[\rho\right]$ with respect to the particle vacuum, i.e.
\begin{align}
    H^{2B}[\rho] =& {\bar h}^{(0)}[\rho] \notag  \\
    &+\frac{1}{(1!)^2} \sum_{\substack{a_1\\b_1}} {\bar h}^{a_1}_{b_1}[\rho] \, A^{a_1}_{b_1} \notag \\
    &+\frac{1}{(2!)^2} \sum_{\substack{a_1a_2\\b_1b_2}} {\bar h}^{a_1a_2}_{b_1b_2}[\rho] \, A^{a_1a_2}_{b_1b_2}  \, . \label{finalapproxH}
\end{align}
In addition to the fact that, by construction, $H^{2B}[\rho]$ does not contain a three-body operator, its structure differs from the original operator expressed in normal order with respect to the particle vacuum (Eq.~\eqref{originalH}) by the fact that it incorporates the pure number ${\bar h}^{(0)}[\rho]$.

\section{Many-body methods and one-body density matrices}
\label{methodsanddensities}

Equations~\eqref{eq:im-int} define a set of nucleus-dependent 0-, 1- and 2-body operators entering $H^{2B}[\rho]$. As in the NO2B approximation, the inclusion of a large part of $W$ into these effective operators, while treating $T$ and $V$ exactly, gives a clear argument that omitting $\mathbf h^{(3)}[\rho]$ leads to small, hopefully small enough, errors. Still, one is left with the question of the optimal character of the one-body density matrix to be employed for a given system and many-body approximation. 

In the hypothesis that exact eigenstates of $H$ in the $A$-body Hilbert space ${\cal H}_{A}$ are known, one may expect that employing the one-body density matrix of the exact ground-state\footnote{This reasoning has of course a chance to be correct only if the target state $| \Psi \rangle$ is a $J^{\Pi}=0^+$ state. If not, the optimal density matrix cannot be equal to $\rho^{\Psi}$ for symmetry reasons as already briefly discussed in the introduction.} is optimal to reproduce the ground-state energy\footnote{One may further think that the density matrix associated with a statistical symmetry-conserving average of a set of exact low-lying states is optimal to best reproduce the low-lying spectroscopy.}. In fact, this intuition is not correct. 

From a formal viewpoint, it would be interesting to find the optimal one-body density matrix to be used in $H^{2B}[\rho]$ to best reproduce, e.g., the energy associated with the (approximate) ground state $| \Psi \rangle$ of the full Hamiltonian $H$. This would however not be of practical use. Consequently, the numerical results displayed in Sec.~\ref{sec:res} rely on testing a set of trial one-body density matrices while obtaining the solution to the $A$-body problem via various approximation methods. As will be concluded, the results are very robust with respect to the employed one-body density matrix as long as the latter respects a minimal set of properties. 

\subsection{Many-body methods}

The many-body methods presently used to solve the $A$-body Schr\"odinger equation for a collection of doubly closed, singly open-shell and doubly open-shell even-even nuclei (to be specified later on) are
\begin{enumerate}
\item axially deformed Hartree-Fock-Bogoliubov (dHFB) theory~\cite{RiSc80,frosini20a},
\item the particle-number- and angular-momentum-projected HFB (PHFB) method~\cite{RiSc80,Bally21a,frosini20a} based on dHFB states,
\item the projected generator coordinate method (PGCM) \cite{RiSc80,frosini20a} mixing PHFB states along the axial quadrupole moment of the underlying dHFB states,
\item quasi-particle random phase approximation for axially deformed and superfluid nuclei (dQRPA) in the finite amplitude method (FAM) formulation~\cite{naka07,beaujeaulty},
\item deformed Bogoliubov many-body perturbation theory\footnote{Whereas BMBPT has already been applied quite systematically to semi-magic spherical nuclei~\cite{Tichai18BMBPT,Tichai2020review}, it is the first time it is performed on top of a deformed Boboliubov state~\cite{frosini20a} in view of describing doubly open-shell nuclei.} at third order (dBMBPT(3))~\cite{Duguet:2015yle,Arthuis:2018yoo,Demol:2020mzd,frosini20a}.
  
\end{enumerate}
Deformed HFB theory constitutes the mean-field baseline that can capture the bulk of static correlations in open-shell nuclei through the spontaneous breaking of U(1) and SU(2) symmetries. Based on it, PHFB, PGCM and dQRPA on the one hand and dBMBPT on the other hand, provide systematic beyond-mean-field extensions whose aim is to capture many-body correlations. While PHFB, PGCM and dQRPA\footnote{While helpful to discuss the performance of a many-body method, the distinction between dynamical and static correlation effects involve a fuzzy boundary, prominently displayed in the dQRPA case. Namely, the dQRPA equations can be derived within different frames, e.g. as a harmonic limit of the GCM equations~\cite{jancovici64a} or via the linearization of time-dependent HFB equations~\cite{khan04a,avez08a}. Depending on these viewpoints, dQRPA either falls in the category of post-HFB extensions grasping static correlations (associated with fluctuation of shapes), or in the category of beyond-mean-field approaches aiming at capturing dynamical correlations (in terms of 2-quasi-particle excitations). In the present work, we make the arbitrary choice to categorize dQRPA among the former class of approaches.} do so via the addition of static correlations associated with the restoration of broken symmetries and the fluctuation of shapes, dBMBPT targets dynamical correlations through the resummation of elementary, i.e. quasi-particle, excitations. Former approaches are well suited to the description of spectroscopy whereas the latter naturally addresses absolute binding energies and associated ground-state observables.

The string of dHFB, PHFB, PGCM and dQRPA calculations can presently be performed with the full inclusion of three-body forces, i.e. employing a realistic nuclear Hamiltonian $H$ without any form of approximation. Such ab initio calculations are the first of their kind~\cite{frosini20a,beaujeaulty} and allow us to benchmark the approximation of $H$ by $H^{2B}[\rho]$ on the basis of non-trivial many-body methods\footnote{PHFB and PGCM calculations based on realistic chiral Hamiltonians have been performed recently for the first time but at the price of approximating three-body operators~\cite{Yao:2018qjv,Yao:2019rck}. The exact treatment of $W$ in realistic PGCM calculations typically increases the CPU time by three orders of magnitude compared to using $H^{2B}[\rho]$~\cite{frosini20a}.}. While it can be envisioned to do so in the future~\cite{Arthuis:2018yoo}, BMBPT is however not implemented yet with full three-body interactions. Deformed BMBPT calculations are thus presently performed with $H^{2B}[\rho]$ for various approximations to $\rho$ and compared to those done earlier~\cite{Tichai18BMBPT,Tichai2020review} on the basis of the PNO2B approximation~\cite{Ripoche2020}. 

\subsection{Trial one-body density matrices}

Employing the many-body schemes introduced above, the goal is to approximate $H$ by $H^{2B}[\rho]$ with $\rho$ computed from various $J^{\Pi}=0^+$ trial states\footnote{Because correlations captured by QRPA do not feedback into the ground-state, there is no non-trivial one-body density matrix $\rho^\text{sQRPA}$ associated with the spherical QRPA solution to be used in the construction of $H^{2B}[\rho]$.}, i.e. 
\begin{enumerate}
\item spherical harmonic oscillator Slater determinant\footnote{In open-shell nuclei, the invariant density matrix is obtained via the use of the equal filling approximation. This approach can be justified on the basis of a specific {\it statistical mixture} of sHO Slater determinants carrying the appropriate number of particles~\cite{PerezMartin08a} or on the basis of a specific {\it linear combination} of sHO Slater determinants carrying different number of particles such that the linear combination has the correct number of particles on average~\cite{Duguet:2020hdm}.} ($\rho^{\text{sHOSD}}$),
\item spherical HF(B) state ($\rho^{\text{sHF(B)}}$),
\item PHFB state ($\rho^{\text{PHFB}}$),
\item PGCM ground-state ($\rho^{\text{PGCM}}$),
\item standard spherical MBPT\footnote{Standard spherical MBPT denotes many-body perturbation theory based on a spherical Slater determinant reference state rather than on a particle-number-breaking Bogoliubov reference state. The former is automatically obtained from the latter in closed-shell nuclei where the dHFB reference state reduces to the spherical HF Slater determinant.} ground-state ($\rho^{\text{sMBPT}}$).
\end{enumerate}
In the numerical results discussed in Sec.~\ref{sec:res}, $\rho^{\text{sHF(B)}}$, $\rho^{\text{PHFB}}$ and $\rho^{\text{PGCM}}$ are extracted from the correspond calculations performed with the {\it full} $H$. Contrarily,  $\rho^{\text{sMBPT}}$ is obtained from a calculation perform with the PNO2B approximation whereas $\rho^{\text{sHOSD}}$ does not require any a priori calculation.

The two options $\rho^{\text{sHF(B)}}$ and $\rho^{\text{sMBPT}}$ originate from {\it symmetry-restricted} HFB and BMBPT calculations, i.e. {\it spherical} HF(B) ensures the $J^{\Pi}=0^+$ character of the state whereas {\it standard spherical} MBPT ensures particle-number conservation\footnote{While it is not a problem to compute $\rho$ from a particle-number-breaking state carrying the correct particle number on average as in sHFB, it happens that ensuring the correct average particle-number requires a non-trivial procedure in BMBPT beyond HFB~\cite{Demol:2020mzd}. For simplicity, we thus presently limit ourselves to nuclei for which dBMBPT automatically reduces to standard spherical MBPT.}. In the latter case, the restriction implies that the use of $\rho^{\text{sMBPT}}$ is limited to doubly closed-shell nuclei.

While the expression of $\rho^{\text{sHF(B)}}$ is textbook material~\cite{RiSc80}, it is not the case for $\rho^{\text{PHFB}}$ and $\rho^{\text{PGCM}}$. Consequently, the derivation of the corresponding expressions are provided in App.~\ref{app:gcmdens}. For the sake of generality and future use~\cite{frosini20a}, the derivation is actually performed for a more general quantity than presently needed, i.e. App.~\ref{app:gcmdens} provides the expression of the {\it transition} one-body density matrix between two arbitrary initial ($J^{\Pi_i}_i$) and final ($J^{\Pi_f}_f$) PGCM states. The result of present interest is then obtained by setting $J^{\Pi_i}_i=J^{\Pi_f}_f=0^{+}$. While the expression for $\rho^{\text{sMBPT}}$ is known material~\cite{strayer73a,Hoppe:2020elo}, the expression of $\rho^{\text{BMBPT}}$ it presently derives from is not. Consequently, the derivation of  $\rho^{\text{BMBPT}}$ is provided in App.~\ref{app_densmat_BMBPT} for the sake of completeness and future use.

\section{Results}
\label{sec:res}

\subsection{Studied nuclei}

A set of properties (i.e. binding energies, matter radii, low-lying spectra as well as electromagnetic properties) are computed for a panel of representative nuclei using the many-body methods and the one-body density matrices introduced in Sec.~\ref{methodsanddensities}. The panel ranges from light to medium-mass nuclei and contains 
\begin{enumerate}
\item doubly closed-shell (\nucl{O}{16}, \nucl{Ca}{40}),
\item singly open-shell (\nucl{O}{18}),
\item doubly open-shell  (\nucl{Ne}{20}, \nucl{Ne}{30}, \nucl{Mg}{24,40}, \nucl{Ar}{42,48}),
\end{enumerate}
systems. The goal is to cover oblate, spherical and prolate representatives among which some nuclei are {\it soft} and others are {\it hard} with respect to axial deformation\footnote{Some of these nuclei, e.g. \nucl{Mg}{24}, display a triaxial minimum if allowed to. Still, present calculations are restricted to axial symmetry.}.

\subsection{Numerical setting}

The numerical solver allowing us to perform dHFB, PHFB, PGCM, dQRPA and dBMBPT calculations based on full two- and three-nucleon interactions will be detailed in two forthcoming publications~\cite{frosini20a,beaujeaulty}. For the present purpose, it is sufficient to specify that the one-body spherical harmonic oscillator basis is employed.  The finite number of oscillator shells is set by the parameter $e_{\text{max}} \equiv \text{max} (2n + \ell)$, where $n$ and $\ell$ denote the principal quantum number and the orbital angular momentum, respectively. The value of the harmonic oscillator frequency $\hbar\omega$ is further needed to fully characterize the working basis. Except if specified otherwise, all calculations are presently performed with $e_{\text{max}}=8$ and $\hbar\omega=20$ MeV. While these values do not permit to generate fully converged calculations of all the nuclei listed above, the conclusions drawn at the end of the paper are independent of them. 

When representing a $n$-body operator, the natural truncation of the tensor-product basis of the $n$-body Hilbert space is set by $e_{n\text{max}}\equiv ne_{\text{max}}$. One and two-body operators are thus represented using $e_{1\text{max}}=e_{\text{max}}$ and $e_{2\text{max}}=2e_{\text{max}}$, respectively. However, $e_{3\text{max}}=8,10,12$ ($\ll3e_{\text{max}}$) is used to represent the three-nucleon interaction given that employing $3e_{\text{max}}$ is largely beyond today's capacities. This truncation will play a key role regarding the quality of the approximation associated with $H^{2B}[\rho]$ in medium-mass and/or neutron-rich nuclei.

The chiral effective field theory Hamiltonian $H$ presently employed combines a two-nucleon interaction at next-to-next-to-next-to-leading order (N3LO)~\cite{Entem03,Machleidt11} with a N$^2$LO three-nucleon interaction~\cite{Navratil07}. It is then evolved to a lower momentum scale $\lambda_{\text{srg}}$ via SRG transformations. While by default results obtained for $\lambda_{\text{srg}}=1.88$\,fm$^{-1}$ are discussed, $\lambda_{\text{srg}}=2.23 $\,fm$^{-1}$ will also be used for comparison. 

\subsection{Ground-state binding energy}
\label{subsec:res-MF}

\subsubsection{Deformed HFB}
\label{dHFBsection}

\begin{figure}
    \centering
    \includegraphics[width=.5\textwidth]{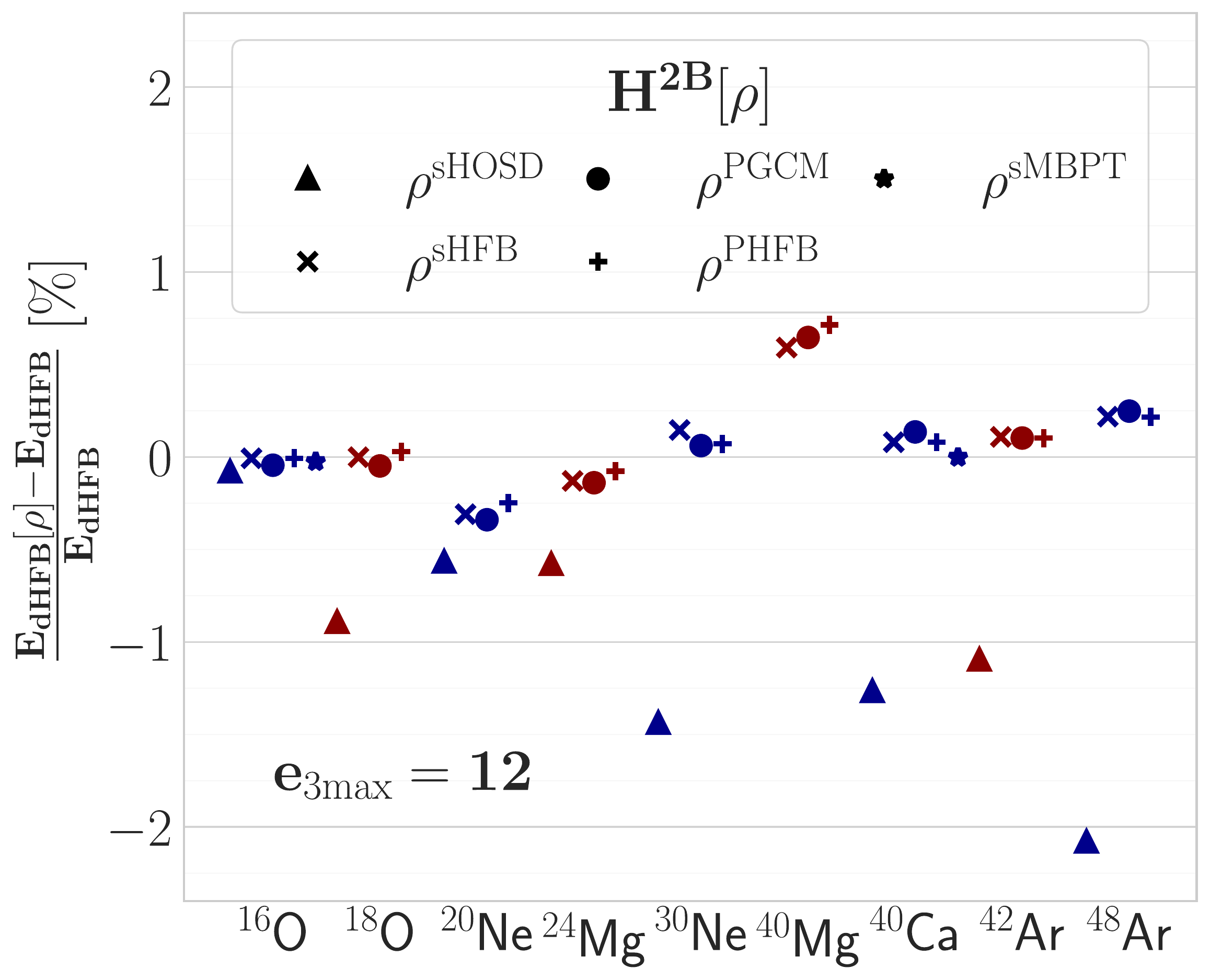}
    \caption{Error (in $\%$) of dHFB ground-state energies obtained with $H^{2B}[\rho]$ for the various test one-body density matrices. The error corresponding to the use of $\rho^\text{sHOSD}$ for $^{40}$Mg amounts to 2.6\% and lies outside the figure. Calculations are performed with $e_{\text{max}}=8$, $e_{3\text{max}}=12$ and $\lambda_{\text{srg}}=1.88$\,fm$^{-1}$.}
    \label{fig:hfb}
\end{figure}

Let us first discuss the use of $H^{2B}[\rho]$ at the mean-field, i.e. dHFB, level. Fig.~\ref{fig:hfb} displays the error (in $\%$) of the corresponding dHFB ground-state energies compared to the reference values obtained from the full $H$. Results are  provided for $\rho = \rho^{\text{sHOSD}}$ (\hosd), $\rho^{\text{sHF(B)}}$ (\shfb), $\rho^{\text{PHFB}}$ (\phfb) and $\rho^{\text{PGCM}}$ (\pgcm), as well as for $\rho = \rho^{\text{sMBPT}}$ (\mbpt) whenever applicable, i.e. in doubly closed-shell nuclei. 

One first observes that $H^{2B}[\rho]$ perform well for the five test one-body density matrices although a notable degradation is visible for $\rho = \rho^{\text{sHOSD}}$. As can be inferred from Tab.~\ref{tab:av-hfb}, the weaker performance of $\rho = \rho^{\text{sHOSD}}$ is systematic but especially pronounced as the mass and/or the isospin-asymmetry of the system increases. While the same trend is at play for $\rho^{\text{sHF(B)}}$, $\rho^{\text{PHFB}}$ and $\rho^{\text{PGCM}}$, the error systematically remains  below $0.2\%$ for these three density matrices, with the exception of $^{40}$Mg whose error lies around $0.7\%$. For $\rho^{\text{sHOSD}}$, the average error over the set is significantly larger ($1.2\%$) throughout the panel and peaks at about $2.6\%$ in $^{40}$Mg (not shown in the figure).

\begin{table*}
    \centering
    \renewcommand{\arraystretch}{1.15}
    \begin{tabular}{l|c|c|c|c|c|c|}
      
     & Closed shell 
     & Open shell
     & Mass \(\le\) 30 
     & Mass \(>\) 30
     & Neutron-rich
     & All
     \\   \cline{1-7}
$\rho^{\text{sHOSD}}$ \bigstrut & 0.67 & 1.32 & 0.71 & 1.76 & 2.04 & 1.18
\\ \cline{1-7}
$\rho^{\text{sHF(B)}}$ \bigstrut & 0.04 & 0.20 & 0.10 &  0.25 &  0.29 & 0.17
\\ \cline{1-7}
$\rho^{\text{PHFB}}$ \bigstrut &  0.04 &  0.21 &  0.09 & 0.28 &  0.33 & 0.17
\\ \cline{1-7}
$\rho^{\text{PGCM}}$ \bigstrut & 0.05 &  0.23 &  0.09 & 0.30 & 0.37 & 0.18 
\\ \cline{1-7}
$\rho^{\text{sMBPT}}$ \bigstrut & 0.04 &  &  &  &  & 
\\ \cline{1-7}
    \end{tabular}
    \caption{Average difference (in \%) between ground-state dHFB energies computed with $H^{2B}[\rho]$ and $H$ for different sub-categories in the test panel and the various test one-body density matrices. The neutron-rich subcategory encompasses \nucl{Ne}{30}, \nucl{Mg}{40} and \nucl{Ar}{48}. See Eq.~\eqref{eq:av-hfb} for the definition of the cost function. Calculations are performed with $e_{\text{max}}=8$, $e_{3\text{max}}=12$ and $\lambda_{\text{srg}}=1.88$\,fm$^{-1}$.}
    \label{tab:av-hfb}
\end{table*}

From a general standpoint, it is not surprising that the error due to the use of $H^{2B}[\rho]$ is small at the mean-field level. To best appreciate this feature, let us focus on doubly closed-shell $^{16}$O and $^{40}$Ca for which dHFB reduces to sHF. As shown in Tab.~\ref{tab:av-hfb}, the error is below $0.05\%$ in these two nuclei for all test one-body density matrices but $\rho^{\text{sHOSD}}$. As explained in App.~\ref{HFfield}, the error would even be strictly zero in such a situation if $\rho$ were equated to the {\it variational} sHF one-body density matrix throughout the sHF iterations based on $H^{2B}[\rho]$. This procedure would be equivalent to working within the NO2B approximation, which is indeed exact at the sHF level, i.e. the NO2B approximation of the Hamiltonian only impacts post-sHF methods by construction. The fact that one rather takes $\rho$ to be a fixed, e.g. $\rho^{\text{sHF}}$ obtained from the full $H$, a priori determined one-body density matrix to build $H^{2B}[\rho]$ induces a marginal error in sHF calculations. 

While the error remains below $0.05\%$ in $^{16}$O and $^{40}$Ca for appropriate density matrices, the distinctly worse result obtained in $^{40}$Ca for $\rho=\rho^{\text{sHOSD}}$ underlines the fact that obtaining a very accurate description is not automatic even in this optimal situation, i.e. it is crucial that the test density matrix contains relevant physical information. Having said that, the results obtained with the other four test density matrices are so similar\footnote{Because sHFB reduces to sHF in doubly closed-shell nuclei, notice that $\rho^{\text{sHFB}}=\rho^{\text{PHFB}}$ in this case such that both test density matrices give identical results by construction.} that no clear characteristic can be easily identified as far as the optimal choice is concerned. Neither the consistency with the employed many-body method nor the degree of correlations encoded in the one-body density matrix seem to constitute a decisive feature. For example, $\rho^{\text{sHFB}}$ performs as well as 
the more advanced $\rho^{\text{sMBPT}}$ that incorporates dynamical correlations beyond the mean field, as can be seen in Tab.~\ref{tab:av-hfb}. We will come back repeatedly to this question throughout the following sections.

\begin{figure}
    \centering
    \includegraphics[width=.5\textwidth]{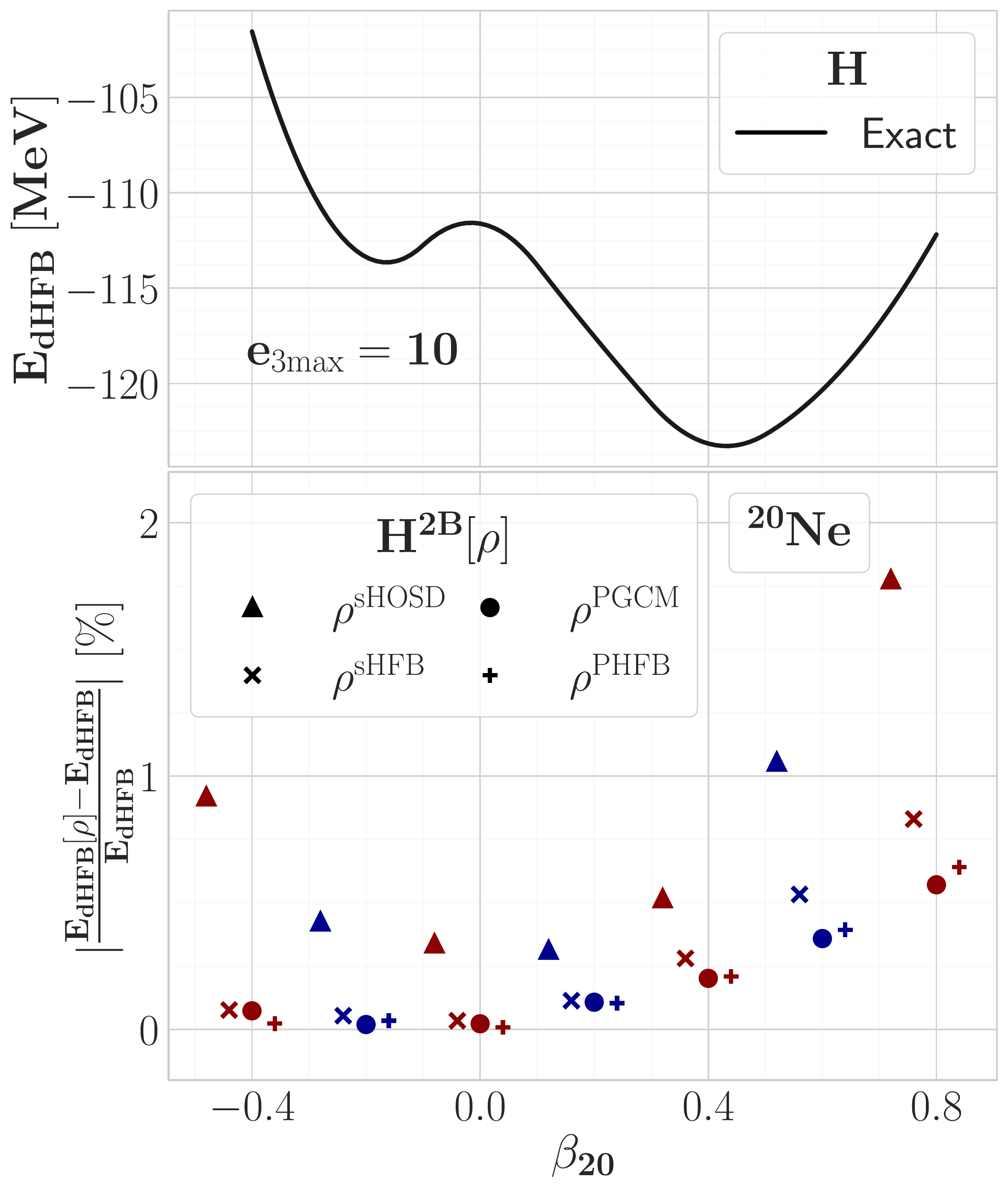}
    \caption{Upper panel: dHFB total energy curve of $^{20}$Ne as a function of the axial quadrupole deformation computed with the full $H$. Lower panel: Error (in $\%$) in the total energy curve when using $H^{2B}[\rho]$ with the various test one-body density matrices. Calculations are performed with $e_{\text{max}}=8$, $e_{3\text{max}}=10$ and $\lambda_{\text{srg}}=1.88$\,fm$^{-1}$.}
    \label{fig:hfb-pes}
\end{figure}

While spherical doubly closed-shell nuclei are particularly amenable to a very accurate description, it is pertinent to investigate the dependence of the approximation on the axial quadrupole deformation of the HFB state. All nuclei in the set but $^{16,18}$O and $^{40}$Ca are doubly open-shell systems and thus spontaneously break rotational symmetry at the dHFB level, $^{20}$Ne and $^{24}$Mg displaying the largest deformation of all.  

The upper panel of Fig.~\ref{fig:hfb-pes} displays the dHFB total energy curve (TEC) calculated in $^{20}$Ne from the full $H$ as a function of the axial quadrupole deformation\footnote{The dimentionless axial quadrupole moment used in the figures is defined as 
\[\beta_{20}\equiv
\frac{4\pi}{3\cdot 1.44A^{\frac53}}
\frac{ \bra\Phi r^2Y_{20} \ket\Phi}{\braket\Phi\Phi}.\]}. This nucleus is significantly deformed, the minimum of the TEC being located at $\beta_{20} = 0.45$. As visible from the lower panel, the error induced by $H^{2B}[\rho]$ is essentially zero at sphericity\footnote{Contrarily to $^{16}$O and $^{40}$Ca, sHFB does not reduce to sHF at sphericity in $^{20}$Ne because neutrons and protons remain superfluid. Consequently, the error due to the use of $H^{2B}[\rho]$ cannot be made strictly equal to zero by any optimization of the test one-body density matrix.}, except for $\rho=\rho^{\text{sHOSD}}$ where it is equal to $0.3\%$, and grows only mildly with the deformation. The error remains below $1\%$ up to a large deformation of $\beta_{20} = 0.8$ ($\beta_{20} = -0.4$) on the prolate (oblate) side\footnote{Note that the edge of the displayed TEC lies 10\,MeV (27\,MeV) above the minimum on the prolate (oblate) side.} for $\rho=\rho^{\text{sHFB}}$, $\rho^{\text{PHFB}}$ or $\rho^{\text{PGCM}}$. For $\rho=\rho^{\text{sHOSD}}$, the error is about twice as large along the TEC.

There exists a trend along the TEC, the results obtained with $\rho^{\text{sHFB}}$ degrading slightly faster with the deformation than those obtained with $\rho^{\text{PGCM}}$ and $\rho^{\text{PHFB}}$. The trend is however not quantitatively significant as can be inferred from the systematic error over open-shell nuclei provided in Tab.~\ref{tab:av-hfb}. Eventually, it is remarkable that all three one-body density matrices give excellent and essentially equivalent results up to large deformations, especially given the fact that $\rho^{\text{sHFB}}$ does not encode any information about deformation properties of $^{20}$Ne. This is a first indication of the robustness of the in-medium 2-body reduction method of 3-body interaction operators.

\begin{figure}
    \centering
    \includegraphics[width=.5\textwidth]{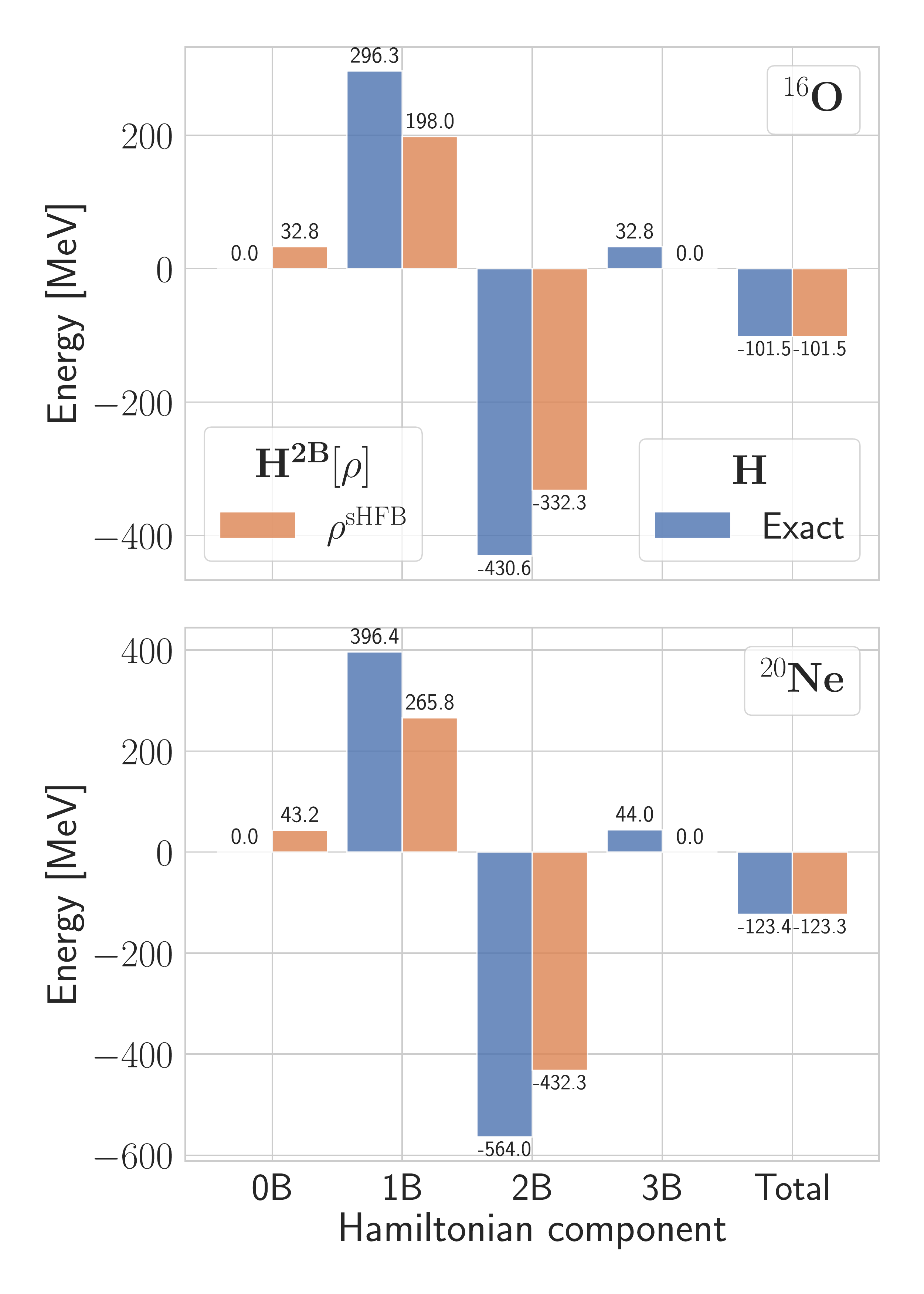}
    \caption{Contributions of the various components of $H^{2B}[\rho]$ and $H$ to the dHFB energy. Upper panel: $^{16}$O. Lower panel: $^{20}$Ne. Calculations are performed with $e_{\text{max}}=6$ and $e_{3\text{max}}=6$.}
    \label{fig:decomposition}
\end{figure}

For orientation, it is interesting to analyze how the various components of $H^{2B}[\rho]$ (Eqs.~\eqref{eq:im-int}-\eqref{finalapproxH}) and $H$ (Eq.~\eqref{originalH}) contribute to the dHFB energy. Figure~\ref{fig:decomposition} decomposes the dHFB total energy accordingly in $^{16}$O and $^{20}$Ne. Results are provided for a schematic model space and $\rho=\rho^{\text{sHFB}}$. Focusing first on $^{16}$O and making the hypothesis that the sHF density matrix is the same in both calculations\footnote{This hypothesis is very well validated in practice, even more so in a small model space such as the one employed in the present calculation.}, the inspection of Eq.~\eqref{eq:im-int} makes clear that (a) the 0-body part of $H^{2B}[\rho]$ is strictly equal to the sHF contribution originating from the three-body interaction in $H$ and that (b) the energy contribution associated with the 1- and 2-body parts of $H^{2B}[\rho]$ originating from the three-body interaction exactly cancel out. These features are indeed observed in the upper panel of Fig.~\ref{fig:decomposition} such that the total sHF energies are identical in both calculations. While this formal analysis does not hold for dHFB in general, the results displayed in the lower panel demonstrate that it remains valid in practice in a well-deformed nucleus such as $^{20}$Ne, which eventually elucidates the high-quality results obtained above over a large set of nuclei.

\subsubsection{Deformed BMBPT}
\label{resultsBMBPT}

While it is satisfying that the error induced by $H^{2B}[\rho]$ is negligible at the mean-field, i.e. dHFB, level, it is to some extent expected and surely not sufficient to claim victory. The performance of $H^{2B}[\rho]$ must thus be tested in beyond mean-field methods where the accurate compensation  observed above between the terms of $H$ and those of $H^{2B}[\rho]$ is not guaranteed to hold. 

\begin{figure}
    \centering
    \includegraphics[width=.5\textwidth]{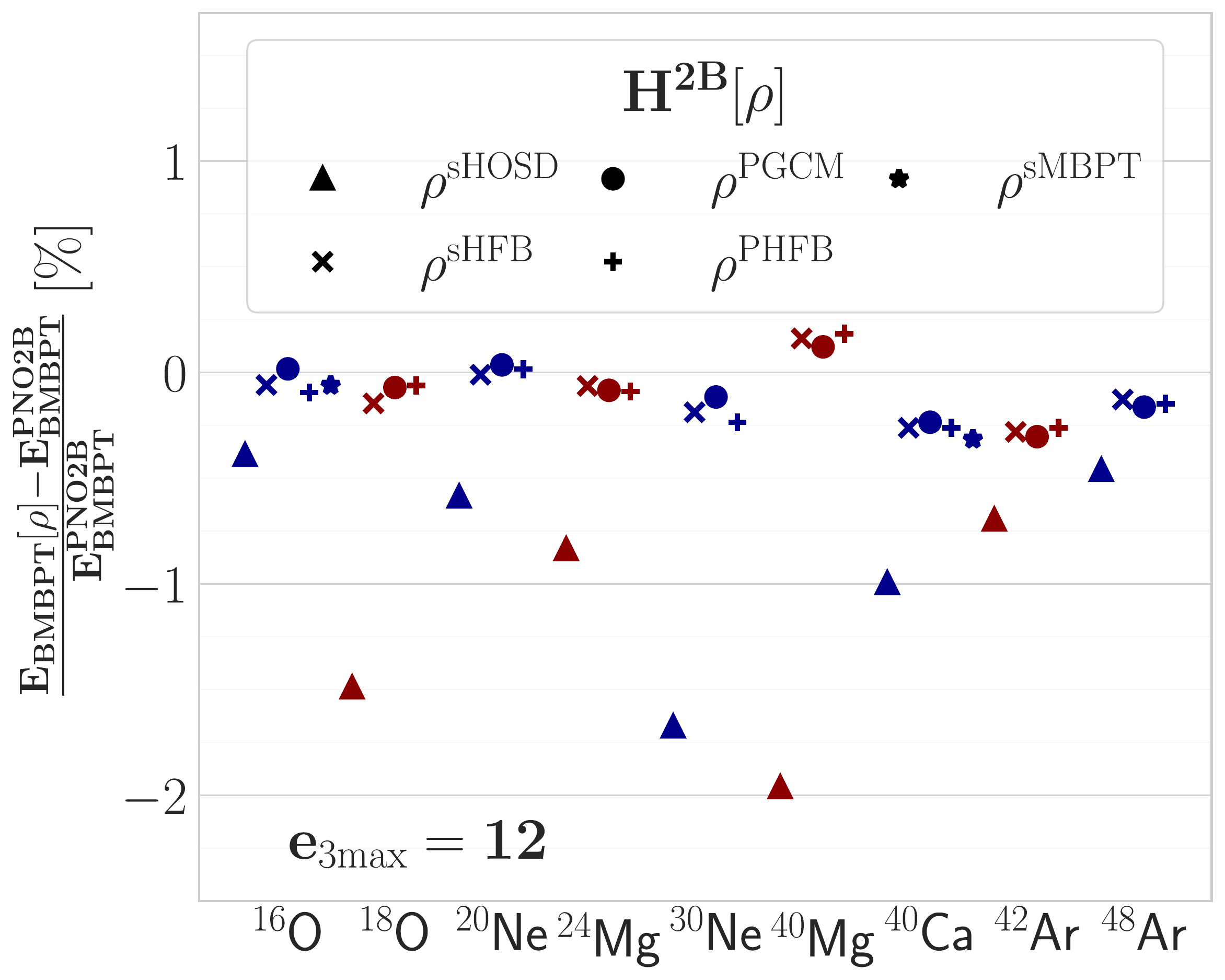}
    \caption{Difference of dBMBPT(3) ground-state energies (in $\%$) obtained with  $H^{2B}[\rho]$ and within the PNO2B approximation of $H$ for the various test one-body density matrices. Calculations are performed with $e_{\text{max}}=8$, $e_{3\text{max}}=12$ and $\lambda_{\text{srg}}=1.88$\,fm$^{-1}$.}
    \label{fig:bmbpt}
\end{figure}

While such a test must be carried out for various ab initio methods, the present section focuses on ground-state energies obtained from dBMBPT that resums dynamical correlations in a perturbative fashion on top of a (possibly deformed and superfluid) HFB state. Present calculations are performed at the BMBPT(3) level that is known to reproduce essentially exact results based on SRG-evolved interactions to better than $2\%$ in oxygen isotopes and those computed from non-perturbative expansion methods to better than $2\%$ in semi-magic nuclei up to the nickel region~\cite{Tichai18BMBPT,Tichai2020review}.

Although envisioned in the future, BMBPT calculations with explicit three-nucleon forces are not available yet. Consequently, calculations with $H^{2B}[\rho]$ are presently benchmarked against those obtained using the PNO2B approximation~\cite{Ripoche2020}, which is the approximation employed so far in all published BMBPT calculations of semi-magic nuclei~\cite{Tichai18BMBPT,Tichai2020review}. In doubly closed-shell nuclei, the PNO2B approximation reduces to NO2B that has itself being benchmarked against the use of full three-body interactions and shown to offer a typical $1-2\%$ accuracy up to $^{16}$O~\cite{RoBi12}. 

\begin{table*}
    \centering
        \renewcommand{\arraystretch}{1.15}
    \begin{tabular}{l|c|c|c|c|c|c|}
      
     & Closed shell 
     & Open shell
     & Mass \(\le\) 30 
     & Mass \(>\) 30
     & Neutron-rich
     & All
     \\   \cline{1-7}
$\rho^{\text{sHOSD}}$ \bigstrut & 0.69 & 1.1 & 1.00 & 1.03 & 1.37 & 1.01
\\\cline{1-7}
$\rho^{\text{sHF(B)}}$ \bigstrut  & 0.16 & 0.14 & 0.09 & 0.21 & 0.16 & 0.14
\\ \cline{1-7}
$\rho^{\text{PHFB}}$ \bigstrut & 0.18 & 0.14 & 0.10 & 0.21 & 0.19 & 0.15
\\ \cline{1-7}
$\rho^{\text{PGCM}}$ \bigstrut & 0.13 & 0.13 & 0.06 & 0.21 & 0.13 & 0.13
\\ \cline{1-7}
$\rho^{\text{sMBPT}}$ \bigstrut  & 0.19 &  &  &  &  & 
\\ \cline{1-7}
    \end{tabular}
    \caption{Average difference (in \%) of ground-state dBMBPT(3) energies obtained with  $H^{2B}[\rho]$ and within the PNO2B approximation of $H$ for different sub-categories in the test panel and the various test one-body density matrices. The neutron-rich subcategory encompasses \nucl{Ne}{30}, \nucl{Mg}{40} and \nucl{Ar}{48}. See Eq.~\eqref{eq:av-bmbpt} for details on the cost function. Calculations are performed with $e_{\text{max}}=8$, $e_{3\text{max}}=12$ and $\lambda_{\text{srg}}=1.88$\,fm$^{-1}$.}
    \label{tab:av-bmbpt}
\end{table*}

Deformed BMBPT(3) binding energy differences (in $\%$) are displayed in Fig.~\ref{fig:bmbpt}. Results produced within both approximations agree to better than $0.3\%$ over the whole set of considered nuclei, except for $\rho = \rho^{\text{sHOSD}}$ where the difference increases up to $2\%$. Just as for the dHFB results discussed above, the use of a one-body density matrix encoding either static or dynamical correlations beyond the mean-field does not have a significant impact on the quality of $H^{2B}[\rho]$ such that the results are essentially equivalent to those obtained with  $\rho=\rho^{\text{sHF(B)}}$. This can be confirmed quantitatively by inspecting the numbers reported in Tab.~\ref{tab:av-bmbpt}. Interestingly, the results also show that the average deviation is independent of the closed- or open-shell character of the nuclei under consideration whereas it slightly increases with the mass even though the deviation remains tiny in all cases.

These remarkable results indicate that the in-medium interaction and PNO2B approximation methods are equivalent as far as quantitative ab initio dBMBPT calculations of mid-mass nuclei are concerned. Given the earlier benchkmarking of the NO2B in doubly-closed shell nuclei, the presently developed in-medium approximation method is well validated in fully-correlated binding energy calculations.

\subsection{PHFB absolute energies and radii}

In the following, we wish to go beyond ground-state energies and test the in-medium approximation method on spectroscopic properties. In order to do so, PHFB, PGCM and dQRPA calculations will be employed. While these techniques resum static correlations associated with the restoration of broken symmetries and the fluctuation of shapes, they do not account for dynamical correlations. 
As a result, whereas \textit{relative} energies and spectroscopic quantities can be well converged and meaningful, \textit{absolute} energies are not realistic, i.e. they are far from converged ab initio values. Still, it is useful to first investigate how these absolute energies differ when computed from $H$ and $H^{2B}[\rho]$.

In this section we thus analyse total (ground- and excited-state) energies obtained at the PHFB level. In addition, corresponding ground-state matter radii are presented. In doing so, the dependence of the results on numerical parameters such as $e_{3\text{max}}$, $e_{\text{max}}$ and $\lambda_{\text{srg}}$ is also investigated.

\subsubsection{\label{sec:systana} Systematic analysis}

\begin{figure*}
    \centering
    \includegraphics[width=\textwidth]{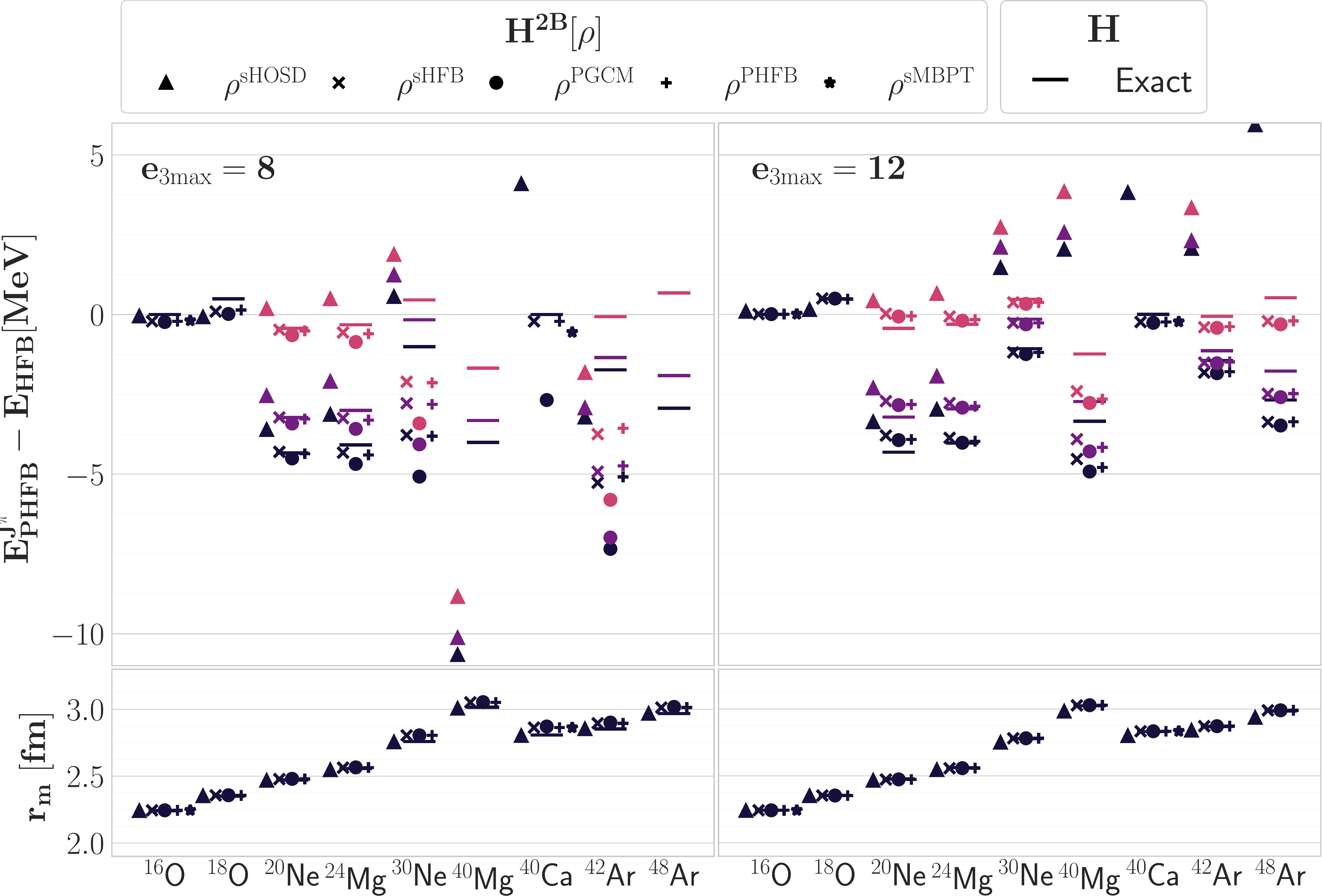}

    \caption{Results of PHFB calculations with $H$ and $H^{2B}[\rho]$ for several test one-body density matrices $\rho$. Left and right panels display results obtained for $e_{3\text{max}}=8$ and $12$, respectively, at fixed $e_{\text{max}}=8$. Upper panel: absolute energies of lowest $J^{\Pi}=0^+,2^+,4^+$ states to which the dHFB energy obtained from $H$ in each nucleus is subtracted. Lower panel: ground-state root-mean-square matter radii. Calculations are performed with $\lambda_{\text{srg}}=1.88$\,fm$^{-1}$.}
    \label{fig:phfb-diff}
\end{figure*}

Upper panels of Fig. \ref{fig:phfb-diff} display binding energies of the lowest-lying $J^{\Pi}=0^+,2^+,4^+$ states obtained via PHFB calculations with $e_{3\text{max}}=8$ and $12$ (at fixed $e_{\text{max}}=8$). The energy of each state obtained from $H$ (\full) is compared to those generated from $H^{2B}[\rho]$ with $\rho = \rho^{\text{sHOSD}}$ (\hosd), $\rho^{\text{sHF(B)}}$ (\shfb), $\rho^{\text{PHFB}}$ (\phfb) and $\rho^{\text{PGCM}}$ (\pgcm), as well as with $\rho = \rho^{\text{sMBPT}}$ (\mbpt) whenever applicable. Energies are shifted by the dHFB value obtained from the full $H$ for the corresponding system such that all nuclei can be displayed on the same figure. 

Reference energies are well reproduced in light nuclei for all test one-body density matrices and both values of \(e_{3\text{max}}\), i.e. absolute deviations remain below 1 MeV until \nucl{Mg}{24}. Increasing the mass and/or isospin asymmetry renders the approximation more and more sensitive to the value of \(e_{3\text{max}}\). Given that ab initio calculations are known to be increasingly more sensitive to \(e_{3\text{max}}\) with the mass and isospin-asymmetry of the system~\cite{Soma:2019bso}, it is not surprising that any approximation of the three-nucleon displays the same feature. Going from $e_{3\text{max}}=8$, through $e_{3\text{max}}=10$ (not shown) and to $e_{3\text{max}}=12$, a clear convergence of the results is observed, although not quite yet for the heaviest and most neutron-rich nuclei of the panel. Eventually, converged results display a similar error in medium-mass nuclei to the one obtained in lighter systems, except for $\rho = \rho^{\text{sHOSD}}$. Below, only results obtained for the largest reachable value of $e_{3\text{max}}$ (typically 12 but not always) are shown.

The lowest panels of Fig. \ref{fig:phfb-diff} display ground-state root-mean-square matter radii (results are similar for excited-states radii). The conclusions are the same as for the energies. Eventually, radii are extremely well reproduced for all nuclei, states and test density matrices, with the exception of $\rho = \rho^{\text{sHOSD}}$ for which a slight underestimation is visible in the heaviest systems.

\begin{table*}
    \centering
    \renewcommand{\arraystretch}{1.15}
    \begin{tabular}{l|c|c|c|c|c|c|}
      
     & Closed-shell 
     & Open shell 
     & Mass \(\le\) 30 
     & Mass \(>\) 30
     & Neutron-rich
     & All
     \\    \cline{1-7}
$\rho^{\text{sHOSD}}$ \bigstrut & 0.67 & 1.36 & 0.62  &1.68 & 2.34 & 1.09
\\ \cline{1-7}
$\rho^{\text{sHF(B)}}$ \bigstrut & 0.04 & 0.22 & 0.13  &0.24 & 0.30 & 0.17
\\ \cline{1-7}
$\rho^{\text{PHFB}}$  \bigstrut & 0.04 & 0.22 & 0.10 & 0.27 & 0.34 & 0.17
\\ \cline{1-7}
$\rho^{\text{PGCM}}$ \bigstrut & 0.05 & 0.24 & 0.11 & 0.29 & 0.38 & 0.19
\\ \cline{1-7}
$\rho^{\text{sMBPT}}$  \bigstrut & 0.04 &        &   &   &  & 
\\ \cline{1-7}
    \end{tabular}
    \caption{Average error (in \%) on absolute PHFB energies of low-lying $J^{\Pi}=0^+, 2^+, 4^+$ and $6^+$ states for different sub-categories in the test panel and the various test one-body density matrices. Calculations are performed with $e_{3\text{max}}=8$, $e_{3\text{max}}=12$ and  $\lambda_{\text{srg}}=1.88$\,fm$^{-1}$. See Eq. ~\eqref{eq:av-phfb} for details on the cost function.}
    \label{tab:av-phfb}
\end{table*}

Focusing on the right panels of Fig. \ref{fig:phfb-diff}, one does notice that the situation regarding the performance of the test one-body density matrices is qualitatively and quantitatively similar to the one encountered in dHFB and dBMBPT(3) calculations.  As soon as the results are converged with respect to \(\etmax\), PHFB energies and radii obtained with $H^{2B}[\rho]$ reproduce the reference results equally well with all employed one-body density matrices but $\rho^{\text{sHOSD}}$, i.e. it seems necessary (compared to $\rho^{\text{sHOSD}}$) and sufficient (compared to $\rho^{\text{PHFB}}$, $\rho^{\text{PGCM}}$ and $\rho^{\text{sMBPT}}$) to employ a test one-body density matrix encoding the information of the spherical mean-field, i.e.  $\rho^{\text{sHF(B)}}$.

The above analysis is put in more quantitative terms via the computation of systematic errors. Corresponding results are shown in Tab.~\ref{tab:av-phfb}. By construction, PHFB results are identical to sHF ones in doubly closed-shell nuclei given that the sole $0^+$ ground-state has been considered for these nuclei and given that the projections on particular number and angular momentum are superfluous for a sHF state. In the other nuclei where the projections typically add few MeV of correlations energy to the ground state, the average error over $J^{\Pi}=0^+, 2^+, 4^+,6^+$ PHFB states is essentially the same as for dHFB ground-state energies, independently of the test one-body density matrix.

\subsubsection{Dependence on $e_{\text{max}}$}

\begin{figure}
    \centering
    \includegraphics[width=.5\textwidth]{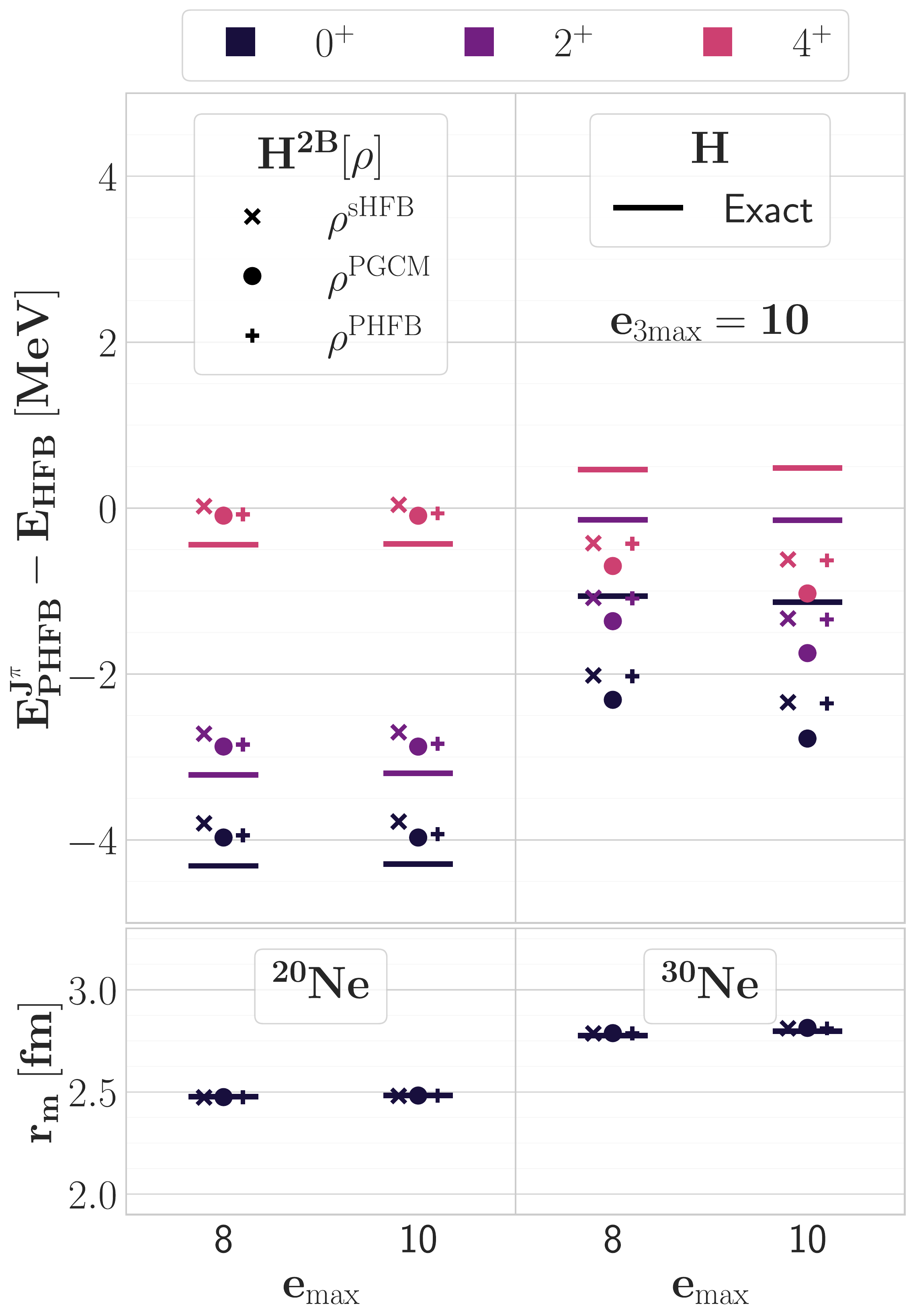}
    \caption{Same as Fig.~\ref{fig:phfb-diff} for \nucl{Ne}{20} (left) and \nucl{Ne}{30} (right) for \(e_{\text{max}}=8\) and $10$ at fixed \(e_{3\text{max}}=10\).}
    \label{fig:phfb-emax}
\end{figure}

Figure~\ref{fig:phfb-emax} probes the dependence of the results on the value of \(e_{\text{max}}\) at fixed \(e_{3\text{max}}\). First, one notices that radii are insensitive to \(e_{\text{max}}\) and are perfectly reproduced. Second, no change is visible in the PHFB energies of \nucl{Ne}{20} when going from \(e_{\text{max}}=8\) to \(e_{\text{max}}=10\). In the more neutron-rich \nucl{Ne}{30} isotope, there exists a slight change of approximate PHFB energies. While the agreement with the reference results are still quantitatively good, the energies degrade slightly when going from \(e_{\text{max}}=8\) to \(e_{\text{max}}=10\). The slight evolution away from the reference results relates in fact to the lack of convergence of the results with respect to  \(e_{3\text{max}}\) discussed earlier. In the present case, \(e_{3\text{max}}\) had to be set to 10 in order to be able to perform PHFB calculations with the explicit 3-body interaction at \(e_{\text{max}}=10\). 

\begin{figure}
    \centering
    \includegraphics[width=.5\textwidth]{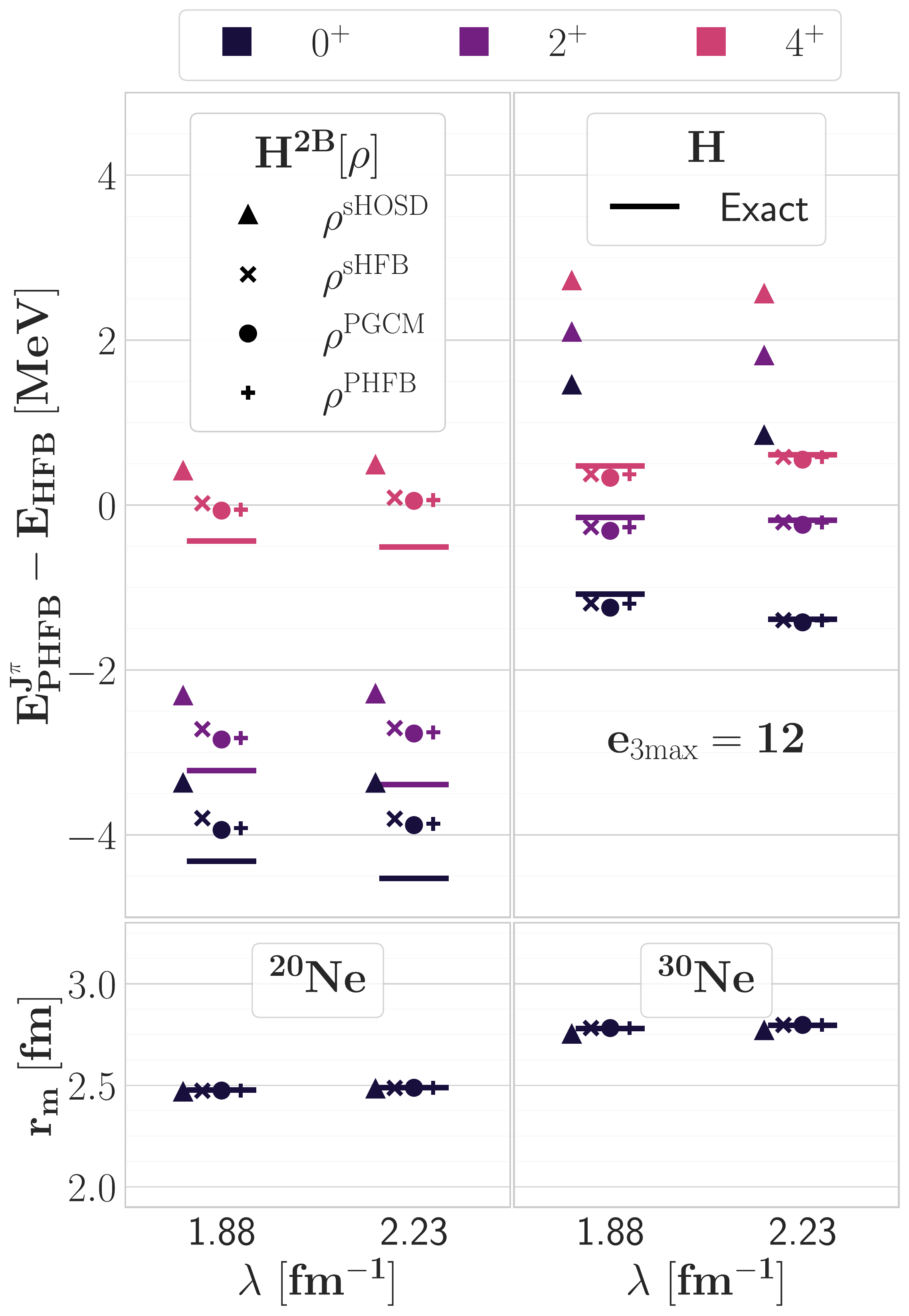}
    \caption{Same as Fig. \ref{fig:phfb-emax} but for two values of the SRG parameter \(\lambda_{\text{srg}}\). Calculations are performed with \(e_{\text{max}}=8\) and \(e_{3\text{max}}=12\).}
    \label{fig:phfb-srg}
\end{figure}

While pushing the calculations to large values of \(e_{3\text{max}}\) would probably improve the agreement further, the overall conclusion is that the high quality of  $H^{2B}[\rho]$ depends only mildly on \(e_{\text{max}}\) as long as the reference calculations themselves are converged enough. Although not shown, the same convergence behavior as a function of \(e_{\text{max}}\) is at play in the BMBPT(3) ground-state energies reported on in Sec.~\ref{resultsBMBPT}.

\subsubsection{Dependence on $\lambda_{\text{srg}}$}

Figure~\ref{fig:phfb-srg} probes the dependence of the results on the smoothness of the Hamiltonian. The SRG Hamiltonian at \(\lambda_{\text{srg}}=2.23\ \mathrm{fm}^{-1}\) is less evolved than the one at \(\lambda_{\text{srg}}=1.88\ \mathrm{fm}^{-1}\) and produces spectra that are slightly less compressed. Still, no significant dependence on the hardness of the Hamiltonian is observed as far as the quality of the results obtained with $H^{2B}[\rho]$ is concerned.

Although not shown, the same convergence behavior as a function of \(\lambda_{\text{srg}}\) is at play in the BMBPT(3) ground-state energies reported in Sec.~\ref{resultsBMBPT}.

\subsection{Spectroscopy}
\label{subsec:res-spectra}

Having analyzed absolute PHFB energies and radii, we are now in position to investigate spectroscopic observables.

\subsubsection{PHFB}

\begin{figure}
    \centering
    \includegraphics[width=.5\textwidth]{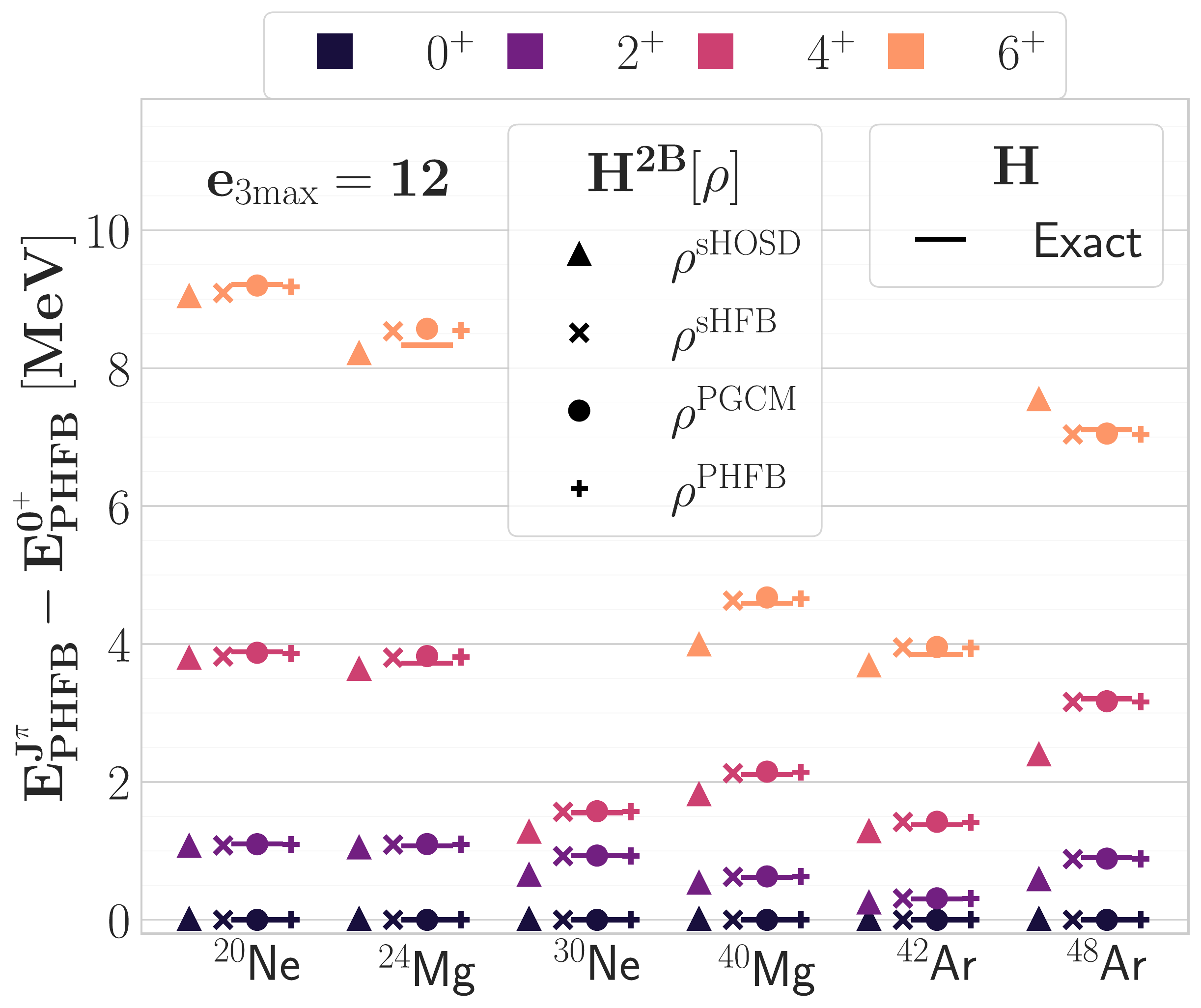}
    \caption{Low-lying PHFB excitation spectra of doubly open-shell nuclei. Reference results calculated from $H$ are compared to those computed from $H^{2B}[\rho]$ using the various one-body test density matrices. Calculations are performed with $e_{\text{max}}=8$, $e_{3\text{max}}=12$ and $\lambda_{\text{srg}}=1.88$\,fm$^{-1}$.}
    \label{fig:phfb-spectrum}
\end{figure}

Low-lying PHFB excitation spectra of doubly open-shell nuclei computed from $H$ and $H^{2B}[\rho]$ are compared in Fig. \ref{fig:phfb-spectrum}. Being based on the minimum of the dHFB TEC, these spectra describe the low-lying part of the ground-state rotational band. 

Reference results are well reproduced for all one-body test densities but $\rho^{\text{sHOSD}}$ for which a degrading arises with increasing mass. Even in nuclei for which absolute PHFB energies were not converged yet with respect to \(e_{3\text{max}}\) (e.g. \nucl{Mg}{40} and \nucl{Ne}{20}), energy {\it differences}  are fully consistent with the reference values. 

\begin{table*}
    \centering
    \renewcommand{\arraystretch}{1.15}
    \begin{tabular}{l|c|c|c|c|}
      
     & Mass \(\le\) 30 
     & Mass \(>\) 30
     & Neutron-rich
     & Total
     \\  \cline{1-5}
$\rho^{\text{sHOSD}}$ \bigstrut  & 5.86 & 15.38 & 16.79 & 10.62
\\ \cline{1-5}
$\rho^{\text{sHF(B)}}$ \bigstrut  & 1.47 &  1.56 &  0.91 &  1.51
\\ \cline{1-5}
$\rho^{\text{PHFB}}$  \bigstrut  & 1.17 &  1.68 &  1.11 &  1.43
\\ \cline{1-5}
$\rho^{\text{PGCM}}$  \bigstrut & 1.27 &  1.90 &  1.22 &  1.51
\\ \cline{1-5}
    \end{tabular}
    \caption{Average error (in \%) on PHFB low-lying excitation energies computed from  $H^{2B}[\rho]$ for various sub-categories of nuclei and test one-body density matrices. See Eq.~\eqref{eq:av-phfb-spec} for details on the cost function. Calculations are performed with $e_{\text{max}}=8$, $e_{3\text{max}}=12$ and $\lambda_{\text{srg}}=1.88$\,fm$^{-1}$.}
    \label{tab:av-phfb-spec}
\end{table*}

As for quantitative measures, systematic results are reported on in Tab.~\ref{tab:av-phfb-spec}. Reference excitation energies are reproduced to better than $2\%$ throughout the whole panel for $\rho^{\text{sHF(B)}}$, $\rho^{\text{PHFB}}$ and $\rho^{\text{PGCM}}$, which amounts to making errors of the order of a few tens of keVs. This is obviously negligible compared to other sources of uncertainties in state-of-the-art ab initio calculations. While this outcome further demonstrates the robustness of the approximation method, the $10\%$ average error obtained for $\rho^{\text{sHOSD}}$ underlines the fact that the employed one-body density matrix must be realistic enough to deliver high accuracy results. Given that the purpose of ab initio PHFB (and PGCM below) calculations is to access excitation energies and not absolute ones, one can be fully satisfied with the performances of $H^{2B}[\rho]$ in the present context.

\subsubsection{PGCM}

While PHFB calculations already provide a good test whenever the system is {\it rigid} with respect to collective variables, the PGCM opens the way to the wider class of so-called {\it soft} nuclei. More generally, it permits to include static correlations induced by shape fluctuations and to access associated vibrational excitations. 

\begin{figure}
    \centering
    \includegraphics[width=.5\textwidth]{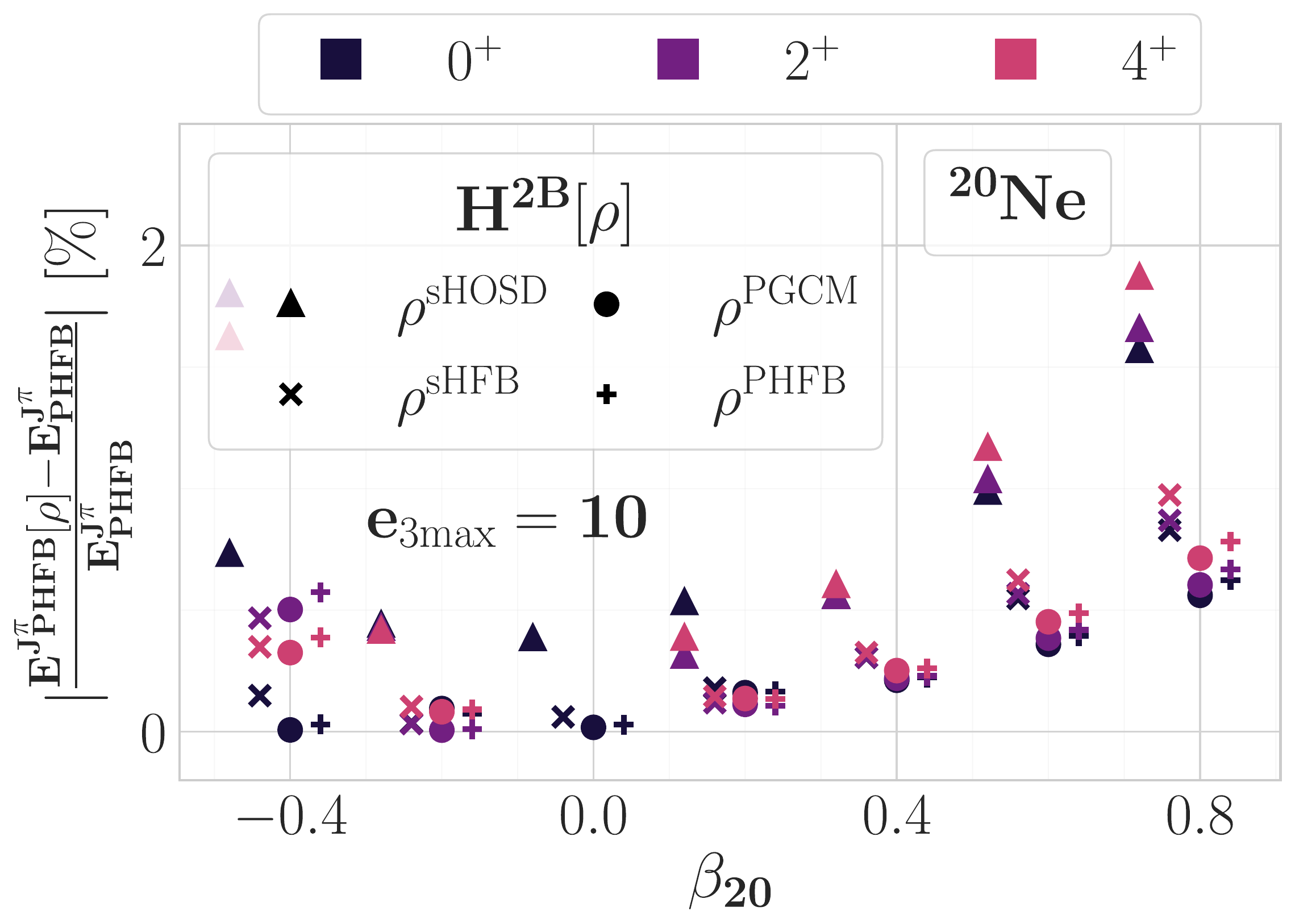}
    \caption{Same as the bottom panel of Fig.~\ref{fig:hfb-pes} for $J^{\Pi}=0^+, 2^+, 4^+$ PHFB states.}
    \label{fig:phfb-pes}
\end{figure}

Presently, PGCM calculations of \nucl{Ne}{20} and \nucl{Ne}{30} along the axial quadrupole coordinate are performed. In order to obtain a first indication of the performance of  $H^{2B}[\rho]$, Fig.~\ref{fig:phfb-pes} extends the study performed at the dHFB level in Sec.~\ref{dHFBsection} by displaying the error obtained for the TEC of $J^{\Pi}=0^+, 2^+, 4^+$ PHFB energies for the various test one-body density matrices. The $J^{\Pi}$ projected TEC constitutes the diagonal part of the Hamiltonian matrix at play in the Hill-Wheeler-Griffin secular equation of the PGCM calculation. The errors obtained along the projected TECs are strictly similar to those displayed in Fig.~\ref{fig:hfb-pes} at the dHFB level. This result gives confidence regarding the quality of the results that can be expected at the PGCM level.

\begin{figure}
    \centering

    \includegraphics[width=.5\textwidth]{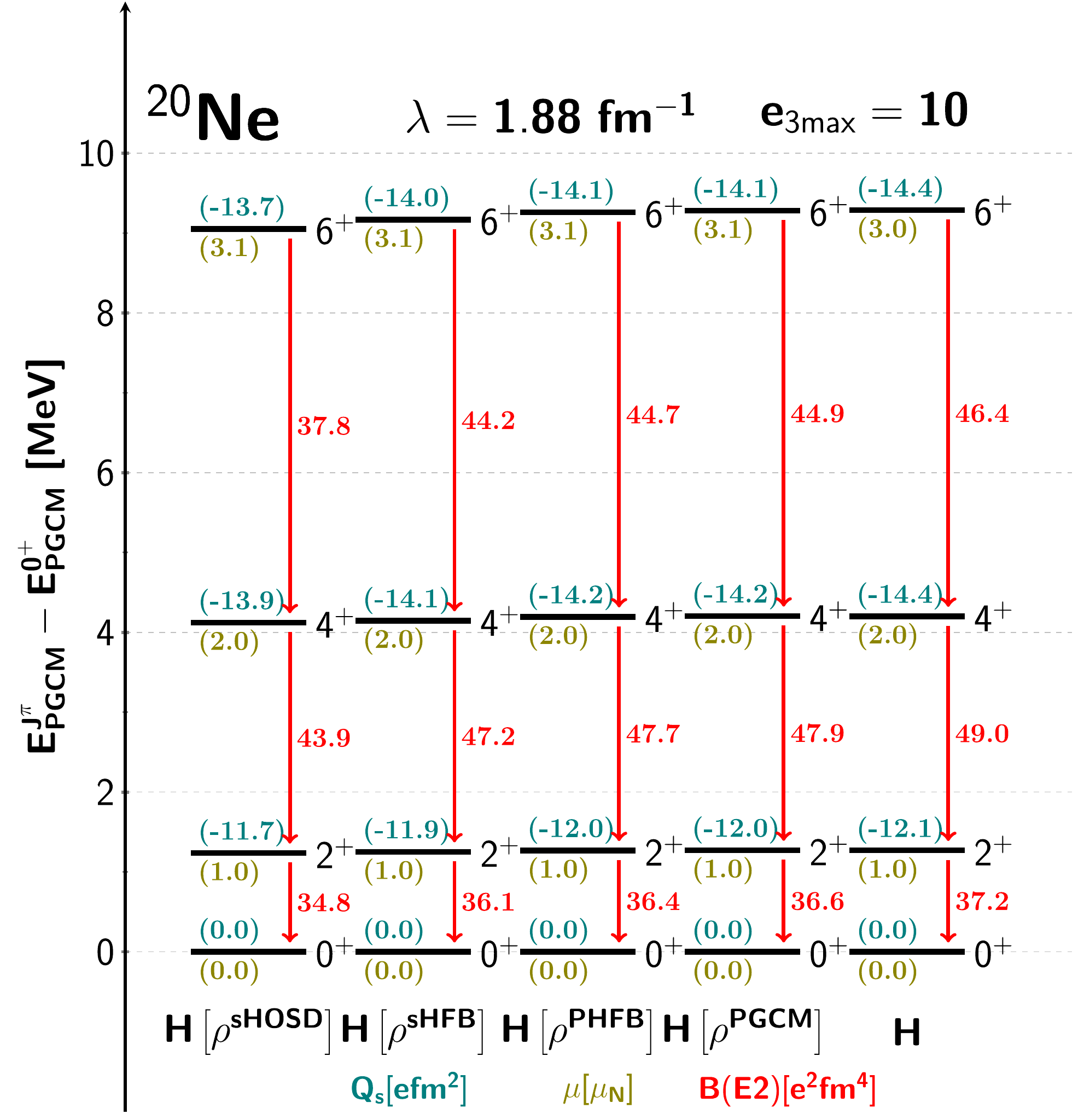}
    \caption{Low-lying part of the PGCM ground-state rotational band of \nucl{Ne}{20}. Reference results calculated from $H$ are compared to those computed from $H^{2B}[\rho]$ using various test one-body density matrices. Each energy level is displayed along with the magnetic dipole (below) and electric quadrupole (above) moments of the associated state. B(E2) transitions strengths are displayed using red arrows. Calculations are performed with \(e_{\text{max}}=8\), \(e_{3\text{max}}=10\) and $\lambda_{\text{srg}}=1.88$\,fm$^{-1}$.}
    \label{fig:gcm-Ne20}
\end{figure}

\begin{table*}
    \centering
    \renewcommand{\arraystretch}{1.25}
    \begin{tabular}{l|c|c|c|c|}
     Nucleus 
     & \multicolumn{2}{c|}{\nucl{Ne}{20}}
     & \multicolumn{2}{c|}{\nucl{Ne}{30}}
     \\\cline{1-5} Quantity
     & Spectrum & Observables
     & Spectrum & Observables
     \\  \cline{1-5}
    $\rho^{\text{sHOSD}}$ \bigstrut  & 2.95  & 1.43 & 4.53 & 2.85
    \\ \cline{1-5}
    $\rho^{\text{sHF(B)}}$ \bigstrut & 1.46  & 0.71   &  2.60 &   2.57
    \\ \cline{1-5}
    $\rho^{\text{PHFB}}$ \bigstrut  &  0.36 &   	0.55 	 &   	2.59 	 &   
    \\ \cline{1-5}
    $\rho^{\text{PGCM}}$ \bigstrut  &  0.26 & 0.48 &   2.98 & 2.85\\ \cline{1-5}
    \end{tabular}
    \caption{Average error (in \%) on PGCM excitation energies and spectroscopic observables computed from $H^{2B}[\rho]$ in \nucl{Ne}{20} and \nucl{Ne}{30} for various test one-body density matrices.  See Eq.~\eqref{eq:av-gcm} for details on the cost function. Calculations are performed with \(e_{\text{max}}=8\), \(e_{3\text{max}}=10\) and $\lambda_{\text{srg}}=1.88$\,fm$^{-1}$.}
    \label{tab:av-pgcm}
\end{table*}

\begin{figure}
    \centering

    \includegraphics[width=.5\textwidth]{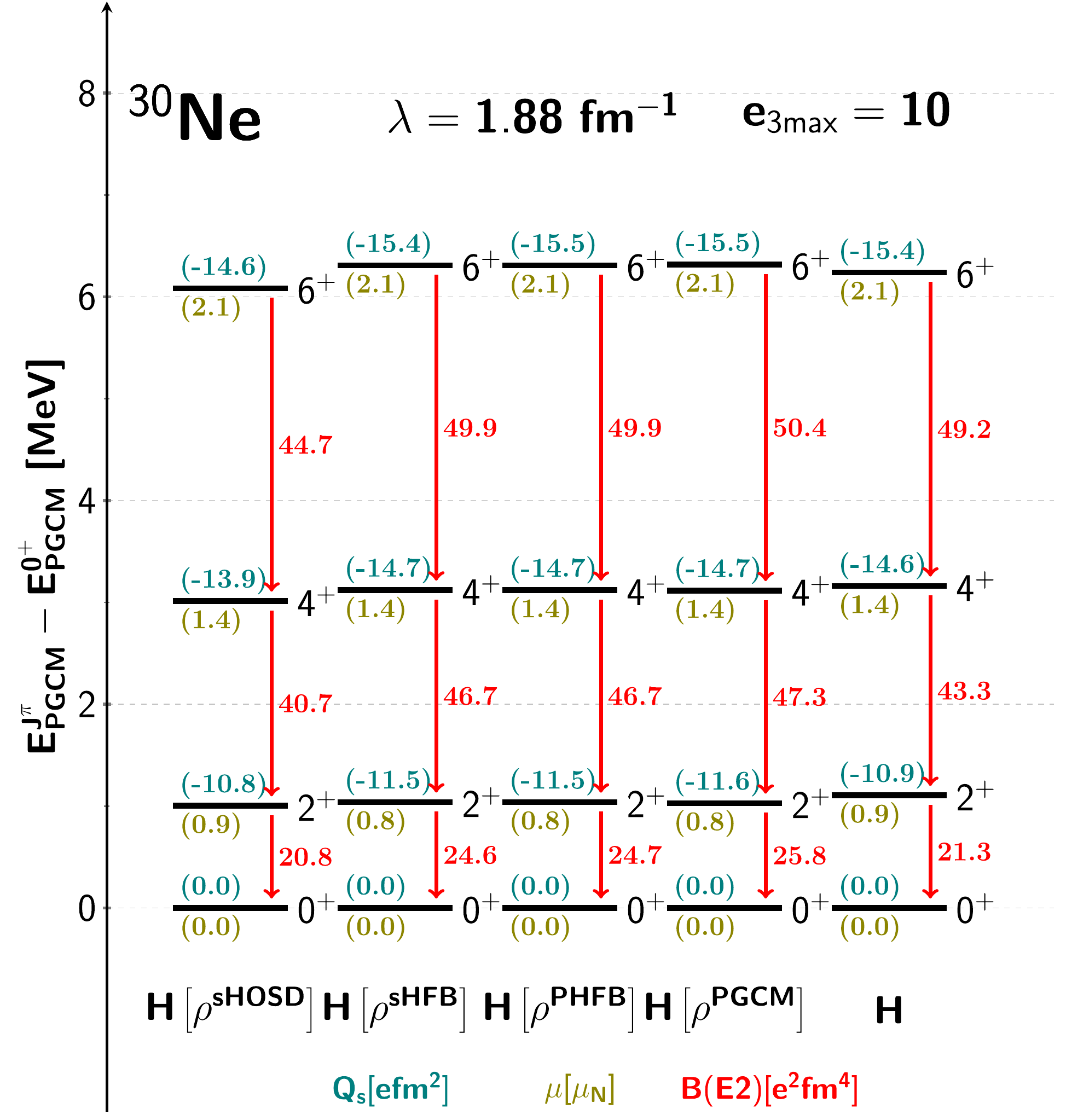}
    \caption{Same as Fig. \ref{fig:gcm-Ne20} for \nucl{Ne}{30}.}
    \label{fig:gcm-Ne30}
\end{figure}

Reference and approximate low-lying PGCM excitation energies of the ground-state rotational band and associated electromagnetic observables are compared for \nucl{Ne}{20} and \nucl{Ne}{30} in Figs.~\ref{fig:gcm-Ne20} and~\ref{fig:gcm-Ne30}, respectively. Due to numerical limitations, only three-body matrix elements up to \(\etmax=10\) could be included in the full calculation, hence hindering the convergence in \nucl{Ne}{30}. Still, building on the results reported in Fig.~\ref{fig:phfb-pes} an excellent agreement emerges in both nuclei for PGCM energies and electromagnetic observables, even more so in \nucl{Ne}{20} where sub-percent accuracy (see Tab.~\ref{tab:av-pgcm}) is achieved. As before, a decent but less optimal reproduction of the reference results is obtained for $\rho=\rho^{\text{sHOSD}}$. The excellent results obtained for electromagnetic observables testify the stability of the PGCM wave-functions themselves with respect to the in-medium approximation of the three-nucleon interaction. 

\subsubsection{dQRPA}

\begin{figure}
    \centering
    \includegraphics[width=.5\textwidth]{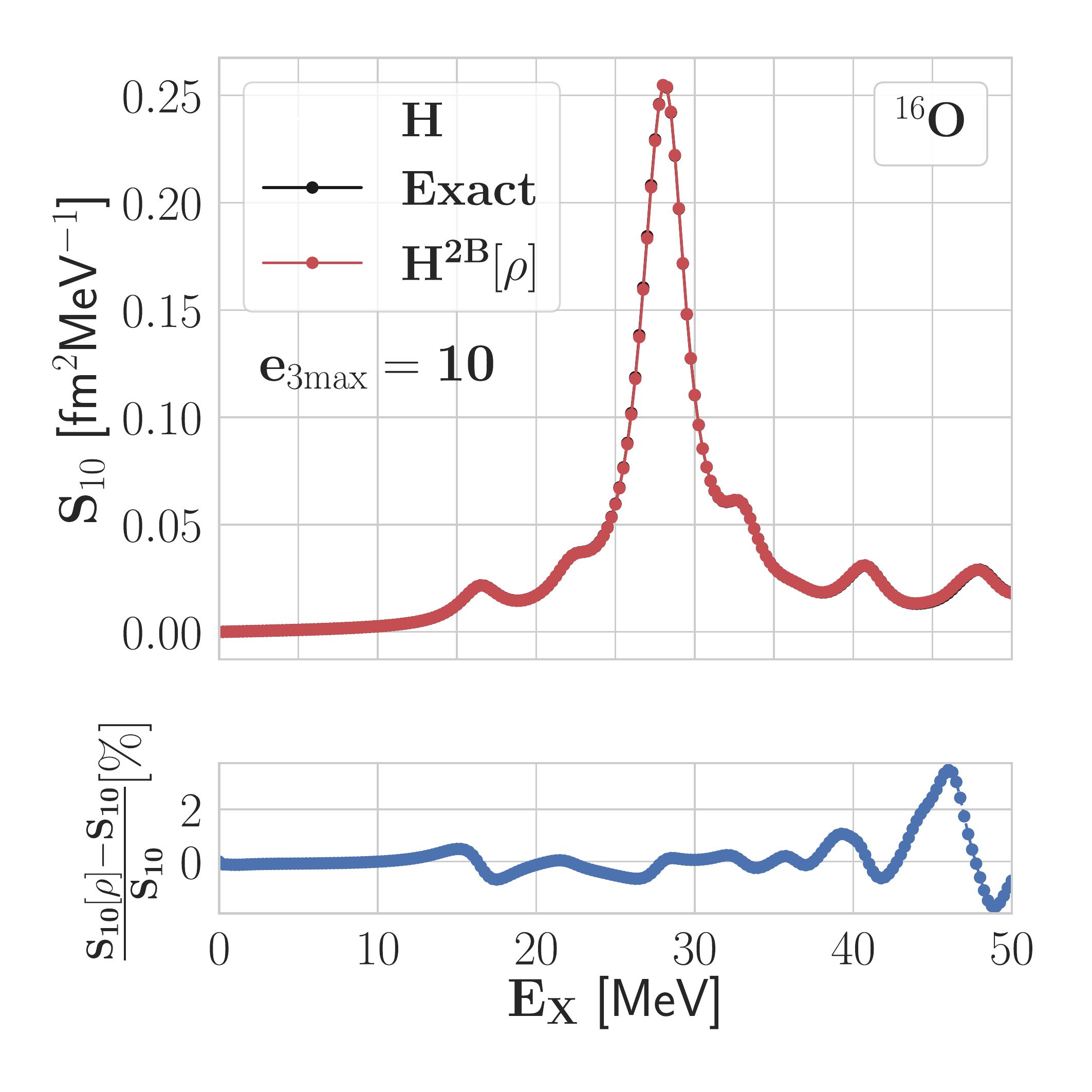}
    \caption{
    Electric isovector dipole strength of \nucl{O}{16} as a function of the excitation energy (upper panel). Reference results calculated from $H$ are compared to those computed from $H^{2B}[\rho]$ using $\rho^\text{sHFB}$. The relative deviation from the strength computed with the exact Hamiltonian is displayed as a function of the excitation energy in the lower panel. Calculations are performed with \(e_{\text{max}}=8\), \(e_{3\text{max}}=10\) and $\lambda_{\text{srg}}=1.88$\,fm$^{-1}$.}
    \label{fig:qfam-O16}
\end{figure}

\begin{figure}
    \centering
    \includegraphics[width=.5\textwidth]{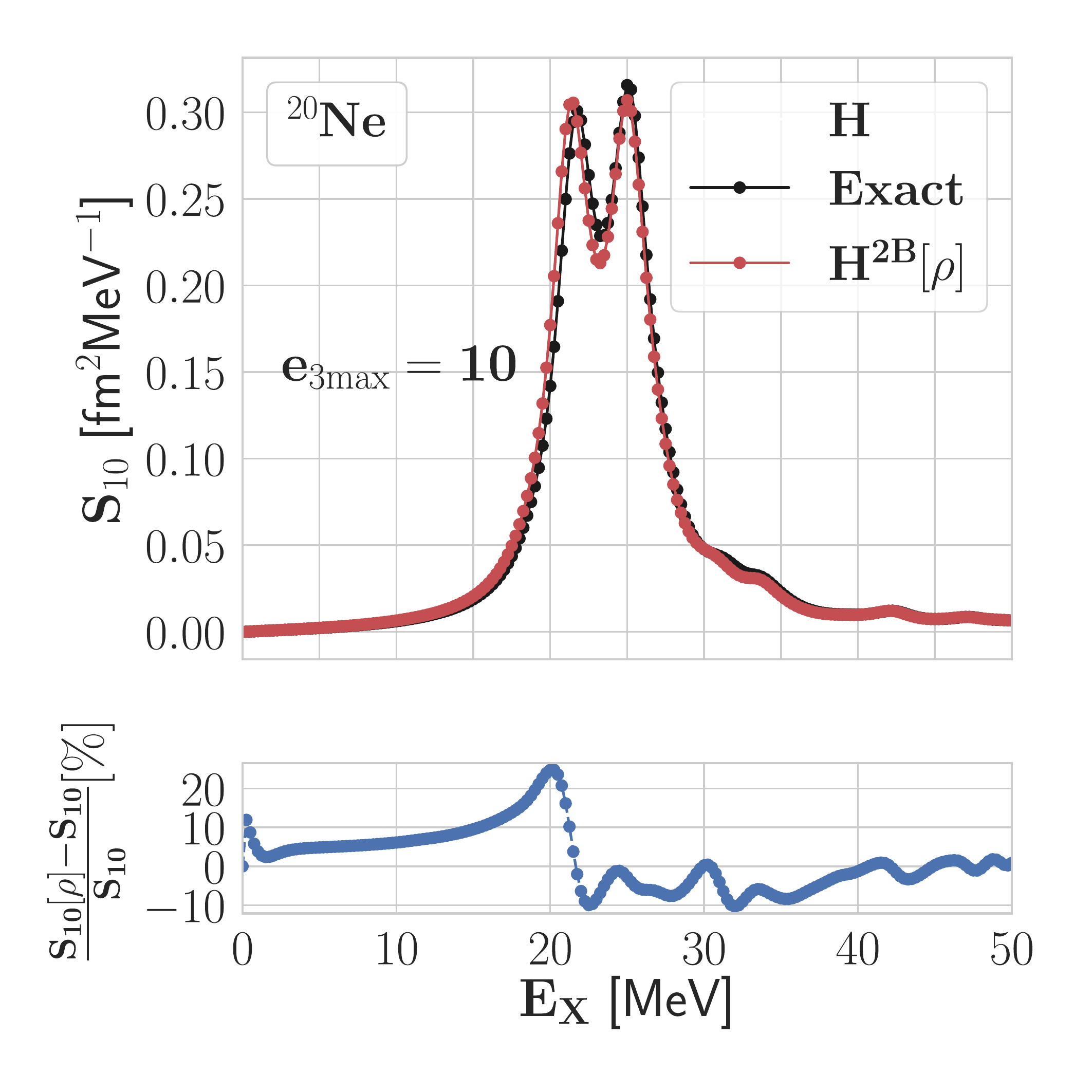}
    \caption{Same as Fig.~\ref{fig:qfam-O16} for \nucl{Ne}{20}.}
    \label{fig:qfam-Ne20}
\end{figure}

QRPA is a method of choice to study excited states of both individual and collective characters, with energies ranging from a few MeV to tens of MeV. The method is appropriate as long as the excited states bear a strong resemblance with the ground-state owing to the harmonic approximation at the heart of the QRPA. In this context, the performance of $H^{2B}[\rho]$ can be assessed by looking at, e.g., electromagnetic strength functions. Figures~\ref{fig:qfam-O16} and~\ref{fig:qfam-Ne20} display the electric isovector dipole ($E1$) strength computed with both $H$ and $H^{2B}[\rho^\text{sHFB}]$, for \nucl{O}{16} and \nucl{Ne}{20} respectively. Similar results are obtained for the other test one-body density matrices and are reported in Table~\ref{tab:err-qfam}. 

In \nucl{O}{16}, where dQRPA reduces to sRPA built on top of a sHF Slater determinant, the difference between the strength functions are hardly noticeable. As in sHF calculations discussed earlier on, this relates to the fact that the sRPA error would actually be strictly zero if $\rho$ entering $H^{2B}[\rho]$ were taken as the {\it variational} sHF density matrix obtained from that approximate Hamiltonian. Because $\rho^\text{sHF}$ coming from the calculation performed with $H$ slightly differs from the variational one obtained from $H^{2B}[\rho^\text{sHF}]$, the error is not strictly zero but remains very tiny. As seen from the bottom panel of Fig.~\ref{fig:qfam-O16} the relative error of the $E1$ strength at each excitation energy does not exceed 4\% over the interval $[0, 50]$\,MeV. This error essentially relates to the horizontal position of the individual dQRPA modes that, as testified in Tab.~\ref{tab:err-qfam}, are shifted by a tiny amount, i.e. 0.05\% on average. Computing the observable total photo-emission cross section by integrating the differential photo-emission cross section deduced from the $E1$ strength, the error on the latter generates a tiny 0.02\% error. While the error remains very small for the other test one-body density matrices, the choice $\rho=\rho^\text{sHFB}$ is optimal in the present case.

\begin{table*}
    \centering
    \renewcommand{\arraystretch}{1.25}
    \begin{tabular}{l|c|c|c|c|}
     Nucleus 
     & \multicolumn{2}{c|}{\nucl{O}{16}}
     & \multicolumn{2}{c|}{\nucl{Ne}{20}}
     \\\cline{1-5} Quantity
     & Excitation energy & Total photo-emission cross section
     & Excitation energy & Total photo-emission cross section
     \\  \cline{1-5}
    $\rho^{\text{sHOSD}}$ \bigstrut  & 0.39  & 0.04 & 0.46 & 0.82
    \\ \cline{1-5}
    $\rho^{\text{sHF(B)}}$ \bigstrut & 0.05  & 0.02   &  1.09 &   0.63
    \\ \cline{1-5}
    $\rho^{\text{PHFB}}$ \bigstrut  &  0.14 & 0.23 	 &   1.13 	 &  0.48 
    \\ \cline{1-5}
    $\rho^{\text{PGCM}}$ \bigstrut  &  0.15 & 0.24 & 1.19 & 0.44\\ \cline{1-5}
    \end{tabular}
    \caption{Average relative error (in \%) on dQRPA excitation energies and on the total photo-emission cross section computed from $H^{2B}[\rho]$ in \nucl{O}{16} and \nucl{Ne}{20} for various test one-body density matrices. Calculations are performed with \(e_{\text{max}}=8\), \(e_{3\text{max}}=10\) and $\lambda_{\text{srg}}=1.88$\,fm$^{-1}$.}
    \label{tab:err-qfam}
\end{table*}

Moving to Fig.~\ref{fig:qfam-Ne20}, the dQRPA dipole strength of \nucl{Ne}{20} obtained from $H^{2B}[\rho^\text{sHFB}]$ at the minimum of the dHFB TEC is also visually very close to the reference one, although slightly deteriorated compared to the \nucl{O}{16} case. Looking at the bottom panel, the situation suddenly appears less favorable with a relative error at fixed excitation energies that can reach nearly 30\% on the left side of the giant resonance. This error relates to the fact that approximating $H$ by $H^{2B}[\rho]$ slightly affects dHFB quasi-particle energies, inducing in turn a small shift (1\% on average) in the position of the dQRPA eigenmodes.  The steep slope of the  $E1$ strength function before and after the giant resonance exacerbates the relative error made at fixed excitation energies. However, propagated to the total photo-emission cross section, this only results in a negligible error of 0.63\% (see Tab.~\ref{tab:err-qfam}). In this case, the optimal character of $\rho=\rho^\text{sHFB}$ is not apparent as all test one-body density matrices deliver similar results.

The above results validates the quality and robustness of $H^{2B}[\rho]$ in the dQRPA context. Although not shown for brevity, essentially identical results hold for other multipolarities of the one-body transition operator. Furthermore, the dependence of the results on $e_\text{3max}$ is along the same line as the one discussed in Sec.~\ref{sec:systana}.  

\subsection{Optimal one-body density matrix}
\label{apprandomdensmat}

The in-medium approximation of three-body interactions proposed in the present work appears to be very robust with respect to the employed symmetry-invariant one-body density matrix. All dHFB, dBMBPT, PHFB PGCM and dQRPA results presented above are of equal (excellent) quality for $\rho=\rho^{\text{sHF(B)}}$, $\rho^{\text{PHFB}}$ and $\rho^{\text{PGCM}}$ but are systematically deteriorated for the more simplistic choice $\rho = \rho^{\text{sHOSD}}$. 

\begin{figure*}
    \centering
    \includegraphics[width=\textwidth]{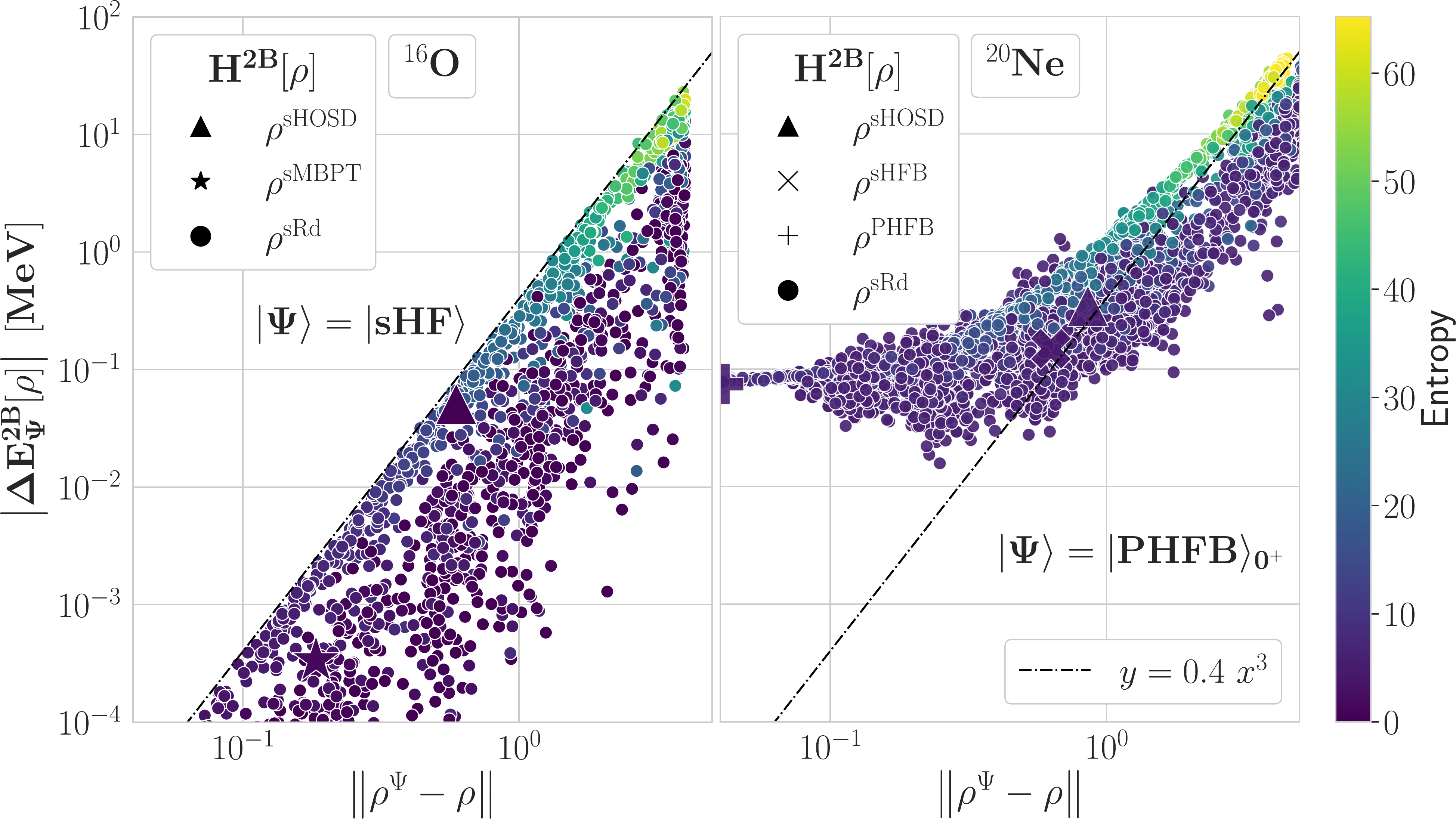}
    \caption{Ground-state energy error $\Delta E^{2B}_{\Psi}[\rho]$ associated with the use of $H^{2B}[\rho]$ as a function of the distance \(\|\rho-\rho^\Psi\|\) between the test one-body density matrix $\rho$ and the actual ground-state one $\rho^\Psi$ in log-log scale. Left panel: sHF ($J^{\Pi}=0^+$) solution for \nucl{O}{16}. Right panel: PHFB ($J^{\Pi}=0^+$) solution for \nucl{Ne}{20}. Data points are for randomly-sampled and physical test one-body density matrices. The color scale characterizes one-body density matrices' von Neumann entropy. The dashed-dotted lines denote the cubic envelop extracted from the left panel and reported on the right panel. Calculations are performed for \(e_{\text{max}}=6\) and \(e_{3\text{max}}=6\).}
    \label{fig:rd}
\end{figure*}

In this context, it is of interest to better assess this robustness and possibly characterize the optimal one-body density matrix to be used in the design of $H^{2B}[\rho]$. For this purpose, trial (symmetry-invariant) one-body density matrices $\{\rho^{\text{sRd}}\}$ are generated by means of the random sampling described in App.~\ref{randomdensmat}.
To evaluate the corresponding approximation $H^{2B}[\rho^\text{sRd}]$, the ground-state energy error
\begin{align}
\Delta E^{2B}_{\Psi}[\rho]  &\equiv \frac{\langle \Psi |  H^{2B}[\rho]  | \Psi \rangle}{\langle \Psi  | \Psi \rangle} - \frac{\langle \Psi |  H  | \Psi \rangle}{\langle \Psi |  \Psi \rangle} \label{diffE} 
\end{align}
is considered; see App.~\ref{randomerrf} for the working expression and a related discussion.
The error function $\Delta E^{2B}_{\Psi}[\rho^{\text{sRd}}]$ computed for a large set of randomly generated matrices is shown in Fig.~\ref{fig:rd} as a function of the distance $\|\rho^{\text{sRd}}-\rho^{\Psi}\|$ between the trial one-body density matrix and the ground-state one $\rho^{\Psi}$ in the many-body calculation of interest. The data points corresponding to the physical one-body density matrices ($\rho^{\text{sHOSD}}$, $\rho^{\text{sHF(B)}}$, $\rho^{\text{PHFB}}$ and $\rho^{\text{PGCM}}$) are also displayed to better make sense of the results obtained so far. 
In addition to the distance to $\rho^{\Psi}$, each trial one-body density matrix is characterized by its von Neumann entropy
\begin{equation}
    S[\rho] \equiv - \text{Tr} \left(\rho \ln \rho\right) \, ,
\end{equation}
which, in the eigenbasis of $\rho$ with eigenvalues \(\{r_a\}\) reads as Shannon's entropy of information theory
\begin{equation}
    S[\rho] \equiv - \sum_a r_a \ln r_a \, .
\end{equation}
In the present context, the size of the entropy essentially characterizes how much the many-body state $\rho$ differs from a Slater determinant for which $S[\rho]=0$, i.e. it is a measure of many-body correlations.

Results for \nucl{O}{16} computed in a small model space at the sHF level are shown in the left panel of Fig.~\ref{fig:rd} while the right panel displays results for \nucl{Ne}{20} computed at the PHFB level. The sHF calculation of \nucl{O}{16} illustrates the situation encountered for an uncorrelated state, i.e. the many-body solution $| \Psi \rangle$ is nothing but a symmetry-conserving Slater determinant. In this particular case, the ground-state energy error (see Eq.~\eqref{diffEexplicit}) takes the simple form
\begin{align}
\Delta E^{2B}_{\text{sHF}}[\rho]
=&  \frac{1}{3!} w^{(3)} \! \cdot \!  \left(\rho-\rho^{\text{sHF}}\right)^{\otimes(3)} \label{errorsHF}
\end{align}
and is thus minimal, actually null, for $\rho=\rho^{\text{sHF}}$. The fact that the optimal one-body density matrix is nothing but the one of the many-body state under scrutiny is confirmed numerically in the left panel of Fig.~\ref{fig:rd}. In absence of genuine correlations, one expects from Eq.~\eqref{errorsHF} that the sampled errors are bounded by a cubic envelope in the variable \(\|\rho-\rho^{\text{sHF}}\|\), which indeed appears clearly in the numerical results. The coefficient ($0.4$) of that cubic envelope extracted from the data is a measure of the employed three-body interaction strength in the utilized model space. 

Besides the null error delivered by $\rho=\rho^{\text{sHF}}$, the errors associated with the physical one-body density matrices $\rho^{\text{sHOSD}}$ and $\rho^{\text{sMBPT}}$ are provided on the figure. Compared to the full range of sampled one-body density matrices\footnote{Given that $\rho^{\text{sHF}}$ relates to a Slater determinant with $16$ particles, the maximum distance is reached for densities associated with Slater determinants obtained by promoting the $16$ particles from hole states into particle states, i.e. $\text{Max}_{\rho} \|\rho-\rho^{\text{sHF}}\| = \sqrt{32} \approx 5.7$, which is indeed the maximum value visible on the left panel of Fig.~\ref{fig:rd}.}  $\rho^{\text{sHOSD}}$ and $\rho^{\text{sMBPT}}$ are rather close to $\rho^{\text{sHF}}$. This is particularly true of $\rho^{\text{sMBPT}}$, which is a sign of the weakly-correlated character of \nucl{O}{16} when eventually going beyond the mean-field. Given the cubic upper-bound, such a proximity between the two density matrices implies a tiny error on the energy obtained for  $\rho=\rho^{\text{sMBPT}}$. In spite of originating from a Slater determinant and thus sharing the same null entropy as $\rho^{\text{sHF}}$, $\rho^{\text{sHOSD}}$ is about $3$ times more distant from it than $\rho^{\text{sMBPT}}$. In agreement with the cubic law governing the error, plus being located closer to the envelope, the associated error is about $170$ times larger. Given the softness of the employed three-body interaction, $\rho^{\text{sHOSD}}$ still provides a small absolute error in the end. Eventually, the sampling provides a fair understanding that, as long as the test one-body density is not too distant from $\rho^{\text{sHF}}$, its detailed properties do not matter much and the error is bound to be small. 

Compared to the previous case, the right panel of Fig.~\ref{fig:rd} allows one to appreciate the qualitatively different situation encountered for a genuinely-correlated state. Indeed, the error $\Delta E^{2B}_{\text{PHFB}}[\rho]$ behaves now differently as a function of the distance\footnote{Because of the log-log scale employed, the point at zero distance associated with $\rho^{\text{PHFB}}$ is artificially placed on the left border of the figure.} $\|\rho-\rho^{\text{PHFB}}\|$. As visible from Eq.~\eqref{diffEexplicit}, $\Delta E^{2B}_{\text{PHFB}}[\rho]$ contains non-zero constant and linear terms in addition to the cubic term encountered in Eq.~\eqref{errorsHF}. 

The constant term delivers the error $\Delta E^{2B}_{\text{PHFB}}[\rho^{\text{PHFB}}]$ associated with the actual ground-state density, i.e. when setting $\rho=\rho^{\text{PHFB}}$. The fact that this error is different from zero is a fingerprint of the fact the PHFB ground-state wave-function carries (a least) genuine three-body correlations. The value of the corresponding error additionally depends on the size of the three-body interaction convoluted with the irreducible three-body density matrix (see Eq.~\eqref{diffEexplicit}). As analyzed in Ref.~\cite{Dyhdalo:2017gyl} in connection with the NO2B approximation, a low-scale Hamiltonian makes the energy contribution from the residual three-body interaction small. For \((e_{\text{max}}=6;e_{3\text{max}}=6)\)  this error is $\Delta E^{2B}_{\text{PHFB}}[\rho^{\text{PHFB}}]\approx 0.1$\,MeV whereas the better converged value obtained earlier on for \((e_{\text{max}}=8;e_{3\text{max}}=12)\) is $0.5$\,MeV ($0.4\%$); i.e. the error is small.

Increasing the distance from $\rho=\rho^{\text{PHFB}}$, one can lower the error such that a minimum $\text{Min}_{\rho} \Delta E^{2B}_{\text{PHFB}}[\rho]$ is found for $\|\rho-\rho^{\text{PHFB}}\|= \text{few}\,10^{-1}$ with a value several times smaller than  $\Delta E^{2B}_{\text{PHFB}}[\rho^{\text{PHFB}}]$. Passed the minimum the error typically increases and is eventually dominated by the cubic terms at large distances such that the cubic envelope extracted from the left panel becomes effective.

The physical density matrices $\rho^{\text{sHOSD}}$ and $\rho^{\text{sHFB}}$ are found right passed the minimum such that their error is small and in fact similar to the one found at the origin. The profile of the error as a function of the distance $\|\rho-\rho^{\text{PHFB}}\|$ rationalizes the fact that small errors can be found over a substantial range of density matrices to which the various physical one-body density matrices one may typically access all belong. This feature provides practitioners with a significant flexibility as far as the choice of the employed one-body density matrix is concerned. Passed that appropriate interval the error rapidly increases with the distance,  as testified by the use of $\rho^{\text{sHOSD}}$ sitting on the edge of it, such that one may not be too cavalier either regarding the choice of $\rho$.

\subsection{Lessons and perspectives}

\subsubsection{Main lessons}

The above results demonstrate the usefulness of the proposed in-medium reduction method of three-body interaction operators in nuclear ab initio calculations. The fact that the method relies on the sole use of a one-body density matrix gives much credit to the {\it simplicity} of the method. Furthermore, the high-quality approximation was shown to be robust with respect to the employed one-body density matrix, which gives much credit to the {\it flexibility} of the method.

These conclusions have been validated for nuclei with closed and open-shell characters, i.e. displaying weak and strong correlations, for light and mid-mass systems as well as for stable and exotic isotopes. While convincingly substantiated via the use of the several many-body methods and for a large class of observables, a further validation of the quality of the approximation on the basis of ab initio methods built on different paradigms are of course welcome in the future. 

\subsubsection{Algorithm}

The independence of the results with respect to a large class of one-body density matrices is of prime importance for practical applications in the future, especially given that ab initio calculations aspire to move up the nuclear chart towards heavy, doubly open-shell nuclei. Specifically, the high-quality results obtained for $\rho=\rho^{\text{sHFB}}$ allow one to build $H^{2B}[\rho]$ at the sole cost of running first a spherical HFB calculation with full three-body forces, thus bypassing the need to run any deformed HFB code followed by projections, which would already be too costly with explicit three-body forces in heavy nuclei requiring large values of \(\etmax\). Eventually, the envisioned working algorithm is
\begin{enumerate}
\item run a spherical HFB calculation with three-nucleon forces to extract $\rho^{\text{sHFB}}$,
\item build $H^{2B}[\rho^{\text{sHFB}}]$,
\item run the many-body method of interest with the two-body  Hamiltonian   $H^{2B}[\rho^{\text{sHFB}}]$,
\end{enumerate}
such that even in (heavy) open-shell nuclei
\begin{itemize}
\item no two-body density matrix has to be extracted,
\item no genuine open-shell calculation with an explicit three-nucleon operator has to be performed.
\end{itemize}

\subsubsection{Odd-even and odd-odd nuclei}

While the method presented in this article relies on the use of symmetry-invariant one-body densities, which can be generated only starting from $J^\Pi = 0^+$ states (or a superposition of such states), it can also be easily used to construct effective $k$-body interactions for odd-even and odd-odd nuclei. For this purpose, one can employ the one-body density generated in a mean-field calculation of a spherical Bogoliubov vacuum constrained to have odd-even or odd-odd numbers of particles on average. In the case of odd-even systems, it was demonstrated in Ref.~\cite{Duguet01a} that such a vacuum represents a good approximation to the true mean-field solution obtained with an odd number parity wave function. 

Of course, the accuracy of the $k$-body interactions constructed in this way will have to be properly checked but there is no reason to believe that they would perform particularly worse than those generated to describe even-even nuclei.

\subsubsection{Further perspectives}

In the longer-term future, the in-medium reduction procedure proposed in the present work can be tested to deal with, e.g., four-body interactions~\cite{Rozpedzik:2006yi,Kruger:2013kua,Kaiser:2015lsa} and/or three-body nuclear currents~\cite{Krebs:2020pii} at a reduced computational cost.

\section{Conclusions}
\label{conclusions}

The present work introduced a novel method to approximate $n$-body operators in terms of $k$-body ones with $k<n$. This is highly pertinent to overcome the steep increase of the computational cost of many-body calculations due to the presence of three-nucleon interactions, especially as ab initio calculations aspire to move to heavier nuclei than presently possible.

The main advantages are that the method is accurate, universal, simple and flexible. The universality of the method not only relates to its applicability to all nuclei, independently of their closed or open-shell character, but also to its independence with respect to the many-body method eventually used to solve Schr\"odinger's equation. The simplicity of the method relates to the fact that it requires the convolution of the, e.g., three-body operator with a sole symmetry-invariant one-body density matrix, even in open-shell nuclei. This it at variance with existing methods that either convolute the three-body operator with one-, two- and three-body density matrices in open-shell systems or with a symmetry-breaking one-body density matrix, thus leading to an approximate operator that explicitly breaks symmetries of the initial Hamiltonian. Eventually, the flexibility of the method relates to the possibility to use various one-body density matrices as an entry. As a matter of fact, the functional form of the error due to the use of the approximate Hamiltonian could be exploited to explain why accurate results can be obtained for a rather large class of one-body density matrices. Such a flexibility can be exploited to use a (not too) simple density matrix in practical calculations, e.g. the density matrix extracted from a spherical Hartree-Fock-Bogoliubov calculation.

Extensive numerical results have demonstrated the high accuracy of the approach over a large panel of nuclei and observables. The approximation method is thus ready to be employed in routine ab initio calculations in the future. Furthermore, the in-medium reduction procedure is ready to be tested on four-nucleon interactions and/or three-body nuclear currents in order to deal with them at a reduced computational cost.

%
\begin{acknowledgements}
We would like to thank A. Tichai for his help with the spherical solver and R. Roth for providing us with the interaction matrix element used in the numerical simulations. Calculations were performed by using HPC resources from GENCI-TGCC (Contract No. A009057392). This project has received funding from the European Union’s Horizon 2020 research and innovation programme under the Marie Skłodowska-Curie grant agreement No.~839847.
\end{acknowledgements}
%
%
%
 
%


\begin{appendices}

\section{Inverse tensor transformations}
\label{twostepsprocedure}

Given the operator $O$, i.e. the set of tensors $\{o^{(n)}; n = 0,\ldots,N\}$, and the one-body density matrix \(\rho\), a second set of matrices is introduced through
\begin{equation}
\label{eqn:wick2app}
\mathbf o^{(k)}[\rho] \equiv \sum_{n=k}^N \inv{(n-k)!} o^{(n)}\!\cdot\! {\rho}^{\otimes(n-k)}
\end{equation}
with $k=0,\ldots, N$. In a second step, the third set of tensors is defined via
\begin{equation}
\label{eqn:nowick_rapp}
{\tilde o}^{(n)}\left[\rho\right] \equiv \sum_{l=n}^N
	\frac{(-1)^{l-n}}{(l-n)!} \mathbf o^{(l)}\left[\rho\right] \!\cdot\! \rho^{\otimes(l-n)} 
\end{equation}
with $n= 0,\ldots,N$.

The third set of tensors is now shown to be the same as the initial one. Noting that
\begin{align}
\mathbf o^{(l)}\left[\rho\right] \cdot \rho^{\otimes(l-n)} 
  &=
      \sum_{k=l}^N \inv{(k-l)!} \left(o^{(k)} \!\cdot\! \rho^{\otimes(k-l)} \right) \!\cdot\!  \rho^{\otimes(l-n)}\nonumber \\
  &=
      \sum_{k=l}^N \inv{(k-l)!} o^{(k)} \!\cdot\! \rho^{\otimes(k-n)},
\end{align}
one obtains
\begin{align}
{\tilde o}^{(n)}\left[\rho\right] &=
    \sum_{l=n}^N\sum_{k=l}^N \frac{(-1)^{l-n}}{(l-n)!}  \inv{(k-l)!} o^{(k)} \cdot \rho ^ {\otimes(k-n)}\nonumber\\
    &=
    \sum_{k=n}^N \left[\sum_{l=n}^k{(-1)^{l-n}}\inv{(l-n)!(k-l)!}\right] o^{(k)} \cdot \rho^{\otimes(k-n)}\nonumber\\
    &=
    \sum_{k=n}^N \left[\sum_{l=n}^k(-1)^{l-n} \binom{k-n}{l-n} \right]
    \inv{(k-n)!} o^{(k)} \cdot \rho ^ {\otimes(k-n)}\nonumber\\
    &=
    \sum_{k=n}^N \delta_{kn} \inv{(k-n)!} o^{(k)} \cdot \rho ^ {\otimes(k-n)}\nonumber\\
    &=
   o^{(n)},\label{eqn:nowick_back}
\end{align}
such that each original tensor $o^{(n)}$, and thus the full original operator \(O\), is recovered through the two-step procedure.

\section{PGCM transition density matrix}
\label{app:gcmdens}

A workable expression for the transition one-body density matrix between two PGCM states is obtained and eventually reduced to the particular case of present interest.

\subsection{Inputs}

\subsubsection{Bogoliubov state}

The deformed Bogoliubov state $| \Phi(q) \rangle$ characterized by the collective deformation $q$ is a vacuum for the set of quasi-particle operators~\cite{RiSc80}
\begin{subequations}
\label{eq:qpbasis}
\begin{align}
\beta^{\dagger}_k(q) &\equiv \sum_a U^{ka}(q) c^{\dagger}_{a} + V_{a}^k(q) c_{a} \, , \\
\beta_k(q) &\equiv \sum_a U^{*}_{ka}(q) c_{a} + V^{*a}_{k}(q) c^{\dagger}_{a} \, .
\end{align}%
\label{eq:bogotrafo}%
\end{subequations}
Equation~\eqref{eq:qpbasis} defines the unitary Bogoliubov transformation\footnote{The covariant indices notation used in this document is extended to \(U\) and \(V\) matrices, such that creator (annihilator) indices are lowered (raised) using complex conjugation, e.g. $U^{*}_{ka}(q) = (U^{ka}(q))^*$.}
that is inverted according to
\begin{subequations}
\label{eq:bogorev}
\begin{align}
c^{\dagger}_a &= \sum_k U^{*ka}(q) \beta^\dag_{k} + V_{a}^{k}(q) \beta_k \, , \\
c_a &= \sum_k U_{ka}(q) \beta_k + V^{*a}_{k}(q) \beta^{\dagger}_{k} \, .
\end{align}%
\end{subequations}

\subsubsection{PGCM state}

Given a set of Bogoliubov states $\{ \ket{\Phi(q)}\}$ differing by the value of the collective deformation parameter $q$, a PGCM state reads as
\begin{align}
		\ket{\mathrm \Psi^{\sigma}_{\mu} }
		&\equiv
		\int\ud q f^{\sigma}_{\mu}(q) P^{\sigma} \ket{\Phi(q)} \, . \label{PGCMstate}
\end{align}
While $\mu$ denotes a principal quantum number, \(\sigma\equiv(\text{J} \text{M} \Pi \text{N} \text{Z})\) collects the set of symmetry quantum numbers labelling the many-body state, i.e. the angular momentum J and its projection M, the parity $\Pi=\pm 1$ as well as neutron N and proton Z numbers. The operator
\begin{align}
P^{\sigma} &\equiv P^{\text{J}}_{\text{M}} P^{\text{N}} P^{\text{Z}} P^{\Pi} \, \label{collecprojector}
\end{align}
collects the projectors on good symmetry quantum numbers
\begin{subequations}
\label{projector}
\begin{align}
P^{\text{J}}_{\text{M}} &\equiv \sum_K g_K(q)  P^{\text{J}}_{\text{MK}}  \nonumber \label{projectorJ} \\
&\equiv \sum_K g_K(q)   \frac{2J+1}{16\pi^2}\int_{[0,4\pi]\times[0,\pi]\times[0,2\pi]} \hspace{-2.1cm}\ud \Omega \, D^{\text{J}*}_{\text{MK}}(\Omega) R_{\vec{J}}(\Omega)   \, ,  \\
P^{\text{N}}  &\equiv  \frac{1}{2\pi}\int_{0}^{2\pi} \ud \varphi_{N} e^{i\varphi_{n} \text{N}} R_{N}(\varphi_{n}) \, , \label{projectorN} \\
P^{\text{Z}}  &\equiv   \frac{1}{2\pi}\int_{0}^{2\pi} \ud \varphi_{Z} e^{i\varphi_{p} \text{Z}} R_{Z}(\varphi_{p})  \, , \label{projectorZ} \\
P^{\Pi}  &\equiv   \frac{1}{2} \sum_{\varphi_\pi=0,\pi} e^{\frac{i}{2}(1-\Pi)\varphi_\pi} \Pi(\varphi_\pi)\, , \label{projectorPi} 
\end{align}
\end{subequations}
such that $\ket{\mathrm \Psi^{\sigma}_{\mu} }$ is an eigenstate of $J^2$, $J_z$, $N$, $Z$ and $\Pi(\pi)$, where the latter denotes the parity operator. The unknown coefficients\footnote{By definition the coefficients $\tilde{f}^{\sigma K}_{\mu q}$ are such that the set of PGCM states $\{\ket{\mathrm \Psi^{\sigma}_{\mu}} ; \mu = 1, 2,\ldots\}$ emerging from a calculation are ortho-normalized.} $\tilde{f}^{\sigma K}_{\mu q}\equiv f^{\sigma}_{\mu}(q)g_K(q)$ are typically obtained by solving Hill-Wheeler-Griffin's equation~\cite{RiSc80}. In Eq.~\eqref{projector}, $\Omega\equiv(\alpha,\beta,\gamma)$,  $\varphi_\pi$ and $\varphi_{n}$ ($\varphi_{p}$) denote Euler, parity and neutron- (proton-) gauge angles, respectively. The rotation operators are given by
\begin{subequations}
\begin{align}
R_{\vec{J}}(\Omega) &\equiv e^{-\imath\alpha J_z} e^{-\imath\beta J_y} e^{-\imath\gamma J_z} \, , \\
R_{N}(\varphi_{n}) &\equiv e^{-i\varphi_{n} N} \, , \\
R_{Z}(\varphi_{p}) &\equiv e^{-i\varphi_{p} Z}  \, , \\
\Pi(\varphi_\pi) &\equiv e^{-i\varphi_{\pi} F} \, ,
\end{align}
\end{subequations}
with the one-body operator  
\begin{align}
F & \equiv \sum_{ab} f_{ab} c^{\dagger}_a c_b
\end{align}
defined through its matrix elements~\cite{egido91a} 
\begin{align}
f_{ab} & \equiv  \frac{1}{2}\left(1- \pi_a \right) \delta_{ab} \, ,
\end{align}
where $\pi_a$ denotes the parity of one-body basis states that are presently assumed to carry a good parity. Last but not least, $D^{\text{J}}_{\text{MK}}(\Omega)\equiv \bra{\text{JM}} R_{\vec{J}}(\Omega) \ket{\text{JK}}$ defines Wigner D-matrices, where $| \text{JM} \rangle$ denotes a generic eigenstate of $J^2$ and $J_z$.

\subsubsection{Off-diagonal one-body density matrix}

Given a state $\ket{\Phi(q)}$ defined through the Bogoliubov transformation $(U(q),V(q))$ and the common index 
\begin{align}
\theta\equiv(\Omega, \varphi_n,\varphi_p, \varphi_\pi)
\end{align}
encompassing all rotation angles, the state obtained through multiple rotations
\begin{align}
\ket{\Phi(q,\theta)} &\equiv R_{\vec{J}}(\Omega)R_{N}(\varphi_{n})R_{Z}(\varphi_{p})\Pi(\varphi_\pi) \ket{\Phi(q)}  \, , \label{rotatedHFB} 
\end{align}
is also a Bogoliubov state whose Bogoliubov transformation  $(U(q,\theta),V(q,\theta))$ can be obtained from  $(U(q),V(q))$ and from the characteristics of the rotation operators~\cite{RiSc80,hara79a}. 

A crucial quantity in terms of which the final results will be expressed is the so-called {\it off-diagonal} one-body density matrix
\begin{align}
\rho(q';q, \theta)^{b}_{a} &\equiv \frac{\bra{\Phi(q')} c_a^\dag c_b  \ket{\Phi(q,\theta)}}{\langle \Phi(q') | \Phi(q,\theta) \rangle}\, , \label{offdiagodensmat} 
\end{align}
which involves two different Bogoliubov states and, as such, can be computed explicitly from the sole knowledge of $(U(q'),V(q'))$ and $(U(q,\theta),V(q,\theta))$~\cite{RiSc80,hara79a}.

\subsection{Definition}

Considering two PGCM states, the one-body {\it transition} density matrix can now be defined through\footnote{In the present derivation, the density matrix is restricted to be diagonal in the isospin quantum number, i.e. single-particle states $a$ and $b$ carry the same isospin projection quantum number.}
\begin{align}
\label{eq:gcm_density}
{\mathbf\rho^{\sigma_f\sigma_i}_{\mu_f\mu_i}}^b_a
  \equiv & \bra{\mathrm\Psi^{\sigma_f}_{\mu_f} }  c_a^\dag c_b \ket{\mathrm\Psi^{\sigma_i}_{\mu_i} } \nonumber \\
  =&
  \int \ud q_f\ud q_i  f_{\mu_f}^{\sigma_f*}(q_f) f_{\mu_i}^{\sigma_i}(q_i) \nonumber \\
	&\,\,\, \times
  \bra {\Phi(q_f)} P^{\sigma_f \dagger} c_a^\dag c_b P^{\sigma_i} \ket{\Phi(q_i)} \nonumber \\
  =&
  \int \ud q_f\ud q_i\sum_{K_fK_i}
    \tilde{f}_{\mu_fq_f}^{\sigma_fK_f*} \tilde{f}_{\mu_iq_i}^{\sigma_iK_i} \delta_{(\Pi_f\pi_a\pi_b)\Pi_i}	\nonumber\\
	&\,\,\, \times \delta_{\text{N}_f\text{N}_i}\delta_{\text{Z}_f\text{Z}_i} \,	{\rho_{q_fq_i}^{\text{J}_f \text{M}_f \text{K}_f\sigma_i\text{K}_i}}^b_a \, , 
\end{align}
where
\begin{equation}
\label{eq:trans_dens}
	{\rho_{q_fq_i}^{\text{J}_f \text{M}_f \text{K}_f\sigma_i\text{K}_i}}^b_a \equiv
	\bra{\Phi(q_f)} P^{\text{J}_f\dagger}_{\text{M}_f\text{K}_f} c_a^\dag c_b  P^{\text{J}_i}_{\text{M}_i\text{K}_i} P^{\text{N}_i} P^{\text{Z}_i} P^{\Pi_i} \ket{\Phi(q_i)} \, ,
\end{equation}
and where the action of $P^{\text{N}_f} P^{\text{Z}_f} P^{\Pi_f}$ was easily resolved.

\subsection{Simplified expressions}

\subsubsection{Expanding the projectors}

In order to evaluate this matrix element, the left angular-momentum projector is expanded according to Eq.~\eqref{projectorJ} such that Eq.~\eqref{eq:trans_dens} is rewritten as
\begin{align}
\label{eq:rho_projx}
	{\rho_{q_fq_i}^{\text{J}_f \text{M}_f \text{K}_f\sigma_i\text{K}_i}}^b_a
  &=\nonumber
	\frac{2\text{J}_f+1}{16\pi^2}
	\int \ud\Omega
	\sum_{M'} D^{\text{J}_f}_{\text{M}_f\text{K}_f}(\Omega)D^{\text{J}_i*}_{\text{M}_i\text{M}'}(\Omega)
	\nonumber\\ & \hspace{-.8cm} \times
  \bra{\Phi(q_f)} c_a^\dag \left[\Omega\right] c_b\left[\Omega\right]
   P^{\text{J}_i}_{\text{M}'\text{K}_i} P^{\text{N}_i} P^{\text{Z}_i} P^{\Pi_i}  \ket{\Phi(q_i)},
\end{align}
where the rotated creation and annihilation operators are defined as
\begin{subequations}
\begin{align}
  c_a^\dag \left[\Omega\right] &\equiv R_{\vec{J}}(\Omega)^\dag c_a^\dag  R_{\vec{J}}(\Omega) \, , \\
  c_b \left[\Omega\right] &\equiv R_{\vec{J}}(\Omega)^\dag c_b  R_{\vec{J}}(\Omega) \, ,
\end{align}
\end{subequations}
and where the identity
\begin{equation}
R_{\vec{J}}(\Omega)^\dag P^{\text{J}_i}_{\text{M}_i\text{K}_i} = \sum_{M'} D^{\text{J}_i*}_{\text{M}_i\text{M}'}(\Omega) P^{\text{J}_i}_{\text{M}'\text{K}_i} \, ,
\end{equation}
has been used.

\subsubsection{Spherical one-body basis}
	
In case one-body basis states carry spherical indices
(\(a\equiv n_a,j_a,m_a,\pi_a,q_a\equiv\alpha_a,j_a,m_a\)), the operators  \(c^\dag_{\alpha_aj_am_a}\) and \((-1)^{m_a-j_a}c_{\alpha_aj_a-m_a}\) 
 transform like the \(m_a^{th}\) component of a rank-\(j_a\) spherical tensor under the action of
\(SU(2)\), which leads to 
\begin{subequations}
\label{eq:rot_creator2}
\begin{align}
  c_{\alpha_aj_am_a}^\dag \left[\Omega\right] &\equiv 
  \sum_{m} D_{m_am}^{j_a*}(\Omega)c_{\alpha_aj_am}^\dag \, , \\
  c_{\alpha_bj_bm_b} \left[\Omega\right] &\equiv \sum_{m}
  D_{m_bm}^{j_b}(\Omega)c_{\alpha_bj_bm} \, .
\end{align}
\end{subequations}
Consequently, the transition density matrix can be simplified as
\begin{align}
	&
{\rho_{q_fq_i}^{\text{J}_f \text{M}_f \text{K}_f\sigma_i\text{K}_i}}^b_a = \nonumber
  \frac{2\text{J}_f+1}{16\pi^2}\sum_{mm'}
  \sum_{M'}
	\nonumber\\&\times
	\left(
  \int\ud \Omega{{D}^{\text{J}_f}_{\text{M}_f\text{K}_f}}(\Omega)D^{\text{J}_i*}_{\text{M}_i \text{M}'}(\Omega)
  {D}^{j_b}_{m_bm}(\Omega)
  {D}^{j_a*}_{m_am'}(\Omega)
	\right)
	\nonumber\\
  &\times
  \bra{\Phi(q_f)}  c_{\alpha_aj_am'}^\dag c_{\alpha_bj_bm}
	P^{\text{J}_i}_{\text{M}'\text{K}_i} P^{\text{N}_i} P^{\text{Z}_i} P^{\Pi_i} \ket{\Phi(q_i)}.\label{eq:rho_proj_2}
\end{align}
The integral over Wigner-D matrices is performed analytically and generates
a sum over Clebsch-Gordan coefficients and 3j-symbols according to
\begin{align}
	&\frac{1}{16\pi^2}\int\ud \Omega{{D}^{\text{J}_f}_{\text{M}_f\text{K}_f}}(\Omega)D^{\text{J}_i*}_{\text{M}_i \text{M}'}(\Omega)
  {D}^{j_b}_{m_bm}(\Omega)
  {D}^{j_a*}_{m_am'}(\Omega)
	\nonumber\\
	&=
  \sum_{\lambda=\mathrm{max}(|J_i-J_f|, |j_a-j_b|)}^{\mathrm{min}(J_i+J_f,
  j_a+j_b)}
  (-1)^{M_f-K_f+J_f-J_i+j_b-j_a}\nonumber\\
  &\nonumber
  \,\,\, \times \threeJ{J_f}{-M_f}{J_i}{M_i}{\lambda}{M_f-M_i}
  \threeJ{j_b}{m_b}{j_a}{-m_a}{\lambda}{M_f-M_i}
  \\&
  \,\,\, \times (-1)^{m'-m_a}
  \cg{J_f}{-K_f}{J_i}{(m-m'+K_f)}{\lambda}{(m-m')}
  \cg{j_b}{m}{j_a}{-m'}{\lambda}{(m-m')}.
	\label{eq:wigner_fact}
\end{align}
The remaining matrix element in Eq.~\eqref{eq:rho_proj_2} is easily obtained in
terms of the off-diagonal one-body density matrix defined through Eq.~\eqref{offdiagodensmat}
\begin{align}
	&
	\bra{\Phi(q_f)}  c_{\alpha_aj_am'}^\dag c_{\alpha_bj_bm}
	P^{\text{J}_i}_{\text{M}'\text{K}_i} P^{\text{N}_i} P^{\text{Z}_i} P^{\Pi_i} \ket{\Phi(q_i)}
	\nonumber\\
	&=
	\frac{2\text{J}_i+1}{16\pi^2}\frac{1}{(2\pi)^2}\frac{1}{2}
	\int \ud \theta D^{\text{J}_i*}_{\text{M}'\text{K}_i}(\Omega) e^{i\varphi_n \text{N}_i} e^{i\varphi_p \text{Z}_i}
	\nonumber\\
	&
	\,\,\,\,\,\, \times  e^{\frac{i}{2}(1-\Pi)\varphi_\pi} \rho(q_f;q_i, \theta)^{\alpha_b j_b m}_{\alpha_aj_a m'} \, \langle \Phi(q_f) | \Phi(q_i,\theta) \rangle
	\, ,
\end{align}
knowing that the overlap $\langle \Phi(q_f) | \Phi(q_i,\theta) \rangle$ between two arbitrary non-orthogonal Bogoliubov states can be computed in several ways~\cite{Robledo:2009yd,Bally:2017nom}.

\subsubsection{Special case of present interest}

One is presently interested in the one-body density matrix of a $\text{J}^{\Pi}=0^{+}$ state. In the above set of equations, it corresponds to setting \(\text{J}_i=\text{J}_f=0\), \(\text{M}_i=\text{M}_f=0\) and \(\Pi_i=+1\). In this case, the triangular inequalities encoded in the 3j-symbols impose that
\begin{subequations}
	\begin{align}
		\lambda &= 0 \, , \\
		m_a &= m_b \, ,\\
		j_a &= j_b \, , \\
		m &= m' \, ,
	\end{align}
\end{subequations}
such that Eq.~\eqref{eq:wigner_fact} becomes
\begin{align}
	&\frac{1}{16\pi^2}\int\ud \Omega{{D}^{0}_{00}}(\Omega)D^{0*}_{00}(\Omega)
  {D}^{j_b}_{m_bm}(\Omega)
  {D}^{j_a*}_{m_am'}(\Omega)
	\nonumber\\&=
	\delta_{m_am_b}\delta_{j_aj_b}\delta_{mm'}
	\inv{2j_a+1} \, .
	\label{eq:wigner_fact0}
\end{align}
The fact that the initial and final states are the same and thus carry the same parity further requires that $\pi_a=\pi_b$. Eventually, Eq.~\eqref{eq:rho_proj_2} reduces to
\begin{align}
	{\rho_{q_fq_i}^{0^+\text{N}_i\text{Z}_i}}^b_a &\equiv
	\delta_{j_aj_b}\delta_{m_am_b}\delta_{\pi_a\pi_b}
	\frac{1}{16\pi^2}\frac{1}{(2\pi)^2}\frac{1}{2}\nonumber\\&\times
	\int \ud \theta  e^{i\varphi_n\text{N}_i} e^{i\varphi_p\text{Z}_i}
	\langle \Phi(q_f) | \Phi(q_i,\theta) \rangle
	\nonumber\\
	&\times
	\inv{2j_a+1}
	\sum_{m=-j_a}^{j_a}
 \rho(q_f;q_i, \theta)^{\alpha_b j_b m}_{\alpha_aj_a m} \, .\label{eq:rho_sph}
\end{align}
The diagonal character of the one-body density matrix in $(j,m)$ and its independence on \(m\) is made clear in Eq.~\eqref{eq:rho_sph} and ends the derivation.

\section{BMBPT transition density matrix}
\label{app_densmat_BMBPT}

The BMBPT(2) transition one-body density matrix is presently derived within the frame of a so-called {\it expectation-value} many-body scheme rather than within a {\it projective} one, i.e. it is computed directly through Eq.~\eqref{deflbodydensmat} for $l=1$ and $| \Theta \rangle \equiv | \Psi^{\text{BMBPT(2)}} \rangle $. The derivation can actually be performed within the larger frame of the Bogoliubov configuration-interaction (BCI) formalism such that BCI-like expansion coefficients are eventually obtained within the frame of BMBPT(2)~\cite{ripoc2019}. 

The present applications are eventually limited to standard MBPT, i.e. calculations are restricted to doubly closed-shell nuclei for which BMBPT reduces to MBPT on top a $J^{\Pi}=0^+$ Slater determinant.

\subsection{Bogoliubov algebra}

In methods based on a Bogoliubov vacuum $| \Phi(q) \rangle$, the grand potential\footnote{In practice two separate Lagrange multipliers $\mu_{\text{N}}$ and $\mu_{\text{Z}}$ are introduced to account for neutron and proton chemical potentials, such that the average neutron and proton numbers are conserved individually.},
\begin{align}
    \Omega \equiv H -\mu A \, ,
\end{align}
must be used rather than the Hamiltonian to control the average particle-number in the system. The many-body algebra is more conveniently worked out by normal-ordering all operators with respect to $| \Phi(q) \rangle$ and by expressing them in terms of quasi-particle operators. Limiting oneself to two-body operators\footnote{This is the case when working with two-body forces only or within the PNO2B or the presently developed 2B approximation.}, the grand potential is thus written as
\begin{align}
\Omega =& \Omega^{00}(q) + \Omega^{20}(q) + \Omega^{11}(q) +\Omega^{02}(q) \nonumber  \\
&+   \Omega^{40}(q) + \Omega^{31}(q) +\Omega^{22}(q) +\Omega^{13}(q) +\Omega^{04}(q) \, ,\nonumber
\end{align} 
where $\Omega^{ij}(q)$ denotes the normal-ordered component involving $i$ ($j$) quasi-particle creation (annihilation) operators, e.g., 
\begin{align}
\Omega^{31}(q) &\equiv \frac{1}{3!}\sum_{k_1 k_2 k_3 k_4}  \Omega^{k_1 k_2 k_3}_{k_4}(q)
   \beta^{\dagger}_{k_1}(q)\beta^{\dagger}_{k_2}(q)\beta^{\dagger}_{k_3}(q)\beta_{k_4}(q) \, , \nonumber 
\end{align} 
where the matrix elements are anti-symmetric with respect to the exchange of any pair of upper or lower indices. For more details about the normal ordering procedure, see Refs.~\cite{Duguet:2015yle,Tichai18BMBPT,Arthuis:2018yoo,Demol:2020mzd,Tichai2020review,frosini20a}.

\subsection{BCI state}

In BCI, many-body states are written as a CI-like expansion on top of the (deformed) Bogoliubov vacuum. Presently truncated to single and double excitations, the BCISD ansatz reads as
\begin{align}
  \ket {\Psi(q)} &\equiv
  \left(1 + 
  \sum_{k_1k_2} C^{k_1k_2}(q) \beta^\dag_{k_1}(q)\beta^\dag_{k_2}(q)\right.
   \nonumber \\
  &+
  \left.
  \sum_{k_1k_2k_3k_4} C^{k_1k_2k_3k_4}(q)
  \beta^\dag_{k_1}(q)\beta^\dag_{k_2}(q)\beta^\dag_{k_3}(q)\beta^\dag_{k_4}(q)
  \right) \nonumber \\
  &\times |\Phi(q)\rangle \,  , \label{BCIstate}
\end{align}
where the unknown coefficients, anti-symmetric with respect to the exchange of any pair of upper indices, can be obtained by diagonalization of $\Omega$ or via BMBPT.

\subsection{Expression in quasi-particle space}

\subsubsection{Definition}

Considering two different BCISD states $\ket{\Psi^i(q)}$ and $\ket{\Psi^f(q)}$, the four transition one-body density matrices defined in terms of quasi-particle operators are given by
\begin{subequations}
\label{defrhoqp}
  \begin{align}
    {\boldsymbol \rho^{fi}}_{k_1}^{k_2}(q) \equiv & 
    \frac{\bra{\Psi^f(q)} \beta^{\dag}_{k_1}(q) \beta_{k_2}(q) \ket{\Psi^i(q)}}
    {\sqrt{\braket{\Psi^f(q)}{\Psi^f(q)}\braket{\Psi^i(q)}{\Psi^i(q)}}} \, ,\\
    {\boldsymbol \kappa^{fi}}^{k_2k_1}(q) \equiv & 
    \frac{\bra{\Psi^f(q)} \beta_{k_1}(q) \beta_{k_2}(q) \ket{\Psi^i(q)}}
    {\sqrt{\braket{\Psi^f(q)}{\Psi^f(q)}\braket{\Psi^i(q)}{\Psi^i(q)}}} \, ,\\
    -{\boldsymbol \kappa^{*fi}}_{k_2k_1}(q) \equiv & 
    \frac{\bra{\Psi^f(q)} \beta^{\dag}_{k_1}(q) \beta^{\dag}_{k_2}(q) \ket{\Psi^i(q)}}
    {\sqrt{\braket{\Psi^f(q)}{\Psi^f(q)}\braket{\Psi^i(q)}{\Psi^i(q)}}} \, ,\\
    -{\boldsymbol \sigma^{*fi}}^{k_1}_{k_2}(q) \equiv & 
    \frac{\bra{\Psi^f(q)} \beta_{k_1}(q) \beta^{\dag}_{k_2}(q) \ket{\Psi^i(q)}}
    {\sqrt{\braket{\Psi^f(q)}{\Psi^f(q)}\braket{\Psi^i(q)}{\Psi^i(q)}}} \, ,
  \end{align}
\end{subequations}
among which the relations
\begin{subequations}
\label{relationdensities}
\begin{align}
  {\boldsymbol\kappa^{*fi}}_{k_2k_1}(q) &= \left({\boldsymbol\kappa^{if}}^{k_2k_1}(q)\right)^* \, ,\\
  {\boldsymbol{\sigma^{*}}^{fi}}^{k_1}_{k_2}(q) &= \left({\boldsymbol\rho^{if}}^{k_2}_{k_1}(q)\right)^*-1 \, ,
\end{align}
\end{subequations}
hold.

\subsubsection{Matrix elements}

Starting from Eqs.~\eqref{BCIstate}-\eqref{defrhoqp} and applying Wick's theorem, one obtains
\begin{align}
{\boldsymbol \rho^{fi}}_{k_1}^{k_2}(q) = & 
    \inv{\sqrt{\braket{\Psi^f(q)}{\Psi^f(q)}\braket{\Psi^i(q)}{\Psi^i(q)}}}
    \nonumber\\
    \times&
  \left[ \inv{2}\sum_{k_3} {C^{f*}}_{k_1k_3}(q){C^i}^{k_2k_3}(q)\right.
  \nonumber\\&+
  \left.
  \inv{12}\sum_{k_3k_4k_5} {C^{f*}}_{k_1k_3k_4k_5}(q) {C^i}^{k_2k_3k_4k_5}(q)
  \right] \, ,  \nonumber
\end{align}
and
\begin{align}
{\boldsymbol \kappa^{fi}}^{k_1k_2}(q) = & 
   \inv  {\sqrt{\braket{\Psi^f(q)}{\Psi^f(q)}\braket{\Psi^i(q)}{\Psi^i(q)}}}
   \nonumber\\
   \times&
   \left[
  \inv{2} {C^i}^{k_1k_2}(q)\right.
  \nonumber\\&+
  \left.
  \inv{4}\sum_{k_3k_4} {C^{f*}}_{k_3k_4}(q) {C^i}^{k_1k_2k_3k_4}(q)\right] \nonumber
.
\end{align}
The expressions of \({\boldsymbol \kappa^{fi}}^*\) and \({\boldsymbol{\sigma^{*}}^{fi}}\) are then deduced via Eq.~\eqref{relationdensities} whereas the norm entering the denominators of the transition one-body density matrices reads, e.g., as
\begin{align}
    \braket {\Psi^i(q)}{\Psi^i(q)} \equiv& 1 \nonumber\\
    &
    + \sum_{k_1k_2} {C^{i*}}_{k_1k_2}(q){C^{i}}^{k_1k_2}(q)
    \nonumber\\
    &
    + \sum_{k_1k_2k_3k_4} {C^{i*}}_{k_1k_2k_3k_4}(q){C^{i}}^{k_1k_2k_3k_4}(q) \, . \nonumber
\end{align}

\subsection{Expression in one-particle space}

Inserting the inverse Bogoliubov transformation (Eq.~\eqref{eq:bogorev}), the normal one-body density matrix expressed in terms of particle operators is obtained under the form
\begin{align}
  {\rho^{fi}}_a^b(q) \equiv & \frac{\bra{ \Psi^f (q)}  c_a^\dag c_b  \ket{\Psi^i(q)}}
  {\sqrt{\braket{\Psi^f(q)}{\Psi^f(q)}\braket{\Psi^i(q)}{\Psi^i(q)}}}
  \nonumber\\=&\sum_{k_1k_2}\left[
  U^{k_2b}(q){\boldsymbol \rho^{fi}}^{k_2}_{k_1} (q){U}_{k_1a}^{*} (q)\right.
  \nonumber\\&
   \phantom{\sum}
  - V^{*b}_{k_2}(q) {\boldsymbol \sigma^{*fi}}^{k_1}_{k_2} (q)V^{k_1}_{a}(q)
  \nonumber\\&
   \phantom{\sum}
  - V^{*b}_{k_2} (q) {\boldsymbol\kappa^{*fi}}_{k_2k_1}(q) U^{*}_{k_1a}(q)
  \nonumber\\&
  \left. \phantom{\sum}
  + U^{k_2b}(q)  {\boldsymbol\kappa^{fi}}^{k_2k_1} (q)V^{k_1}_{a}(q)\right]
\, . \label{transonebodydensBCI}
\end{align}

\subsection{One-body density matrix}

In the present work, one is interested in the case where \(\ket {\Psi^i(q)}=\ket {\Psi^f(q)}\equiv
\ket {\Psi(q)}\) such that Eq.~\eqref{transonebodydensBCI} reduces to
\begin{align}
 {\rho^{\Psi}}_a^b(q) =
 &\sum_{k_1k_2}\left[
  V^{*b}_{k_2}(q) V^{k_1}_{a}(q)\right.
  \nonumber\\&
   \phantom{\sum}
  +U^{k_2b}(q){\boldsymbol \rho^{\Psi}}^{k_2}_{k_1} (q){U}_{k_1a}^{*} (q)
  \nonumber\\&
   \phantom{\sum}
  - \left(V_{b}^{k_2}(q)  {\boldsymbol\rho^{\Psi}}^{k_2}_{k_1} (q)V_{k_1}^{*a}(q)\right)^*
  \nonumber\\&
   \phantom{\sum}
  - \left(V_{b}^{k_2} (q){\boldsymbol \kappa^{\Psi}}^{k_2k_1}(q) U^{*k_1a}(q)\right)^*
  \nonumber\\&
  \left. \phantom{\sum}
  + U^{k_2b}(q) {\boldsymbol \kappa^{\Psi}}^{k_2k_1} (q)V^{k_1}_{a}(q)\right]
\, .
\end{align}
When time-reversal symmetry is preserved, the one-body density matrix can be chosen
to be real such that the final expression reads as
\begin{align}
{\rho^{\Psi}}_a^b(q) =&
    {\rho^\Phi}_a^b(q) \nonumber \\
 &+ \sum_{k_1k_2}\left[
  U^{k_2b}(q){\boldsymbol \rho^{\Psi}}^{k_2}_{k_1} (q){U}_{k_1a}^{*} (q)\right.
  \nonumber\\&
   \phantom{\sum_{k_1k_2}}
  - V_{b}^{k_2}(q) {\boldsymbol \rho^{\Psi}}^{k_2}_{k_1} (q)V_{k_1}^{*a}(q)
  \nonumber\\&
   \phantom{\sum_{k_1k_2}}
  - V_{b}^{k_2} (q){\boldsymbol \kappa^{\Psi}}^{k_2k_1}(q) U^{*k_1a}(q)
  \nonumber\\&
  \left. \phantom{\sum_{k_1k_2}}
  + U^{k_2b}(q) {\boldsymbol \kappa^{\Psi}}^{k_2k_1}(q)V^{k_1}_{a}(q)\right]
\, ,
\end{align}
where ${\rho^\Phi}(q)$ denotes the one-body density matrix of the reference state $| \Phi(q) \rangle$ such that the additional terms relate to BCISD corrections on top of it.

\subsection{BMBPT coefficients}
\label{sec:bmbpt-bci}

The coefficients of the BCISD state obtained at first- and second-order in BMBPT are now explicitly provided~\cite{ripoc2019}. Given that the present application is limited to a $J^{\Pi}~=~0^+$ Slater determinant reference state  $| \Phi \rangle$, the resulting one-body density matrix ${\rho^\Psi}^b_a$ is actually proportional to $\delta_{j_aj_b}$ and relates to a many-body state that is an eigenvector of the particle-number operator.

\subsubsection{Partitioning}

To formulate BMBPT with respect to the Bogoliubov reference state $| \Phi(q) \rangle$, the grand potential is split into an unperturbed part $\Omega_{0}$ and a residual part $\Omega_1$~\cite{Duguet:2015yle,Tichai18BMBPT,Arthuis:2018yoo}
\begin{equation}
\label{split1}
\Omega = \Omega_{0}(q) + \Omega_{1}(q) \ ,
\end{equation}
such that
\begin{align*}
\Omega_{0}(q) &\equiv \Omega^{00}(q)+\bar{\Omega}^{11}(q) \ , \\
\Omega_{1}(q) &\equiv \Omega^{20}(q) + \breve{\Omega}^{11}(q) + \Omega^{02}(q) \nonumber \\
  &\phantom{\equiv } + \Omega^{40}(q) + \Omega^{31}(q) + \Omega^{22}(q) +  \Omega^{13}(q) + \Omega^{04}(q)  \ ,
 \label{e:perturbation}
\end{align*}
with $\breve{\Omega}^{11}(q)\equiv\Omega^{11}(q)- \bar{\Omega}^{11}(q)$ and where the normal-ordered one-body part of $\Omega_{0}(q)$ is diagonal, i.e.,
\begin{equation*}
\bar{\Omega}^{11}(q) \equiv \sum_{k} E_k(q) \beta^{\dagger}_k(q) \beta_k(q) \, , \label{onebodypiece}
\end{equation*}
with $E_k(q) > 0$ for all $k$.

\subsubsection{First-order correction}

At first order in BMBPT,  singles and doubles BCI-like coefficients read as
\begin{subequations}
\begin{align}
    {C^{(1)}}^{k_1k_2} (q)\equiv& -\frac{{\Omega}^{k_1k_2}(q)}{E_{k_1k_2}(q)} \, ,  \\
    {C^{(1)}}^{k_1k_2k_3k_4}(q) \equiv& -\frac{{\Omega}^{k_1k_2k_3k_4}(q)}{E_{k_1k_2k_3k_4}(q)} \, ,
\end{align}
\end{subequations}
where $E_{k_1 k_2 \ldots}(q) \equiv E_{k_1}(q) + E_{k_2}(q) +\ldots$.

\subsubsection{Second-order correction}

At second order in BMBPT, singles and doubles BCI-like coefficients read
\begin{subequations}
\begin{align}
    {C^{(2)}}^{k_1k_2} (q)\equiv& \inv 6 P(k_1/k_2) \sum_{k_3k_4k_5} \frac{{C^{(1)}}^{k_1k_4k_5k_3}(q){\Omega}^{k_2}_{k_4k_5k_3}(q)}{E_{k_1k_2}(q)}
    \nonumber\\&
    +\inv2 \sum_{k_3k_4} \frac{{C^{(1)}}^{k_1k_2k_3k_4}(q){\Omega}_{k_3k_4}(q)}{E_{k_1k_2}(q)}
    \nonumber\\&
    +\inv2 \sum_{k_3k_4} \frac{{C^{(1)}}^{k_3k_4}(q){\Omega}_{k_3k_4}^{k_1k_2}(q)}{E_{k_1k_2}(q)}
    \nonumber\\&
    +P(k_1 / k_2) \sum_{k_3} \frac{{C^{(1)}}^{k_1k_3}(q)
    {\breve\Omega}^{k_2}_{k_3}(q)}{E_{k_1k_2}(q)} \, , 
    \\
    {C^{(2)}}^{k_1k_2k_3k_4}(q) \equiv& 
    \inv2 P(k_1k_2 / k_3k_4)
    \sum_{k_5k_6} \frac{{C^{(1)}}^{k_1k_2k_5k_6}(q){\Omega}_{k_5k_6}^{k_3k_4}(q)}
    {E_{k_1k_2k_3k_4}(q)}
    \nonumber\\&
    +P(k_4 / k_1k_2k_3)
    \sum_{k_5} \frac{{C^{(1)}}^{k_1k_2k_3k_5}(q)
    {\breve\Omega}^{k_4}_{k_5}(q)}{E_{k_1k_2k_3k_4}(q)}
    \nonumber\\&
    +P(k_1 / k_2k_3k_4)
    \sum_{k_5} \frac{{C^{(1)}}^{k_1k_5}(q)
    {\Omega}^{k_2k_3k_4}_{k_5}(q)}{E_{k_1k_2k_3k_4}(q)}
    \nonumber\\&
    + P(k_1k_2 / k_3k_4) 
    \left[{C^{(1)}}^{k_1k_2}(q)
    {C^{(1)}}^{k_3k_4}(q)\right] \, ,
\end{align}
\end{subequations}
where anti-symmetrizing operators \(P(\cdots/\cdots)\) are expressed in terms of the of the permutation operator as
\begin{subequations}
\begin{align}
    P(k_1 / k_2)  \equiv& 1 - P_{k_1k_2}\, ,\\
    P(k_1 / k_2k_3k_4)  \equiv &1 - P_{k_1k_2} - P_{k_1k_3} - P_{k_1k_4}\, ,\\
    P(k_1k_2 / k_3k_4) \equiv  &1 - P_{k_1k_3} - P_{k_1k_4} - P_{k_2k_3} - P_{k_2k_4} 
    \nonumber\\&
    + P_{k_1k_3}P_{k_2k_4}\, .
\end{align}
\end{subequations}

\section{Error-function sampling}
\label{randomdensmat}

\subsection{Error function}
\label{randomerrf}

The error function introduced in Eq.~\eqref{diffE} can be interpreted as the first-order correction to the energy due to the perturbation $\delta H[\rho] \equiv H^{2B}[\rho]-H$.
Knowing that for a generic $n$-body operator $O^{(nn)}$
\begin{align}
\frac{\langle \Psi |  O^{(nn)}  | \Psi \rangle}{\langle \Psi  | \Psi \rangle}  &= \frac{1}{n!} \frac{1}{n!}  
	\sum_{\substack{a_1\cdots a_i\\b_1\cdots b_i}}
	o^{a_1\cdots a_n}_{b_1\cdots b_n} \, 
	\frac{\langle \Psi |  A^{a_1\cdots a_n}_{b_1\cdots b_n}  | \Psi \rangle}{\langle \Psi  | \Psi \rangle}  \nonumber \\
	&= \left(\frac{1}{n!}\right)^2 o^{(n)} \! \cdot \! \rho^{(n)\Psi}\, ,
\end{align}
Eq.~\eqref{diffE} can be written as
\begin{align}
\Delta E^{2B}_{\Psi}[\rho]
=& -\left(\frac{1}{3!}\right)^2 w^{(3)} \! \cdot \! \rho^{(3)\Psi}  \nonumber \\
& +\left(\frac{1}{2!}\right)^2 \left(w^{(3)} \! \cdot \! \rho^{(2)\Psi}\right) \! \cdot \! \rho  \nonumber \\
& - \frac{1}{2!} \left(w^{(3)} \! \cdot \! \rho^{\Psi}\right) \! \cdot \! \rho^{\otimes(2)}   \nonumber \\
& + \frac{1}{3!} w^{(3)} \! \cdot \! \rho^{\otimes(3)} \nonumber \\
=&  -\left(\frac{1}{3!}\right)^2 w^{(3)} \! \cdot \! \lambda^{(3)\Psi} \nonumber  \\
&  +\left(\frac{1}{2!}\right)^2 \left(w^{(3)} \! \cdot \! \lambda^{(2)\Psi}\right) \! \cdot \! \left(\rho-\rho^{\Psi}\right) \nonumber  \\
& + \frac{1}{3!} w^{(3)} \! \cdot \!  \left(\rho-\rho^{\Psi}\right)^{\otimes(3)} \label{diffEexplicit}
\end{align}
where $\lambda^{(n)\Psi}$ denotes the {\it irreducible} $n$-body density matrix (or cumulants)~\cite{mazziotti98a,mazziotti98b} that, for $n\geq 2$, encodes genuine $n$-body correlations in $| \Psi \rangle$. Whenever $| \Psi \rangle$ reduces to a Slater determinant, one has $\lambda^{(n)\Psi}=0$ for $n\geq 2$. 

Inspecting Eq.~\eqref{diffEexplicit}, one observes that the error
\begin{enumerate}
\item only depends on the three-nucleon interaction and involves up to the 3-body (irreducible) density matrix of $| \Psi \rangle$,
\item can be written as a cubic polynomial in the variable $\rho-\rho^{\Psi}$ whenever involving irreducible density matrices of $| \Psi \rangle$,
\item is zero (and thus minimal in absolute value) for $\rho=\rho^{\Psi}$ whenever $| \Psi \rangle$ contains {\it at most} genuine 2-body correlations, which is notably the case whenever $| \Psi \rangle$ reduces to a Slater determinant,
\item is in general non-zero for $\rho=\rho^{\Psi}$ and measures in that case genuine 3-body correlations encoded into $\lambda^{(3)\Psi}$. While the error does not minimize for $\rho=\rho^{\Psi}$ in general, the fact that $\rho=\rho^{\Psi}$ is the optimal solution whenever $| \Psi \rangle$ contains at most genuine 2-body correlations indicates that the optimal $\rho$ cannot be very different from $\rho^{\Psi}$ whenever $| \Psi \rangle$ is a weakly correlated $J^{\Pi}=0^+$ state.
\end{enumerate}

\subsection{Random one-body density matrices}
\label{randomrho}

The goal is to sample the error function $\Delta E^{2B}_{\Psi}[\rho]$ within the space of one-body density matrices $\{\rho\}$ associated with $J^{\Pi}=0^+$ states\footnote{Strickly speaking, and as the procedure detailed in Sec.~\ref{novelapprox} makes clear, the one-body density matrix employed in the construction of $H^{2B}[\rho]$ does not have to be actually related to a many-body state, i.e. it does not have to be {\it N-representable}. It is at least mandatory to use trial one-body density matrices carrying the fingerprints of the symmetry constraints associated with a true state in order for $H^{2B}[\rho]$ to display appropriate symmetries, which translates into Eqs.~\eqref{constraintsrhoA}, \eqref{constraintsrhoB} and \eqref{constraintsrhoC} as far as hermiticity, angular momentum, parity and particle number are concerned.}. Thus, a large set of density matrices $\{\rho^{\text{sRd}}\}$ is randomly generated in the sHO basis  $\{| a \rangle ; a \equiv \alpha_a j_a m_a \}$ under the constraints that
\begin{subequations}
\label{constraintsrho}
\begin{align}
&\rho^{b}_{a} = (\rho^{a}_{b})^{\ast} \, , \label{constraintsrhoA} \\
&\rho^{b}_{a} \equiv  \delta_{j_aj_b}\delta_{m_am_b} \delta_{\pi_a\pi_b} \varrho^{\alpha_b}_{\alpha_a} \, , \label{constraintsrhoB} \\
&\text{Tr}\rho = \mathbbm{1}^{(1)}  \! \cdot \! \rho = \text{A}  \, , \label{constraintsrhoC} \\
&0\leq \text{diag}(\rho)^{b}_{b} \leq 1 \, , \, \forall b  \, , \label{constraintsrhoD}
\end{align}
\end{subequations}
where $\mathbbm{1}^{(1)}$ denotes the identity operator on ${\cal H}_1$ and where $\text{diag}(\rho)$ gathers the eigenvalues. More specifically, the procedure works as follows
\begin{enumerate}
\item choice of a reference one-body density matrix $\rho^{\text{ref}}$,
\item diagonalization of $\rho^{\text{ref}}$
\begin{equation}
\rho^{\text{ref}} \equiv L^T \text{diag(r)} \,  L \, ,
\end{equation}
where $L$ denotes an orthogonal matrix.
\item choice of two coefficients \(\alpha_d,\ \alpha_o\) characterizing the amplitude of the random perturbation to be performed next.
\item sampling of a random perturbation \(\delta r\)  of the diagonal matrix elements of \(r\) verifying
\begin{subequations}
\begin{align}
    \sum_a \delta r_a &= 0\,, \\
    r_a +\delta r_a &\in [0,1] \, ,  \, \forall a\,,\\
    |\delta r_a| &\le \alpha_d \,  ,  \, \forall a\,.
\end{align}
\end{subequations}
\item sampling of a random skew-symmetric matrix \(\delta l\) with all upper-diagonal coefficients chosen via a normal distribution \(\mathcal N(0,1)\).
\item exponentiation of \(\delta l\) to obtain an orthogonal matrix
\begin{equation}
    \delta L \equiv \exp\left[\alpha_o \delta l\right] \, .
\end{equation}
\item computation of the random neighbour of \(\rho^{\text{ref}}\)
\begin{equation}
    \rho^{sRd} \equiv \left(L \delta L\right)^T \text{diag}(r+\delta r) \, L \delta L \, .
\end{equation}
\end{enumerate}
Although the sampling is not uniform, all densities with the required properties can in principle be obtained via this method.

\section{Spherical Hartree-Fock field}
\label{HFfield}

Given a test one-body density matrix $\rho$ and considering that the many-body state of interest $| \Psi \rangle$ is a Slater determinant, the one-body Hamiltonian at play in the HF minimization problem based on $H^{2B}[\rho]$ is, given Eq.~\eqref{diffEexplicit},
\begin{align}
\left[h^{\text{HF}(2B)}[\rho^{\Psi};\rho]\right]^{a}_{b}\equiv & \frac{\delta\langle \Psi |  H^{2B}[\rho]  | \Psi \rangle}{\delta [\rho^{\Psi}]^{b}_{a}}  \label{aqHFfield}   \\
=&\frac{\delta \langle \Psi |  H  | \Psi \rangle}{\delta [\rho^{\Psi}]^{b}_{a}} + \frac{\delta \Delta E^{2B}_{\Psi}[\rho]}{\delta [\rho^{\Psi}]^{b}_{a}} \nonumber \\
 =& \left[h^{\text{HF}}[\rho^{\Psi}]\right]^{a}_{b} - \frac{1}{2!} \left[w^{(3)} \! \cdot \!  \left(\rho^{\Psi}\!-\!\rho\right)^{\otimes(2)}\right]^{a}_{b} \, . \nonumber
\end{align}
In Eq.~\eqref{aqHFfield}, $h^{\text{HF}}[\rho^{\Psi}]$ denotes the one-body HF Hamiltonian  obtained from the full $H$ whose associated solution is $\rho^{\text{sHF}}$. Equation~\eqref{aqHFfield} allows one to appreciate the implications of using $H^{2B}[\rho]$ at the sHF level, i.e. in the mean-field calculation of a doubly closed-shell nucleus such as $^{16}$O and $^{40}$Ca. One observes that
\begin{itemize}
\item in general, the use of $H^{2B}[\rho]$ generates an additional term on top of $h^{\text{HF}}[\rho^{\Psi}]$, eventually leading to $\rho^{\Psi} \neq  \rho^{\text{sHF}}$ and $\Delta E^{2B}_{\Psi}[\rho^{\text{sHF}}]\neq 0$ at convergence,
\item even when using $H^{2B}[\rho^{\text{sHF}}]$, the additional term differs from zero such that $\rho^{\Psi}\neq  \rho^{\text{sHF}}$ and $\Delta E^{2B}_{\Psi}[\rho^{\text{sHF}}]\neq 0$ at convergence,
\item only if one were to set $\rho=\rho^{\Psi}$ in $H^{2B}[\rho]$ throughout the iterative procedure, thus modifying the approximate Hamiltonian along the way, would the correction term  vanish in Eq.~\eqref{aqHFfield} at convergence and the sHF solution based on $H^{2B}[\rho]$ be the same as the one obtained from $H$. This particular case is equivalent to constructing $H^{2B}[\rho]$ through Wick's theorem with respect to the self-consistent sHF Slater determinant itself and is thus identical to the NO2B procedure, which indeed does not lead to any approximation at the HF level.
\end{itemize}

\section{Measure of the systematic deviations}
\label{sec:cost-function}

In order to assess quantitatively the errors induced by the approximation and compare the different effective interactions, measures of the average deviation between the results obtained with $H^{2B}[\rho]$ and $H$ are introduced for each method, i.e. 
\begin{subequations}
\begin{align}
    r_{\text{HFB}} &\equiv \inv{n_{\text{data}}}\sum_{n_{\text{nuclei}}}\left|\frac{E_{\text{HFB}}[\rho] - E_{\text{HFB}}}{ E_{\text{HFB}}}\right| ,\label{eq:av-hfb} \\
    r_{\text{PHFB}} &\equiv \inv{n_{\text{data}}}\sum_{n_{\text{nuclei}}}\sum_{J}\left|\frac{E^{J^\Pi}_{\text{PHFB}}[\rho] - E^{J^\Pi}_{\text{PHFB}}}{ E^{J^\Pi}_{\text{PHFB}}}\right| ,\label{eq:av-phfb} \\
    r_{\text{BMBPT}} &\equiv  \inv{n_{\text{data}}}\sum_{n_{\text{nuclei}}} \left|\frac{E_{\text{BMBPT}}[\rho] - E_{\text{BMBPT}}^{\text{PNO2B}}}{ E_{\text{BMBPT}}^{\text{PNO2B}}}\right|,\label{eq:av-bmbpt}\\
    r_{\text{PGCM}} &\equiv  \inv{n_{\text{data}}}\sum_{n_{\text{nuclei}}} \sum_J \sum_O \left|\frac{O^{J^\Pi}_{\text{PGCM}}[\rho] - O^{J^\Pi}_{\text{PGCM}}}{ O^{J^\Pi}_{\text{PGCM}}}\right|,\label{eq:av-gcm}
\end{align}
\end{subequations}
where in practice states up to \(J=6\) are taken into account when applicable and where $O$ denotes any observable computed within the PGCM formalism (energy, electric and magnetic moments, transitions and radii). In each case, $n_{\text{data}}$ denotes the number of terms in the sum(s).

The deviation on PHFB excitation energies is given by the formula
\begin{equation}
    r_{\text{PHFB-S}} \equiv\inv{n_{\text{data}}}\sum_{n_{\text{nuclei}}} \sum_{J} \left|\frac{\delta E^{J^\Pi}_{\text{PHFB}}[\rho] - \delta E^{J^\Pi}_{\text{PHFB}}}{
    \delta E^{J^\Pi}_{\text{PHFB}}}\right|.\label{eq:av-phfb-spec}
\end{equation}

\end{appendices}

\bibliography{bibliography.bib}

\end{document}